\documentclass[12pt]{article}
\usepackage{latexsym,amsmath,amsfonts,amssymb}
\usepackage[latin1]{inputenc}
\usepackage[american]{babel}
\usepackage[dvips]{graphicx}
\usepackage{bbm}
\newcommand{\be}{\begin{equation}}
\newcommand{\ee}{\end{equation}}
\newcommand{\beq}{\begin{equation}}
\newcommand{\eeq}{\end{equation}}
\newcommand{\bea}{\begin{eqnarray}}
\newcommand{\eea}{\end{eqnarray}}
\newcommand{\nn}{\nonumber}

\newcommand{\ba}{\begin{eqnarray}}
\newcommand{\ea}{\end{eqnarray}}
\begin{document}
\baselineskip=15.5pt
\pagestyle{plain}
\setcounter{page}{1}


\def\del{{\partial}}
\def\vev#1{\left\langle #1 \right\rangle}
\def\cn{{\cal N}}
\def\co{{\cal O}}
\def\IC{{\mathbb C}}
\def\IR{{\mathbb R}}
\def\IZ{{\mathbb Z}}
\def\RP{{\bf RP}}
\def\CP{{\bf CP}}
\def\Poincare{{Poincar\'e }}
\def\tr{{\rm tr}}
\def\tp{{\tilde \Phi}}

\def\TL{\hfil$\displaystyle{##}$}
\def\TR{$\displaystyle{{}##}$\hfil}
\def\TC{\hfil$\displaystyle{##}$\hfil}
\def\TT{\hbox{##}}
\def\HLINE{\noalign{\vskip1\jot}\hline\noalign{\vskip1\jot}}
\def\seqalign#1#2{\vcenter{\openup1\jot
  \halign{\strut #1\cr #2 \cr}}}
\def\lbldef#1#2{\expandafter\gdef\csname #1\endcsname {#2}}
\def\eqn#1#2{\lbldef{#1}{(\ref{#1})}%
\begin{equation} #2 \label{#1} \end{equation}}
\def\eqalign#1{\vcenter{\openup1\jot
    \halign{\strut\span\TL & \span\TR\cr #1 \cr
   }}}
\def\eno#1{(\ref{#1})}
\def\href#1#2{#2}
\def\half{{1 \over 2}}

\def\ads{{\it AdS}}
\def\adsp{{\it AdS}$_{p+2}$}
\def\cft{{\it CFT}}

\newcommand{\ber}{\begin{eqnarray}}
\newcommand{\eer}{\end{eqnarray}}

\newcommand{\beqar}{\begin{eqnarray}}
\newcommand{\cN}{{\cal N}}
\newcommand{\cO}{{\cal O}}
\newcommand{\cA}{{\cal A}}
\newcommand{\cT}{{\cal T}}
\newcommand{\cF}{{\cal F}}
\newcommand{\cC}{{\cal C}}
\newcommand{\cR}{{\cal R}}
\newcommand{\cW}{{\cal W}}
\newcommand{\eeqar}{\end{eqnarray}}
\newcommand{\tht}{\thteta}
\newcommand{\lm}{\lambda}\newcommand{\Lm}{\Lambda}
\newcommand{\eps}{\epsilon}


\newcommand{\nonu}{\nonumber}
\newcommand{\oh}{\displaystyle{\frac{1}{2}}}
\newcommand{\dsl}
  {\kern.06em\hbox{\raise.15ex\hbox{$/$}\kern-.56em\hbox{$\partial$}}}
\newcommand{\id}{i\!\!\not\!\partial}
\newcommand{\as}{\not\!\! A}
\newcommand{\ps}{\not\! p}
\newcommand{\ks}{\not\! k}
\newcommand{\D}{{\cal{D}}}
\newcommand{\dv}{d^2x}
\newcommand{\Z}{{\cal Z}}
\newcommand{\N}{{\cal N}}
\newcommand{\Dsl}{\not\!\! D}
\newcommand{\Bsl}{\not\!\! B}
\newcommand{\Psl}{\not\!\! P}
\newcommand{\eeqarr}{\end{eqnarray}}
\newcommand{\ZZ}{{\rm \kern 0.275em Z \kern -0.92em Z}\;}

                                                                                                    
\def\del{{\delta^{\hbox{\sevenrm B}}}} \def\ex{{\hbox{\rm e}}}
\def\azb{A_{\bar z}} \def\az{A_z} \def\bzb{B_{\bar z}} \def\bz{B_z}
\def\czb{C_{\bar z}} \def\cz{C_z} \def\dzb{D_{\bar z}} \def\dz{D_z}
\def\im{{\hbox{\rm Im}}} \def\mod{{\hbox{\rm mod}}} \def\tr{{\hbox{\rm Tr}}}
\def\ch{{\hbox{\rm ch}}} \def\imp{{\hbox{\sevenrm Im}}}
\def\trp{{\hbox{\sevenrm Tr}}} \def\vol{{\hbox{\rm Vol}}}
\def\rl{\Lambda_{\hbox{\sevenrm R}}} \def\wl{\Lambda_{\hbox{\sevenrm W}}}
\def\fc{{\cal F}_{k+\cox}} \def\vev{vacuum expectation value}
\def\nodiv{\mid{\hbox{\hskip-7.8pt/}}}
\def\ie{{\em i.e.}}
\def\ie{\hbox{\it i.e.}}

\def\CC{{\mathchoice
{\rm C\mkern-8mu\vrule height1.45ex depth-.05ex
width.05em\mkern9mu\kern-.05em}
{\rm C\mkern-8mu\vrule height1.45ex depth-.05ex
width.05em\mkern9mu\kern-.05em}
{\rm C\mkern-8mu\vrule height1ex depth-.07ex
width.035em\mkern9mu\kern-.035em}
{\rm C\mkern-8mu\vrule height.65ex depth-.1ex
width.025em\mkern8mu\kern-.025em}}}
                                                                                                    
\def\RR{{\rm I\kern-1.6pt {\rm R}}}
\def\NN{{\rm I\!N}}
\def\ZZ{{\rm Z}\kern-3.8pt {\rm Z} \kern2pt}
\def\IB{\relax{\rm I\kern-.18em B}}
\def\ID{\relax{\rm I\kern-.18em D}}
\def\II{\relax{\rm I\kern-.18em I}}
\def\IP{\relax{\rm I\kern-.18em P}}
\newcommand{\CS}{{\scriptstyle {\rm CS}}}
\newcommand{\CSs}{{\scriptscriptstyle {\rm CS}}}
\newcommand{\rc}{\nonumber\\}
\newcommand{\bear}{\begin{eqnarray}}
\newcommand{\eear}{\end{eqnarray}}
\newcommand{\W}{{\cal W}}
\newcommand{\F}{{\cal F}}
\newcommand{\x}{{\cal O}}
\newcommand{\LL}{{\cal L}}
                                                                                                    
\def\mani{{\cal M}}
\def\calo{{\cal O}}
\def\calb{{\cal B}}
\def\calw{{\cal W}}
\def\calz{{\cal Z}}
\def\cald{{\cal D}}
\def\calc{{\cal C}}
\def\to{\rightarrow}
\def\ele{{\hbox{\sevenrm L}}}
\def\ere{{\hbox{\sevenrm R}}}
\def\zb{{\bar z}}
\def\wb{{\bar w}}
\def\nodiv{\mid{\hbox{\hskip-7.8pt/}}}
\def\menos{\hbox{\hskip-2.9pt}}
\def\dr{\dot R_}
\def\drr{\dot r_}
\def\ds{\dot s_}
\def\da{\dot A_}
\def\dga{\dot \gamma_}
\def\ga{\gamma_}
\def\dal{\dot\alpha_}
\def\al{\alpha_}
\def\cl{{closed}}
\def\cls{{closing}}
\def\vev{vacuum expectation value}
\def\tr{{\rm Tr}}
\def\to{\rightarrow}
\def\too{\longrightarrow}


\def\a{\alpha}
\def\b{\beta}
\def\c{\gamma}
\def\d{\delta}
\def\e{\epsilon}           
\def\f{\phi}               
\def\vf{\varphi}  \def\tvf{\tilde{\varphi}}
\def\vp{\varphi}
\def\g{\gamma}
\def\h{\eta}
\def\i{\iota}
\def\j{\psi}
\def\k{\kappa}                    
\def\l{\lambda}
\def\m{\mu}
\def\n{\nu}
\def\o{\omega}  \def\w{\omega}
\def\q{\theta}  \def\th{\theta}                  
\def\r{\rho}                                     
\def\s{\sigma}                                   
\def\t{\tau}
\def\u{\upsilon}
\def\x{\xi}
\def\z{\zeta}
\def\pt{\tilde{\varphi}}
\def\tt{\tilde{\theta}}
\def\lab{\label}  
\def\6{\partial}
\def\wg{\wedge}
\def\atanh{{\rm arctanh}}
\def\bpsi{\bar{\psi}}
\def\bt{\bar{\theta}}
\def\bvf{\bar{\varphi}}

%
                                                                                                    
\newfont{\namefont}{cmr10}
\newfont{\addfont}{cmti7 scaled 1440}
\newfont{\boldmathfont}{cmbx10}
\newfont{\headfontb}{cmbx10 scaled 1728}
\newcommand{\re}{\,\mathbb{R}\mbox{e}\,}
\newcommand{\hyph}[1]{$#1$\nobreakdash-\hspace{0pt}}
\providecommand{\abs}[1]{\lvert#1\rvert}
\newcommand{\Nugual}[1]{$\mathcal{N}= #1 $}
\newcommand{\sub}[2]{#1_\text{#2}}
\newcommand{\partfrac}[2]{\frac{\partial #1}{\partial #2}}
\newcommand{\bsp}[1]{\begin{equation} \begin{split} #1 \end{split} \end{equation}}

\numberwithin{equation}{section}

\newcommand{\Tr}{\mbox{Tr}}    


%
\renewcommand{\theequation}{{\rm\thesection.\arabic{equation}}}
\begin{titlepage}
\rightline{SISSA 78/2006/EP}  
\rightline{US-FT-5/06}
\vspace{0.1in}

\begin{center} 
\Large \bf Unquenched Flavors in the Klebanov-Witten Model
\end{center}
\vskip 0.2truein
\begin{center}
Francesco Benini${}^{*}$\footnote{benini@sissa.it}, 
Felipe Canoura$^{**}$\footnote{canoura@fpaxp1.usc.es}, Stefano 
Cremonesi${}^{*}$\footnote{cremones@sissa.it}, 
Carlos 
N\'u\~nez${}^{\dagger}$\footnote{c.nunez@swansea.ac.uk} \\ and 
Alfonso V. Ramallo ${}^{**}$\footnote{alfonso@fpaxp1.usc.es}
\vspace{0.3in}\\
${}^{*}$ \it{ SISSA/ISAS and INFN-Sezione di Trieste\\ Via Beirut 2; 
I-34014\\Trieste, Italy}
\vspace{0.3in}
\vskip 0.1truein
${}^{**}$ 
\it{
Departamento de  Fisica de Particulas, Universidade 
de Santiago de 
Compostela\\and\\Instituto Galego de Fisica de Altas 
Enerxias (IGFAE)\\E-15782, Santiago de Compostela, Spain
}
\vspace{0.3in}
\vskip 0.1truein
${}^{\dagger}$ \it{Department of Physics\\ University of Swansea, Singleton 
Park\\
Swansea SA2 8PP, United Kingdom.}
\vspace{0.3in}
\end{center}
\vspace{0.3in}
\centerline{{\bf Abstract}}
Using AdS/CFT, we study the addition of an arbitrary number of backreacting
flavors to the Klebanov-Witten theory, making many checks of consistency
between our new Type IIB plus branes solution and expectations from field
theory. We study generalizations of our method for adding flavors to all \Nugual{1}
SCFTs that can be realized on D3-branes at the tip of a Calabi-Yau cone. Also,
general guidelines suitable for the addition of massive flavor branes are
developed.

\smallskip
\end{titlepage}
\setcounter{footnote}{0}
\tableofcontents

\newpage
\section{Introduction}

The AdS/CFT conjecture originally proposed by Maldacena \cite{Maldacena:1997re}
and refined in \cite{Gubser:1998bc, Witten:1998qj} is one of the most 
powerful analytic tools for studying strong coupling effects in gauge 
theories. There are many examples that go beyond the initially 
conjectured duality and first steps in generalizing it 
to non-conformal models 
were taken in \cite{Itzhaki:1998dd}. 
Later, very interesting developments led to the construction 
of the gauge-string 
duality in phenomenologically more relevant 
theories \ie\, minimally or non-supersymmetric gauge theories 
\cite{Girardello:1999bd}. 

Conceptually, a clear setup for duals to 
theories with few SUSY's is obtained by breaking  conformality and 
(perhaps partially) 
supersymmetry, deforming ${\cal N}=4$ SYM with relevant operators or 
VEV's.  
The models put forward in \cite{Girardello:1999bd} are very good examples 
of this.

Even when there are important technical differences, in the same line of 
thought, we can consider the model(s) developed by Klebanov and 
a distinguished list of physicists: Witten \cite{Klebanov:1998hh}, Nekrasov 
\cite{Klebanov:1999rd}, Tseytlin \cite{Klebanov:2000nc}, 
Strassler \cite{Klebanov:2000hb}, Herzog and Gubser \cite{Gubser:2004qj}
and Dymarsky and Seiberg \cite{Dymarsky:2005xt}.
In these papers (and many extensions of them), a far reaching idea has 
been developed, namely to flow to a confining field theory with minimal 
SUSY starting from an \Nugual{1} SCFT with a product gauge group 
$SU(N_c)\times SU(N_c)$, bifundamental chiral matter and a 
quartic superpotential for the chiral superfields.%
\footnote{ It is obvious 
that such a field theory is non-renormalizable 
and must be thought of as the IR  of some UV well defined theory. In 
\cite{Klebanov:1998hh} 
a UV completion in terms of an orbifolded \Nugual{2} field theory is 
given.} 
The superconformal field theory described above rules
the low energy dynamics of $N_c$ D3-branes 
at the tip of the conifold. Then conformality is broken by the addition 
of fractional branes, that effectively unbalance the ranks of the gauge 
groups \cite{Klebanov:1999rd,Klebanov:2000nc}. A 
``duality cascade''  starts and the flow to the IR leaves us 
with a confining field theory \cite{Klebanov:2000hb}. Subtleties related 
to the last steps of the 
cascade have been discussed in \cite{Gubser:2004qj,Butti:2004pk}. All this interesting physics is very nicely described 
with great detail in \cite{Strassler:2005qs}.

In this paper we will concentrate on a nonconformal theory without cascade. The starting point is a Type IIB solution dual to an $SU(N_c)\times SU(N_c)$ \Nugual{1} SCFT 
also known as the Klebanov-Witten field theory/geometry.
One of the aims of the paper is to add an arbitrary large number of flavors
to each of the gauge groups. The addition of fundamental degrees of 
freedom is an important step toward the understanding of QCD-like dynamics, in 
different regions of the space of parameters.

A very fructiferous idea used to  add flavors to different field 
theories (using the string dual) was described in \cite{Karch:2002sh} and 
then applied to various backgrounds, `flavoring' different dual field 
theories, in many 
subsequent publications (for a complete list see citations 
to \cite{Karch:2002sh}). As it was clearly stated in the original paper, 
the procedure spelled out in \cite{Karch:2002sh}
consists in the addition of a finite number $N_f$ of spacetime filling flavor D7-branes to the $N_c\to\infty$ color D3-branes extending in the Minkowski directions, and the usual decoupling limit ($g_s\to 0$, $N_c\to\infty$, $g_s N_c$ fixed) of the D3-branes is performed, keeping the number $N_f$ of flavor branes fixed. Then the D3-branes generate the geometry and the flavor branes only minimize their worldvolume Dirac-Born-Infeld action in this background without deforming it. This is the probe limit.  
In the dual description they are considering the addition of a finite number of flavors $N_f$ to the large $N_c$ gauge theory, in the strict double scaling 't Hooft limit ($g_{YM}\to 0$, $N_c\to\infty$, $\lambda=g_{YM}^2 N_c$ fixed). In the lattice literature this is called the `quenched' approximation: the dynamics of the colors and its effect on the flavors is completely taken into account, but the backreaction of the flavors onto the colors is neglected. In the probe limit this approximation becomes exact.

It is interesting to go beyond this `quenched' or 
`non-backreacting' approximation and see what happens when one adds a 
large number of flavors, of the same order of the number of colors, 
and the backreaction effects of the flavor branes 
are considered. Indeed, many phenomena that cannot be captured by the 
quenched approximation, might be apparent when a string 
backreacted background is found.

In this paper we will propose a Type IIB dual to the field theory of 
Klebanov and Witten, in the case in which a large number 
of flavors ($N_f \sim N_c$) is added to each gauge group. We will also present interesting 
generalizations of this to cases describing different duals to \Nugual{1} SCFT's constructed from D3-branes placed at singularities.

Let us briefly describe the procedure we will follow, inspired mostly by 
the papers  \cite{Klebanov:2004ya, Bigazzi:2005md,Casero:2006pt} and more recently 
\cite{Paredes:2006wb, Murthy:2006xt}.
In those papers (dealing with the addition of many fundamentals in the 
non-critical string and Type IIB string respectively),  flavors are 
added into the dynamics of the dual background via the introduction of 
$N_f$ spacetime filling flavor branes, whose dynamics is given by a 
Dirac-Born-Infeld action. This dynamics is intertwined with the usual
Einstein-like action of IIB and a new solution is found, up to 
technical subtleties described below.

\subsection{Generalities of the Procedure Used}
To illustrate the way flavor branes will be added, let us  start 
by considering the background of Type IIB supergravity that 
is conjectured to be dual to the Klebanov-Witten field theory: an \Nugual{1} 
SCFT with gauge group $SU(N_c)\times SU(N_c)$, 
two chiral multiplets of bifundamental matter  $A_i, B_i,\; i=1,2$ and 
a (classically irrelevant) quartic superpotential 
\beq \label{superpote}
W= \lambda \Tr(A_i B_j A_k B_l) \: \epsilon^{ik}\epsilon^{jl} \;.
\eeq
The dual Type IIB background reads
\begin{align} \label{kw}
ds^2 &= h(r)^{-1/2} dx_{1,3}^2 + 
h(r)^{1/2} \bigg\{ dr^2 +
\frac{r^2}{6} \sum_{i=1,2} ( d\theta_i^2 + \sin^2 \theta_i \, d\varphi_i^2)  + \frac{r^2}{9} (d\psi + \sum_{i=1,2} \cos\theta_i \, d\varphi_i)^2 \bigg\} \nonumber \\
F_5 &= \frac{1}{g_s} (1+\ast) \, d^4x \wedge dh(r)^{-1} \nonumber \\
h(r) &= \frac{27 \pi g_s N_c \alpha'^2}{4 r^4}
\end{align}
with constant dilaton and all the other fields in Type IIB supergravity vanishing. The 
set of coordinates that will be used in the rest of the paper
is $ x^M= \{x^0, x^1, x^2, x^3, r, \theta_1, \varphi_1,\theta_2, 
\varphi_2,\psi\}$. For the sake of brevity, in the following we will take units is which $g_s=1$, $\alpha'=1$.

We will add $N_f$ spacetime filling D7-branes to this geometry, in a way that preserves some amount of supersymmetry. This problem was 
studied in 
\cite{Arean:2004mm,Ouyang:2003df} for the conformal case and in \cite{Kuperstein:2004hy} for the cascading theory. These authors found calibrated embeddings of D7-branes which preserve (at least some fraction of) the supersymmetry of the background.
We will choose to put two sets of D7-branes on  the surfaces 
parametrized by
\begin{align} \label{surfaceskappa}
\xi^\alpha_1 &= \{x^0,x^1,x^2,x^3,r,\theta_2,\varphi_2,\psi\} \qquad 
\theta_1=\text{const.} \qquad \varphi_1=\text{const.} \;, \nonumber\\
\xi^\alpha_2 &= \{x^0,x^1,x^2,x^3,r,\theta_1,\varphi_1,\psi\} 
\qquad \theta_2=\text{const.} \qquad \varphi_2=\text{const.} \;.
\end{align}
Note that these two configurations 
are mutually supersymmetric with the background. 
Moreover, since the two embeddings are noncompact, the gauge theory supported on the D7's has vanishing 4d effective coupling on the Minkowski directions; therefore the gauge symmetry on them is seen as a flavor symmetry by the 4d gauge theory of interest.
The two sets of flavor branes introduce 
a $U(N_f) \times U(N_f)$ symmetry%
\footnote{The diagonal axial $U(1)$ is anomalous},
the expected flavor symmetry with 
massless flavors. The configuration with two sets (two branches) 
can be deformed 
to a single set, shifted from the origin, that represents massive flavors, 
and realizes the explicit breaking of the flavor 
symmetry to the diagonal vector-like $U(N_f)$. Our configuration 
(eq. (\ref{surfaceskappa})) for probes  is nothing else than 
the $z_1=0$ holomorphic embedding of \cite{Ouyang:2003df}.

We will then write an action for a system consisting of type IIB 
supergravity%
\footnote{The problems with writing an action for Type 
IIB that includes the self-duality condition are well known. Here, we 
just mean a Lagrangian from which the equations of motion of Type IIB 
supergravity are derived. The self-duality condition is imposed 
on the solutions.}
plus D7-branes described by their 
Dirac-Born-Infeld action (in Einstein frame):
\begin{equation} \label{actionnotsmear}
\begin{split} 
S &= \frac{1}{2\kappa_{10}^2} \int d^{10}x \, \sqrt{-G} 
\Big[ R - \frac{1}{2} \partial_M \phi \partial^M \phi -\frac{1}{2} 
e^{2\phi} |F_1|^2 -\frac{1}{4} |F_5|^2 \Big] + \\ 
 & \qquad - T_7\sum^{N_f} \int d^{8}\xi \, e^\phi \Big[ 
\sqrt{-\hat{G}_8^{(1)}} + \sqrt{-\hat{G}_8^{(2)}} \Big] 
 + T_7 \sum^{N_f} \int  \hat C_8 \;,
\end{split}
\end{equation}
where we have chosen the normalization $|F_p|^2 = \frac{1}{p!} F_p F_p (G^{-1})^p$.\\
Notice that we did not excite the worldvolume gauge fields, but this is a 
freedom of the approach we adopted. Otherwise one may need to find new suitable \hyph{\kappa}symmetric embeddings. 

These two sets of D7-branes are localized in their two 
transverse directions, hence
the equations of motion derived from (\ref{actionnotsmear})
will be quite complicated to solve, due to the presence of source terms 
(Dirac delta functions). 

But we can take some advantage of the fact that we are adding lots of 
flavors. Indeed, since we will 
have many ($N_f\sim N_c$) flavor branes, we might think
about distributing them in a homogeneous way on their respective 
transverse directions.
This `smearing procedure' boils down to approximating
\begin{align}
T_7\sum^{N_f} \int d^{8}\xi\, e^\phi 
\sqrt{-\hat{G}_8^{(i)}} \quad &\to \quad \frac{T_7 N_f}{4\pi}\int d^{10}x \, 
e^\phi \,\sin\theta_i 
\sqrt{-\hat{G}_8^{(i)}} \nonumber\\
T_7 \sum^{N_f} \int  \hat C_8 \quad &\to \quad \frac{T_7 N_f}{4\pi}\int  \,
\Big[ Vol(Y_1) + Vol(Y_2) \Big] \wedge C_8 \;,
\label{smearingproc}
\end{align}
with 
$Vol(Y_i)=\sin \theta_i \, d\theta_i\wedge d\varphi_i$ the volume form of the $S^2$'s.

This effectively generates a ten dimensional action
\begin{equation} \label{action}
\begin{split} 
S &= \frac{1}{2\kappa_{10}^2} \int d^{10}x \, \sqrt{-G} \Big[ R - 
\frac{1}{2} 
\partial_M \phi \partial^M \phi -\frac{1}{2} e^{2\phi} |F_1|^2 
-\frac{1}{4} |F_5|^2 \Big] + \\ 
&\qquad -\frac{T_7 N_f}{4\pi} \int d^{10}x \, e^\phi \sum_{i=1,2} \sin\theta_i \,
\sqrt{-\hat{G}_8^{(i)}} 
+ \frac{T_7 N_f}{4\pi} \int \Big[ Vol(Y_1)+Vol(Y_2) \Big] \wedge C_8 \;.
\end{split}
\end{equation}

We can derive in the smeared case the following (not so involved) equations of motion, coming 
from the action (\ref{action}):
\begin{align}
\begin{split}
R_{MN} - \frac{1}{2}G_{MN} R &= 
\frac{1}{2} \Big( \partial_M \phi \partial_N \phi -\frac{1}{2} G_{MN} 
\partial_P \phi \partial^P \phi \Big)
+ \frac{1}{2} e^{2\phi} \Big( F_M^{(1)} F_N^{(1)} 
-\frac{1}{2} G_{MN} |F^{(1)}|^2 \Big) + \nonumber\\
& \qquad + \frac{1}{96} F_{MPQRS}^{(5)} F_N^{(5)PQRS} + T_{MN}
\end{split} \nonumber\\
D^M \partial_M \phi &= e^{2\phi} |F_1|^2 
+ \frac{2\kappa_{10}^2 T_7}{\sqrt{-G}} \frac{N_f}{4\pi} e^\phi 
\sum_{i=1,2} \sin\theta_i \,
\sqrt{-\hat{G}_8^{(i)}} \nonumber \\
d\Big(e^{2\phi} \ast F_1 \Big) &= 0 \nonumber\\
dF_1 &= -2\kappa_{10}^2 T_7 \frac{N_f}{4\pi} 
\Big[ Vol(Y_1)+Vol(Y_2) \Big]  \nonumber\\
dF_5 &= 0 \;.
\label{eqs2nd}
\end{align}
The modified Bianchi identity is obtained through 
$F_1 = - e^{-2\phi} \ast F_9$, and comes from the WZ part of the action 
(\ref{action}).
The contribution to the stress-energy tensor coming from the two 
sets of $N_f$ D7 flavor 
branes is given by
\begin{equation} \label{tmunu}
\begin{split}
T^{MN} &= \frac{2\kappa_{10}^2}{\sqrt{-G}} 
\frac{\delta S^{flavor}}{\delta G_{MN}} = -\frac{N_f}{4\pi} \frac{e^\phi}{\sqrt{-G}} 
\sum_{i=1,2} \sin \theta_i \frac{1}{2} \sqrt{-\hat{G}_8^{(i)}} 
\hat{G}_8^{(i)\alpha\beta} \delta_\alpha^M \delta_\beta^N \;,
\end{split} \end{equation}
where $\alpha,\beta$ are coordinate indices on the D7. 
In the subsequent sections we will solve the equations of motion 
\eqref{eqs2nd}-\eqref{tmunu} and will 
propose that this Type IIB background is dual to the Klebanov-Witten field 
theory when two sets of $N_f$ flavors are added for each gauge group. We 
will actually find BPS 
equations for the purely bosonic background, by 
imposing that the variations of the dilatino and gravitino vanish. We will verify that these 
BPS first-order equations solve all the equations of motion \eqref{eqs2nd}.

Let us add some remarks on some  important  points about the 
resolution of the system. First of all, it is clear from the Bianchi 
identity of $F_1$ in (\ref{eqs2nd}) that we will not be able to define the axion field $C_0$
on open subsets.

Regarding the solution of the equations of motion, we will proceed by 
proposing a {\it deformed background} ansatz of the form
\begin{align} \label{configuration}
ds^2 &= h^{-1/2} dx_{1,3}^2 + 
h^{1/2} \bigg\{ dr^2 +
\frac{e^{2g}}{6} \sum_{i=1,2} ( d\theta_i^2 + \sin^2 \theta_i \, d\varphi_i^2)  + \frac{e^{2f}}{9} (d\psi + \sum_{i=1,2} \cos\theta_i \, d\varphi_i)^2 \bigg\} \nonumber \\
F_5 &= (1+\ast) \, d^4x \wedge K\, dr \\
F_1 &= \frac{N_f}{4\pi} ( d\psi + \cos\theta_2\, d\varphi_2 + 
\cos\theta_1\, d\varphi_1 ) \;. \nonumber 
\end{align}
Thanks to the smearing procedure, all the unknown function $h$, $f$, $g$, $K$ and the dilaton $\phi$ only depend on the radial coordinate $r$.

The Bianchi identity for the five-form field-strength gives
\begin{equation}
K\,h^2\,e^{4g+f}=  27\pi N_c \;,
\end{equation} 
and we will obtain solutions to \eqref{eqs2nd} by imposing that the BPS equations 
derived from the vanishing of the gravitino and gaugino variations and 
the Bianchi identities are satisfied. These will produce ordinary first-order equations 
for $f(r)$, $g(r)$, $h(r)$, $K(r)$, $\phi(r)$. We will also be able to derive these 
BPS equations from a superpotential in the reduction of Type IIB supergravity.

We will study in detail the dual field theory to the supergravity
solutions mentioned above, making a considerable number of matchings. 
The field theories turn out to have positive \hyph{\beta}function along the
flow, exhibiting a Landau pole in the UV. In the IR we still have a strongly coupled
field theory, which is ``almost conformal''.
We will also generalize all these results to the interesting case
of a large class of different \Nugual{1} SCFTs, deformed by the addition of flavors.
In particular we will be able to add flavors to every
gauge theory whose dual is $AdS_5 \times M_5$, where $M_5$ is a five-dimensional Sasaki-Einstein manifold. New solutions will be found that describe the `unflavored' case,
making contact with old results. Finally, a possible
way of handling the massive flavor case is undertaken.

We have explained the strategy we adopt to add flavors, so this is perhaps a 
good place to discuss some interesting issues. The 
reader might be wondering about the `smearing procedure' discussed above, 
what is its significance and effect on the dual gauge theory, among other 
questions. It is clear that we smear the flavor branes just to be able to 
write a 10-dimensional action that will produce ordinary (in contrast to 
partial) differential equations without Dirac delta functions source terms.

The results we will show and the experience obtained in 
\cite{Casero:2006pt,Paredes:2006wb} show that many 
properties of the flavored field theory are still well captured by the 
solutions obtained following the procedure described above. It is not 
clear what important phenomena on the gauge 
theory we are losing in smearing, but see below for an important subtlety. 

\begin{figure}[ht]
\begin{center}
\includegraphics[width=0.95\textwidth]{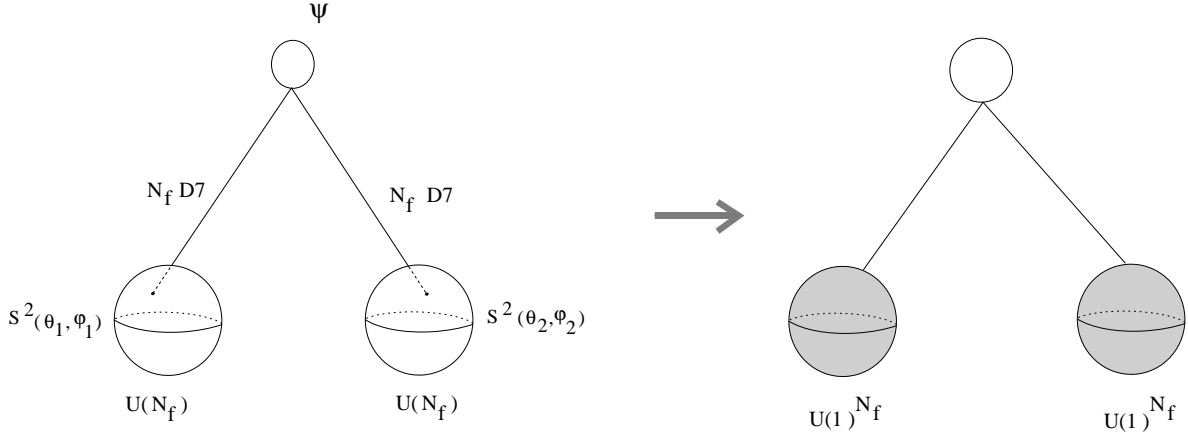}
\end{center}
\caption[smearingfig]{We see on the left side the two stacks of $N_f$ flavor-branes localized on
each of their respective $S^2$'s (they wrap the other $S^2$). The flavor
group is clearly $U(N_f) \times U(N_f)$. After the smearing on the right
side of the figure, this global symmetry is broken to $U(1)^{N_f - 1}\times
U(1)^{N_f -1} \times U(1)_B \times U(1)_A$. \label{smearingfig}} 
\end{figure}

One relevant point to discuss is related to global symmetries. Let us go 
back to the 
weak coupling ($g_s N_c\to 0$) limit, in which we have branes living on a 
spacetime that is the product of four Minkowski directions and the 
conifold. When all the flavor branes of the two separate 
stacks (\ref{surfaceskappa}) are on top 
of each other, the gauge symmetry on the D7's worldvolume 
is given by the product $U(N_f) \times U(N_f)$. When we take the 
decoupling limit for the D3-branes  $\alpha'\to 0$, with fixed $g_s N_c$ and keeping constant the energies of the excitations on the branes, we are left 
with a solution of Type IIB supergravity that we propose is dual to 
the Klebanov-Witten field theory with $N_f$ flavors for both gauge groups \cite{Ouyang:2003df}. In this case the flavor symmetry is $U(N_f)\times U(N_f)$, where the axial $U(1)$ is anomalous. 
This background would be for sure very involved, since it would 
depend on the coordinates $(r,\theta_1,\theta_2)$, if the embeddings of the two stacks of D7-branes are $\theta_1=0$ and $\theta_2=0$, respectively. When we smear the 
$N_f$ D7-branes, we are breaking $U(N_f)\to U(1)^{N_f}$ (see Figure \ref{smearingfig}).

There is one important point to contrast with \cite{Casero:2006pt}. In 
that paper, a smearing is also proposed but it is argued that the 
dual field theory (a version of \Nugual{1} SQCD with a quartic superpotential 
in the quark superfields) possesses $U(N_f)$ global flavor 
symmetry. 
As in all backgrounds constructed on wrapped branes, the effects of the Kaluza-Klein modes play an important r\^ole
and the dual field theory behaves as 4-dimensional only in the far IR.%
\footnote{For a detailed study of the role and dynamics of the KK 
modes in wrapped brane setups, see \cite{Gursoy:2005cn}.} 
In this regime, when the internal manifold shrinks to small size, 
for energies below this inverse size
we do not effectively see the breaking $U(N_f) \to U(1)^{N_f}$. 
In contrast, the backgrounds obtained by placing D-branes at conical singularities, like \cite{Klebanov:1998hh}-\cite{Gubser:2004qj} as well as our solution, describe a four 
dimensional field theory all along the flow. 

It might be interesting for the reader to note that the papers in the 
line of \cite{Karch:2002sh} are working in the context of
't Hooft expansion \cite{'tHooft:1973jz}.
When the ratio $N_f/N_c$ is very small, one can ignore the backreaction 
effects of the flavor branes on the geometry. 
This  is the dual version to the suppression of 
effects that include the running of fundamentals in internal loops.
Even when these fundamentals are massless, their effects 
while running in loops are suppressed by a  factor of ${\cal O}(N_f/N_c)$.
But in the strict 't Hooft limit, if the number of flavors is kept fixed, 
the corrections due to the quantum dynamics of quarks exactly vanish.
In the cases considered in \cite{Casero:2006pt,Paredes:2006wb}, the ratio above is of order one and we are 
working on the so called Veneziano's topological expansion 
\cite{Veneziano:1976wm}.
New physics (beyond the 't Hooft limit)
 is captured by Veneziano's proposal: we 
will be able to see this in the present paper that considers the backreaction of the flavor branes, 
just like was observed in \cite{Casero:2006pt,Paredes:2006wb}, in contrast with the papers that 
worked in the 't Hooft approximation as proposed in \cite{Karch:2002sh}.

Another point that is worth elaborating on is whether there is a limit on the 
number of D7-branes that can be added. Indeed, since a D7-brane is a 
codimension-two object (like a vortex in $2+1$ dimensions) its gravity 
solution will generate a deficit angle; having many seven branes, will 
basically ``eat-up'' the transverse space. This led to the conclusion that 
solutions that can be globally extended cannot have more than a maximum 
number of twelve D7-branes \cite{Greene:1989ya} (and exactly twenty-four in compact 
spaces). 
In this paper we are adding a number $N_f\to\infty$ of D7-branes, 
certainly larger that the bound mentioned above.
Like in the papers \cite{Aharony:1998xz,Grana:2001xn}, 
we will adopt the attitude of analyzing the behavior of our solutions and 
we will see that they give sensible results. 
But there is more than that: the smearing procedure distributes the D7's all over this 2-dimensional compact space, in such a way that the equation for the axion-dilaton is not the one in the vacuum at any point. This avoids the constraint on the number of D7-branes, which came from solving the equation of motion for the axion-dilaton outside sources.

Finally, we must emphasize that this is not the first paper that deals 
with the D3/D7 system in the context of ``AdS/CFT with flavors''. 
Indeed, very good papers have been written where this problem was faced looking for 
a solution where the flavor branes are replaced by fluxes
in terms of the 
Type IIB supergravity fields, dilaton 
and an axion ($\phi$, $C_0$). The BPS equations for the D3/D7 system in 
cases preserving 8 supercharges were written in \cite{Aharony:1998xz, Grana:2001xn}, 
a partially explicit solution of the equations of motion in the presence of sources was found in \cite{Bertolini:2001qa} for the orbifold case,
more interesting geometrical aspects were discussed 
in  \cite{Burrington:2004id} and an involved solution was found in
\cite{Kirsch:2005uy}, where some matching with gauge theory behavior was 
attempted.%
\footnote{Even though slightly unrelated to the D3/D7 system, we 
cannot resist here to mention the beautiful solution found by Cherkis and Hashimoto for a 
localized D2/D6 system \cite{Cherkis:2002ir}.} 

The papers \cite{Aharony:1998xz,Grana:2001xn,Bertolini:2001qa,Burrington:2004id,Kirsch:2005uy}
were written with the idea of letting the flavor branes backreact. 
One qualitative difference with respect to what 
we explained above is that the authors of \cite{Aharony:1998xz,Grana:2001xn,Burrington:2004id,Kirsch:2005uy}
consider the case in which D3-branes are added in the background produced by D7-branes and solve the 
Laplace equation, in this case for the deformation introduced by the D3's. 
In contrast, we consider  the 
background produced by the D3-branes and we deform it to take into account 
the ``smeared'' backreaction of the D7-branes.  The two procedures are different.

One advantage of the approach  
proposed in \cite{Casero:2006pt} is that the flavor degrees of freedom 
explicitly appear in the DBI action that allows the introduction of $SU(N_f)$ gauge fields in the bulk that are dual to the global symmetry in the dual field theory, while it is 
difficult to see how they will appear in a Type IIB 
solution that only includes RR fluxes. Our approach produces a 
simple SUSY solution to (\ref{eqs2nd}) and the analysis of 
gauge theory effects is simple to do. 
Besides, the proposal of \cite{Casero:2006pt} used in 
the present work is the natural continuation of the many successful results obtained in papers in the line of \cite{Karch:2002sh}. Indeed, we are just following the idea of 
\cite{Karch:2002sh} for a large number of flavor branes.

\subsection{Organization of This Paper}

This paper is organized in two main parts. In  part I we will 
present the addition of flavors to the Klebanov-Witten solution. A 
detailed analysis of the supergravity plus branes solutions and the study 
of the dual gauge theory, as read from the above mentioned solutions, is performed.
A reader mainly interested with the line of research, but who does not 
want to go in full details, should be happy reading this introduction, part I and Appendix \ref{appAlt}.

The readers who intend to work on this subject and want to study these 
results in more technical detail or want to appreciate the beauty and generality in our formalism are referred to part II. Also in part II the reader will find a sketch of how to deal with massive flavors using these techniques.

Those readers who are not attracted by the physics of flavor using 
AdS/CFT techniques, but just want to learn about some new solutions 
(born out of our `deformed backgrounds' as described above), should read the 
introduction and the appendix B. 

Some other appendices complement nicely our presentation.

The section of conclusions includes also a summary of results and proposes 
future directions that the interested reader might want to pursue.

\section{Part I: Adding Flavors to the Klebanov-Witten Field Theory }
\label{sect2}

\subsection{What to Expect from Field Theory Considerations}

In this first part we will address in detail the problem of adding a large number of backreacting non-compact D7-branes to the Klebanov-Witten Type IIB supergravity solution, which describes D3-branes at the tip of the conifold. Before presenting the solution and describing how it is obtained, we would like to have a look at the dual field theory, and sketch which are the features we expect.

For this purpose, we consider the case of probe D7-branes, and mainly summarize what was pointed out in \cite{Ouyang:2003df}. The conifold is a non-compact Calabi-Yau 3-fold, defined by one equation in $\mathbb{C}^4$:
\begin{equation} \label{conifold}
z_1 z_2 - z_3 z_4 = 0 \;.
\end{equation} 
Since this equation is invariant under a real rescaling of the variables, the conifold is a real cone, whose base is the Sasaki-Einstein space $T^{1,1}$ \cite{Klebanov:1998hh,Candelas:1989js}. It can be shown that $T^{1,1}$ is a $U(1)$ bundle over the K\"ahler-Einstein space $S^2\times S^2$, and that its isometry group is $SU(2)\times SU(2)\times U(1)$.

Klebanov and Witten \cite{Klebanov:1998hh} obtained an interesting example of gauge/gravity duality by placing a stack of $N_c$ D3-branes at the apex of the conifold. The branes source the RR 5-form flux and warp the geometry, giving the Type IIB supergravity solution \eqref{kw}. The dual field theory, describing the IR dynamics on the worldvolume of the branes, has gauge group $SU(N_c)\times SU(N_c)$ and matter fields $A_i$, $B_i$, $i=1,2$ which transform in the bifundamental representations $(N_c,\overline{N_c})$ and $(\overline{N_c},N_c)$ respectively. The theory has also a quartic superpotential $W_{KW} = \lambda\, \Tr(A_i B_j A_k B_l) \: \epsilon^{ik}\epsilon^{jl}$.
The field theory is \Nugual{1} superconformal, and the anomaly-free $U(1)$ R-symmetry of the superconformal algebra is dual to the $U(1)$ isometry of the fiber in $T^{1,1}$, generated by the so-called Reeb vector. In the algebraic definition \eqref{conifold} it is realized as a common phase rotation of the four coordinates: $z_i \to e^{-i\alpha} z_i$.

The addition of flavors, transforming in the fundamental and antifundamental representations of the gauge groups, can be addressed by including probe D7-branes into the geometry, following the procedure proposed in \cite{Karch:2002sh}.
This was done in \cite{Ouyang:2003df}, where the embedding of the flavor branes and the corresponding superpotential for the fundamental and antifundamental superfields were found. The D7-branes have four Minkowski directions parallel to the stack of D3-branes transverse to the conifold, whereas the other four directions are embedded holomorphically in the conifold. In particular, D7-branes describing massless flavors can be introduced by considering the holomorphic noncompact embedding $z_1=0$. The flavors, which correspond to 3-7 and 7-3 strings, are massless because the D7-branes intersect the D3-branes. Note that the D7-branes have two branches, described by $z_1=z_3=0$ and $z_1=z_4=0$, each one corresponding to a stack. The presence of two branches is required by RR tadpole cancellation: in the field theory this amounts to adding flavors in vector-like representations to each gauge group, hence preventing gauge anomalies.
The fundamental and antifundamental chiral superfields of the two gauge groups will be denoted as $q$, $\tilde{q}$ and $Q$, $\tilde{Q}$ respectively, and the gauge invariant and flavor invariant superpotential proposed in \cite{Ouyang:2003df} is 
\begin{equation}
W=W_{KW}+W_f\;,
\end{equation}
where
\begin{equation}
W_{KW}=\lambda \: \Tr(A_i B_k A_j B_l) \, \epsilon^{ij} \epsilon^{kl}
\end{equation}
is the $SU(2)\times SU(2)$ invariant Klebanov-Witten superpotential for the bifundamental fields.
For a stack of flavor branes, it is conventional to take the coupling between bifundamentals and quarks at a given point of $S^2$ as
\begin{equation}
W_f = h_1 \:\tilde{q}^a A_1 Q_a + h_2 \:\tilde{Q}^a B_1 q_a \;.
\end{equation}
This coupling between bifundamental fields and the fundamental and antifundamental flavors arises from the D7 embedding $z_1=0$. The explicit indices are flavor indices. This superpotential, as well as the holomorphic embedding $z_1=0$, explicitly breaks the $SU(2)\times SU(2)$ global symmetry (this global symmetry will be recovered after the smearing).

\begin{figure}[ht]
\begin{center}
\includegraphics[width=0.4\textwidth]{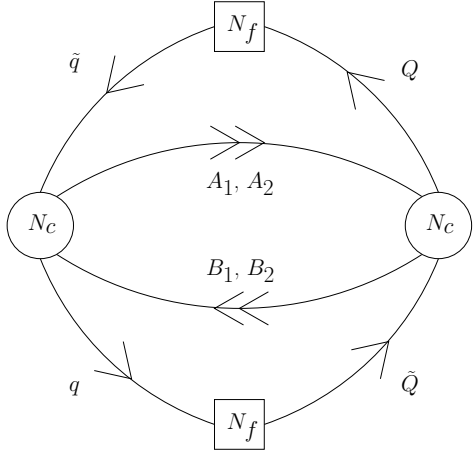}
\end{center}
\caption[Quiver diagram]{Quiver diagram of the Klebanov-Witten gauge theory with flavors. Circles are gauge groups while squares are non-dynamical flavor groups. \label{quiverdiagram}} 
\end{figure}

The field content and the relevant gauge and flavor symmetries of the theory are summarized in Table \ref{fieldcontent} and depicted in the quiver diagram in Figure \ref{quiverdiagram}.

\begin{table}[ht]
\begin{center}
\begin{tabular}{c|c|c|c|c|c|c|}
 & $SU(N_c)^2$  & $SU(N_f)^2$ &  $SU(2)^2$ & $U(1)_R$ & $U(1)_B$ & $U(1)_{B'}$ \\
\hline &&&&&&\\
$A$ & $(N_c,\overline{N_c})$  & $(1,1)$ & $(2,1)$ & $1/2$ & $0$ & $1$ \\
$B$ & $(\overline{N_c},N_c)$  & $(1,1)$ & $(1,2)$ & $1/2$ & $0$ & $-1$ \\
$q$ & $(N_c,1)$  & $(\overline{N_f},1)$ & $(1,1)$ & $3/4$ & $1$ & $1$ \\
$\tilde{q}$ & $(\overline{N_c},1)$  & $(1,N_f)$ & $(1,1)$ & $3/4$ & $-1$ & $-1$ \\
$Q$ & $(1,N_c)$  & $(1,\overline{N_f})$ & $(1,1)$ & $3/4$ & $1$ & $0$ \\
$\tilde{Q}$ & $(1,\overline{N_c})$  & $(N_f,1)$ & $(1,1)$ & $3/4$ & $-1$ & $0$ \\
\end{tabular}
\end{center}
\caption[Field content]{Field content and symmetries of the KW field theory with massless flavors. \label{fieldcontent}} 
\end{table}

The $U(1)_R$ R-symmetry is preserved at the classical level by the inclusion of  D7-branes embedded in such a way to describe massless flavors, as can be seen from the fact that the equation $z_1=0$ is invariant under the rotation $z_i\to e^{-i\alpha}z_i$ and the D7 wrap the R-symmetry circle. Nevertheless the $U(1)_R$ turns out to be anomalous after the addition of flavors, due to the nontrivial $C_0$ gauge potential sourced by the D7. The baryonic symmetry $U(1)_B$ inside the flavor group is anomaly free, being vector-like.

As was noted in \cite{Ouyang:2003df}, the theory including D7-brane probes is also invariant under a rescaling $z_i\to \beta z_i$, therefore the field theory is scale invariant in the probe approximation. In this limit the scaling dimension of the bifundamental fields is $3/4$ and the one of the flavor fields is $9/8$, as required by power counting in the superpotential.
Then the beta function for the holomorphic gauge couplings in the Wilsonian scheme is 
\begin{equation} \label{betas}
\beta_{\frac{8\pi^2}{g_i^2}} = -\frac{16\pi^2}{g_i^3} \beta_{g_i}=-\frac{3}{4} N_f \qquad \qquad \beta_{\lambda_i} = \frac{1}{(4\pi)^2} \frac{3N_f}{2N_c} \lambda_i^2 \;,
\end{equation}
with $\lambda_i= g_i^2 N_c$ the 't Hooft couplings.
In the strict planar 't Hooft limit (zero order in $N_f/N_c$), the field theory has a fixed point specified by the afore-mentioned choice of scaling dimensions, because the beta functions of the superpotential couplings and the 't Hooft couplings are zero. As soon as $N_f/N_c$ corrections are taken into account, the field theory has no fixed points for nontrivial values of all couplings. Rather it displays a ``near conformal point'' with vanishing beta functions for the superpotential couplings, but non-vanishing beta functions for the 't Hooft couplings.
In a $N_f/N_c$ expansion, formula \eqref{betas} holds at order $N_f/N_c$ \textit{if} the anomalous dimensions of the bifundamental fields $A_j$ and $B_j$ do not get corrections at this order. \emph{A priori} it is difficult to expect such a behavior from string theory, since the energy-momentum tensor of the flavor branes will induce backreaction effects on the geometry at linear order in $N_f/N_c$, differently from the fluxes, which will backreact at order $(N_f/N_c)^2$.

Moreover, since we are adding flavors to a conformal theory, we can naively expect a Landau pole to appear in the UV. Conversely, we expect the theory to be slightly away from conformality in the far IR.

\subsection{The Setup and the BPS Equations}

The starting point for adding backreacting branes to a given background is
the identification of the supersymmetric embeddings in that background,
that is the analysis of probe branes.  In \cite{Arean:2004mm}, by imposing
\hyph{\kappa}symmetry on the brane world-volume, the following
supersymmetric embeddings for D7-branes on the Klebanov-Witten background
were found:
\bsp{
\xi^\alpha_1 &= \{x^0,x^1,x^2,x^3,r,\theta_2,\varphi_2,\psi\} \qquad
\theta_1=\text{const.} \qquad \varphi_1=\text{const.} \\
\xi^\alpha_2 &= \{x^0,x^1,x^2,x^3,r,\theta_1,\varphi_1,\psi\} \qquad
\theta_2=\text{const.} \qquad \varphi_2=\text{const.}
}
They are precisely the two branches of the supersymmetric embedding
$z_1=0$ first proposed in \cite{Ouyang:2003df}. Each branch realizes a
$U(N_f)$ symmetry group, giving the total flavor symmetry group
$U(N_f)\times U(N_f)$ of massless flavors (a diagonal axial $U(1)_A$ is
anomalous in field theory, which is dual to the corresponding gauge field
getting massive in string theory through Green-Schwarz mechanism).
We choose these embeddings because of the following properties: they reach the
tip of the cone and intersect the color D3-branes; wrap the $U(1)_R$
circle corresponding to rotations $\psi\to\psi+\alpha$; are invariant
under radial rescalings. So they realize in field theory massless flavors,
without breaking explicitly the $U(1)_R$ and the conformal symmetry.
Actually, they are both broken by quantum effects. Moreover the
configuration does not break the $\mathbb{Z}_2$ symmetry of the conifold
solution which corresponds to exchanging the two gauge groups.

The fact that we must include both the branches is due to D7-charge
tadpole cancellation, which is dual to the absence of gauge anomalies in
field theory. An example of a (non-singular) 2-submanifold in the conifold
geometry is $\mathcal{D}_2=\{\theta_1=\theta_2, \, \varphi_1=2\pi-\varphi_2, \, \psi=\text{const},
\, r=\text{const}\}$. The charge distributions of the two branches are
\begin{equation}
\omega^{(1)} = \sum\nolimits_{N_f} \delta^{(2)}(\theta_1,\varphi_1) \, d\theta_1\wedge d\varphi_1 \qquad \qquad \omega^{(2)} = \sum\nolimits_{N_f} \delta^{(2)}(\theta_2,\varphi_2) \, d\theta_2\wedge d\varphi_2 \;,
\end{equation}
where the sum is over the various D7-branes, possibly localized at different points, and a correctly normalized scalar delta function (localized on an 8-submanifold) is $\delta^{(2)}(x) \sqrt{-\hat{G}_8}/\sqrt{-G}$.
Integrating the two D7-charges on the 2-submanifold we get:
\begin{equation}
\int_{\mathcal{D}_2} \omega^{(1)} = - N_f \qquad \qquad \int_{\mathcal{D}_2} \omega^{(2)} = N_f \;.
\end{equation} 
Thus, whilst the two branches have separately non-vanishing tadpole, putting an equal number of them on the two sides the total D7-charge cancels. This remains valid for all (non-singular) 2-submanifolds.

The embedding can be deformed into a single D7 that only reaches a minimum
radius, and realizes a merging of the two branches. This corresponds to
giving mass to flavors and explicitly breaking the flavor symmetry to
$SU(N_f)$ and the R-symmetry completely. These embeddings were also found
in \cite{Arean:2004mm}.

Each embedding preserves the same four supercharges, irrespectively to
where the branes are located on the two 2-spheres parameterized by
$(\theta_1,\varphi_1)$ and $(\theta_2,\varphi_2)$. Thus we can smear the
distribution and still preserve the same amount of supersymmetry. The
2-form charge distribution is readily obtained to be the same as the
volume forms on the two 2-spheres in the geometry, and through the
modified Bianchi identity it sources the flux $F_1$.%
\footnote{The modified Bianchi identity of $F_1$ is obtained from the
Wess-Zumino action term with $F_1 = -e^{-2\phi}\ast F_9$.}
We expect to obtain a solution where all the functions have only radial
dependence. Moreover we were careful in never breaking the $\mathbb{Z}_2$
symmetry that exchanges the two spheres. The natural ansatz is:
\begin{align}
\begin{split}
ds^2 &= h(r)^{-1/2} dx_{1,3}^2 + h(r)^{1/2} \biggl\{ dr^2 + \\
&\qquad+\,\frac{e^{2g(r)}}{6} \sum_{i=1,2} ( d\theta_i^2 + \sin^2 \theta_i \,
d\varphi_i^2)  + \frac{e^{2f(r)}}{9} (d\psi + \sum_{i=1,2} \cos\theta_i \,
d\varphi_i)^2 \biggr\} 
\end{split} \\
\phi &= \phi(r) \\
F_5 &= K(r) \, h(r)^{3/4} \Big( e^{x^0x^1x^2x^3r} -
e^{\theta_1\varphi_1\theta_2\varphi_2\psi}
\Big) \\
F_1 &= \frac{N_f}{4\pi} \bigl( d\psi + \cos\theta_1\,
d\varphi_1 + \cos\theta_2\, d\varphi_2 \bigr) = \frac{3N_f}{4\pi} \, h(r)^{-1/4} e^{-f(r)} \, e^{\psi} \\
dF_1 &= -\frac{N_f}{4\pi} \bigl(\sin\theta_1 \, d\theta_1\wedge d\varphi_1 +
\sin\theta_2 \, d\theta_2\wedge d\varphi_2 \bigr) \;,
\end{align}
where the unknown functions are $h(r)$, $g(r)$, $f(r)$, $\phi(r)$ and
$K(r)$. The angular coordinates $\theta_i$ are defined in $[0,\pi]$ while the others have fundamental domain $\varphi_i\in [0,2\pi)$ and $\psi\in[0,4\pi)$ with appropriate patching rules\footnote{\label{patching}The correct patching rules on $T^{1,1}$ in the
coordinates of \eqref{kw} are:
\begin{equation}
\psi \equiv \psi + 4\pi \;, \qquad \binom{\varphi_1}{\psi} \equiv
\binom{\varphi_1+2\pi}{\psi+2\pi} \;, \qquad \binom{\varphi_2}{\psi} \equiv
\binom{\varphi_2+2\pi}{\psi+2\pi} \nn \;.
\end{equation}
In fact the space is a $U(1)$ fibration over $S^2\times S^2$. The first identification is just the one of the fiber. On the base 2-spheres we must identify the angular variables according to $\varphi_i \equiv \varphi_i + 2\pi$, but this could be accompanied by a shift in the fiber. To understand it, draw the very short (in proper length) path around the point $\theta_1=0$: $\theta_1\ll 1$, $\varphi_1=t=4\pi-\psi$ with $t\in[0,2\pi]$ a parameter along the path. To make it closed, a rotation in $\varphi_1$ must be accompanied by an half-rotation in $\psi$. This gives the second identification.
}.
The vielbein is:
\begin{align} \label{T11frame}
\begin{split}
e^{x^i} &= h^{-1/4} \, dx^i \\
e^{\theta_i} &= \frac{1}{\sqrt{6}} h^{1/4} e^g \, d\theta_i \\
e^{\psi} &= \frac{1}{3} h^{1/4} e^f \,
( d\psi + \cos\theta_1\, d\varphi_1 + \cos\theta_2\, d\varphi_2) \;.
\end{split} \begin{split}
e^r &= h^{1/4} \, dr \\
e^{\varphi_i} &= \frac{1}{\sqrt{6}} h^{1/4} e^g \, \sin\theta_i d\varphi_i \\
\phantom{X}&
\end{split} \end{align}
With this ansatz the field equation $d\big(e^{2\phi}\ast F_1)=0$ is
automatically satisfied, as well as the self-duality condition $F_5=\ast
F_5$. The Bianchi identity $dF_5=0$ gives:
\begin{equation}
K \, h^2 \, e^{4g+f} = 27\pi N_c\;,
\end{equation}
and $K(r)$ can be solved. The previous normalization comes from Dirac
quantization of the D3-brane charge:
\begin{equation}
\int_{T^{1,1}} F_5 = 2\kappa_{10}^2 T_3 \, N_c = (2\pi)^4 N_c\;,
\label{F5quantization}
\end{equation}
using a suitable orientation for the volume form of the $T^{1,1}$  space and 
the fact that  $Vol(T^{1,1})= \frac{16}{27}\pi^3$.

We impose that the ansatz preserves the same four supersymmetries as the
probe D7-branes on the Klebanov-Witten solution. With this purpose, let us 
write the supersymmetric variations of the
dilatino and gravitino in type IIB supergravity. For a background of the
type we are analyzing, these variations are:
\begin{align} \label{complexvariations}
\delta_{\epsilon}\,\lambda &= \frac{1}{2}\, \Gamma^M\, \Bigl(\,
\partial_M\,\phi\,-\,ie^{\phi}\, F^{(1)}_M\,\Bigr)\, {\epsilon} \nonumber \\
\delta_{\epsilon}\,\psi_M &= \nabla_M\,\epsilon\,+\,i\,\frac{e^\phi}{4}
F^{(1)}_M\,\epsilon\,+\,\frac{i}{1920} F^{(5)}_{PQRST}\, 
\Gamma^{PQRST}\,
\Gamma_M\, 
{\epsilon} \;,
\end{align}
where we have adopted the formalism in which $\epsilon$ is a complex
Weyl spinor of fixed ten-dimensional chirality (see Appendix \ref{SUSYapp}). It
turns out (see Section \ref{SEBPS}) that the Killing spinors $\epsilon$ (which
solve the equations $\delta_\epsilon \, \lambda = \delta_\epsilon \, \psi_M = 0$) in  the
frame basis \eqref{T11frame} can be written as:
\begin{equation}
\epsilon = h^{-\frac{1}{8}}\,\,e^{-\frac{i}{2}\psi}\,\,\eta
\end{equation}
where $\eta$ is a constant spinor which satisfies
\begin{align}
&\Gamma_{x^0x^1x^2x^3}\,\eta = -i\eta \nonumber\\
&\Gamma_{\theta^1\varphi^1}\,\eta =
\Gamma_{\theta^2\varphi^2}\,\eta =i\eta\,\,,\qquad\qquad
\Gamma_{r\psi} \, \eta = -i\eta \;.
\end{align}
Moreover, from (\ref{complexvariations}) we get the following system  of first-order BPS
differential equations:
\begin{equation} \label{BPSsystem}
\left\{ \begin{aligned}
g' &= e^{f-2g} \\
f' &= e^{-f} (3-2e^{2f-2g}) - \frac{3N_f}{8\pi} e^{\phi-f} \\
\phi' &= \frac{3N_f}{4\pi} e^{\phi-f} \\
h' &= -27\pi N_c \, e^{-f-4g}
\end{aligned} \right.
\end{equation}

Notice that taking $N_f=0$ in the BPS system \eqref{BPSsystem} we simply
get equations for a deformation of the Klebanov-Witten solution without
any addition of flavor branes. Solving the system we find both the
original KW background and the solution for D3-branes at a conifold
singularity, as well as other solutions which correspond on the gauge
theory side to giving VEV to dimension 6 operators. These solutions 
were considered in \cite{PandoZayas:2001iw,Benvenuti:2005qb}, 
and are shown to follow from our system in appendix \ref{appendixb}.

In order to be sure that the BPS equations  (\ref{BPSsystem}) capture the correct dynamics, we have to check that the Einstein, Maxwell 
and dilaton equations are solved. This can be done even before finding
actual solutions of the BPS system. We checked that the first-order system
\eqref{BPSsystem} (and the Bianchi identity) in fact \textit{implies} the
second order Einstein, Maxwell  and dilaton differential equations. An analytic
general proof will be given in Section \ref{BPS-Einstein}. 
 In the coordinate basis the stress-energy tensor
\eqref{tmunu} is computed  to be:
\begin{align}
\begin{split}
T_{\mu\nu} &= -\frac{3N_f}{2\pi} h^{-1} e^{\phi-2g} \, \eta_{\mu\nu} \\
T_{rr} &= -\frac{3N_f}{2\pi} e^{\phi-2g} \\
T_{\theta_i \theta_i} &= -\frac{N_f}{8\pi} e^\phi
\end{split}
\begin{split}
T_{\varphi_i \varphi_i} &= -\frac{N_f}{24\pi} e^{\phi-2g} \Big[
4e^{2f}\cos^2\theta_i + 3e^{2g}\sin^2\theta_i \Big] \\
T_{\varphi_1 \varphi_2} &= - \frac{N_f}{6\pi} e^{\phi+2f-2g} \cos\theta_1
\cos\theta_2 \\
T_{\varphi_i \psi} &= -\frac{N_f}{6\pi} e^{\phi+2f-2g} \cos\theta_i \\
T_{\psi\psi} &= -\frac{N_f}{6\pi} e^{\phi+2f-2g} \;.
\end{split}
\label{Tmunu}
\end{align}
It is correctly linear in $N_f$.
 We did not explicitly check the Dirac-Born-Infeld
equations for the D7-brane distribution. We expect them to be solved
because of \hyph{\kappa}symmetry (supersymmetry) on their world-volume.

\subsubsection*{Solution with General Couplings}

We can generalize our set of solutions by switching on non-vanishing VEVs for the bulk gauge potentials $C_2$ and $B_2$. We show that this can be done without modifying the previous set of equations, and the two parameters are present for every solution of them. The condition is that the gauge potentials are flat, that is with vanishing field-strength. They correspond thus to (higher rank) Wilson lines for the corresponding bundles.

Let us switch on the following fields:
\begin{equation}
C_2 = c \, \omega_2 \qquad \qquad B_2 = b \, \omega_2 \;,
\end{equation}
where the 2-form $\omega_2$ is Poincar\'e dual to the 2-cycle $\mathcal{D}_2$:
\begin{gather}
\mathcal{D}_2 = \{\theta_1=\theta_2,\: \varphi_1=2\pi-\varphi_2,\: \psi=\text{const},\:  r=\text{const} \} \\
\omega_2 = \frac{1}{8\pi} \big( \sin\theta_1 \, d\theta_1\wedge d\varphi_1 - \sin\theta_2 \, d\theta_2\wedge d\varphi_2 \big) \,\,,
\qquad \qquad \int_{\mathcal{D}_2}\omega_2 = 1 \;.
\end{gather}
We see that $F_{(3)}=0$ and $H_{(3)}=0$. So the supersymmetry variations are not modified, neither are the gauge invariant field-strength definitions. In particular the BPS system \eqref{BPSsystem} does not change.

Consider the effects on the action (the argument is valid both for localized and smeared branes). It can be written as a bulk term plus the D7-brane terms:
\begin{equation}
S = S_{bulk} - T_7 \int d^8\xi \, e^\phi \, \sqrt{-\det(\hat G_8 + \mathcal{F})} + T_7 \int \Big[ \sum\nolimits_q \hat C_q \wedge e^\mathcal{F} \Big]_8 \;,
\end{equation} 
with $\mathcal{F}=\hat B_2 + 2\pi\alpha'\, F$ is the D7 gauge invariant field-strength, and hat means pulled-back quantities. To get solutions of the \hyph{\kappa}symmetry conditions and of the equations of motion, we must take $F$ such that
\begin{equation}
\mathcal{F}=\hat B_2 + 2\pi\alpha'\, F =0\;.
\end{equation}
Notice that there is a solution for $F$ because $B_2$ is flat: $d\hat B_2=\widehat{dB_2}=0$. With this choice \hyph{\kappa}symmetry is preserved as before, since it depends on the combination $\mathcal{F}$. The dilaton equation is fulfilled. The Bianchi identities and the bulk field-strength equations of motion are not modified, since the WZ term only sources $C_8$. The energy momentum tensor is not modified, so the Einstein equations are fulfilled. The last steps are the equations of $B_2$ and $A_1$ (the gauge potential on the D7). For this notice that they can be written:
\begin{align}
d \frac{\delta S}{\delta F} &= 2\pi\alpha' \: d \frac{\delta S_{brane}}{\delta \mathcal{F}} =0\\
\frac{\delta S}{\delta B_2} &= \frac{\delta S_{bulk}}{\delta B_2} + \frac{\delta S_{brane}}{\delta \mathcal{F}} =0\:.
\end{align}
The first is solved by $\mathcal{F}=0$ since in the equation all the terms are linear or higher order in $\mathcal{F}$. This is because the brane action does not contain terms linear in $\mathcal{F}$, and this is true provided $C_6=0$ (which in turn is possible only if $C_2$ is flat). The second equation then reduces to $\frac{\delta S_{bulk}}{\delta B_2}=0$, which amounts to $d(e^{-\phi}\ast H_3)=0$ and is solved.

As we will see in Section \ref{field theory analysis}, being able to switch on arbitrary constant values $c$ and $b$ for the (flat) gauge potentials, we can freely tune the two gauge couplings (actually the two renormalization invariant scales $\Lambda$'s) and the two theta angles \cite{Klebanov:1998hh,Morrison:1998cs}. This turns out to break the $\mathbb{Z}_2$ symmetry that exchanges the two gauge groups, even if the breaking is mild and only affects $C_2$ and $B_2$, while the metric and all the field-strength continue to have that symmetry. However this does not modify the behavior of the gauge theory.

\subsection{The Solution in Type IIB Supergravity}
\label{solKW}

The BPS system \eqref{BPSsystem} can be solved through the change of
radial variable
\begin{equation}
e^f \frac{d}{dr} \equiv \frac{d}{d\rho} \qquad \Rightarrow \qquad e^{-f}dr
= d\rho \;.
\end{equation}
We get the new system:
\begin{align}
\dot{g} &= e^{2f-2g} \label{eqng} \\
\dot{f} &= 3 - 2 e^{2f-2g} - \frac{3N_f}{8\pi} e^\phi \label{eqnf} \\
\dot{\phi} &= \frac{3N_f}{4\pi} e^\phi \label{eqnphi} \\
\dot{h} &= -27\pi N_c \, e^{-4g} \label{eqnh} \;,
\end{align}
where derivatives are taken with respect to $\rho$.

Equation \eqref{eqnphi} can be solved first. By absorbing an integration
constant in a shift of the radial coordinate $\rho$, we get
\begin{equation}
e^{\phi} = - \frac{4\pi}{3 N_f} \frac{1}{\rho} \qquad \qquad \Rightarrow
\qquad \qquad \rho<0 \;.
\label{dilatonKW}
\end{equation}
The solution is thus defined only up to a maximal radius
$\rho_\text{MAX}=0$ where the dilaton diverges. As we will see, it
corresponds to a Landau pole in the ultraviolet (UV) of the gauge theory.
On the contrary for $\rho\to -\infty$, which corresponds in the gauge
theory to the infrared (IR), the string coupling goes to zero. Note
however that the solution could stop at a finite negative
$\rho_\text{MIN}$ due to integration constants or, for example, more dynamically,  due to the presence of massive flavors. Then define
\begin{equation}
u = 2f-2g \qquad \qquad \Rightarrow \qquad \qquad \dot{u} = 6(1-e^u) +
\frac{1}{\rho}\;,
\end{equation}
whose solution is
\begin{equation}
e^u = \frac{-6\rho \, e^{6\rho}}{(1-6\rho) e^{6\rho} + c_1} \;.
\end{equation}

The constant of integration $c_1$ cannot be reabsorbed, and according to
its value the solution dramatically changes in the IR. A systematic
analysis of the various behaviors is presented in Section \ref{asymp analysis}. The
value of $c_1$ determines whether there is a (negative) minimum value for
the radial coordinate $\rho$. The requirement that the function $e^u$ be
positive defines three cases:
\begin{equation} \nn \begin{split}
-1<c_1<0 \qquad &\rightarrow \qquad \sub{\rho}{MIN} \leq \rho \leq 0 \\
c_1 =0 \qquad &\rightarrow \qquad -\infty < \rho \leq 0 \\
c_1 > 0 \qquad &\rightarrow \qquad -\infty < \rho \leq 0 \;.
\end{split} \end{equation}
In the case $-1<c_1<0$, the minimum value $\rho_\text{MIN}$ is given by an
implicit equation. It can be useful to plot this value as a function of
$c_1$:
\\[0.3cm]
\parbox[c]{0.55\textwidth}{\quad\includegraphics[width=0.45\textwidth]{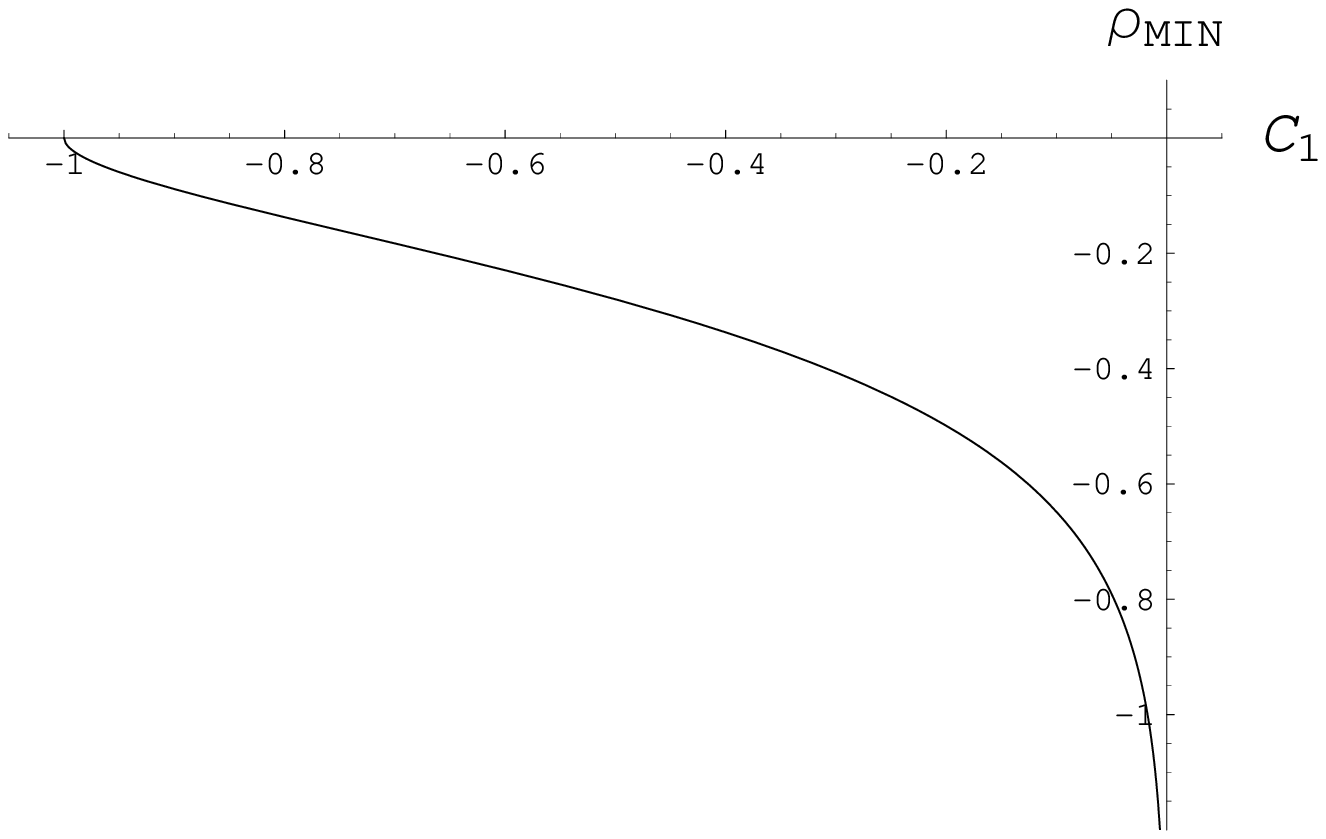}}
\parbox[c]{0.3\textwidth}{\begin{equation}
0 = (1-6\rho_\text{MIN}) \, e^{6\rho_\text{MIN}} + c_1 \nonumber
\end{equation}
}
\\[0.3cm]
As it is clear from the graph, as $c_1\to-1^+$ the range of the solution
in $\rho$ between the IR and the UV Landau pole shrinks to zero size,
while in the limit $c_1\to 0^-$ we no longer have a minimum radius.

The functions $g(\rho)$ and $f(\rho)$ can be analytically integrated,
while the warp factor $h(\rho)$ and the original radial coordinate
$r(\rho)$ cannot (in the particular case $c_1=0$ we found an explicit
expression for the warp factor). By absorbing an irrelevant integration
constant into a rescaling of $r$ and $x^{0,1,2,3}$, we get:
\begin{align}
e^g &= \Big[ (1-6\rho)e^{6\rho} + c_1 \Big]^{1/6} \label{egKW}\\
e^f &= \sqrt{-6\rho} \, e^{3\rho} \Big[ (1-6\rho)e^{6\rho} + c_1
\Big]^{-1/3} \\
h(\rho) &= -27\pi N_c \int_0^\rho e^{-4g} + c_2 \label{warp}\\
r(\rho) &= \int^\rho e^f \;.
\end{align}
This solution is a very important  result of our paper. We accomplished in finding a
supergravity solution describing a (large) $N_f$ number of backreacting
D7-branes, smeared on the background produced by D3-branes at the tip of a
conifold geometry.

The constant $c_1$ and $c_2$ correspond in field theory to switching on VEV's for relevant operators, as we will see in Section \ref{irtheory}. Moreover, in the new radial coordinate $\rho$, the metric reads
\begin{equation} \begin{split}
ds^2 &= h^{-\frac{1}{2}} dx_{1,3}^2 + h^{\frac{1}{2}} \, e^{2f} \, \bigg\{ d\rho^2 +
\frac{e^{2g-2f}}{6} 
\sum_{i=1,2} ( d\theta_i^2 + \sin^2 \theta_i \, d\varphi_i^2)
+ \frac{1}{9} (d\psi + \sum_{i=1,2} \cos\theta_i \, d\varphi_i)^2\bigg\} \;.
\end{split} \end{equation}

\subsection{Analysis of the Solution: Asymptotics and Singularities} \label{asymp analysis}

We perform here a systematic analysis of the possible solutions of the BPS system, and study the asymptotics in the IR and in the UV. 
In this section we will make use of the following formula for the Ricci scalar curvature, which can be obtained for solutions of the BPS system:
\begin{equation} \label{ricci scalar}
R=-2 \, \frac{3N_f}{4\pi} \, h^{-1/2} e^{-2g+\frac{1}{2}\phi} \, \bigg[7 + 4 \frac{3N_f}{4\pi} e^{2g-2f+\phi}  \bigg]\;.
\end{equation}

\subsubsection{The Solution with $c_1=0$}
Although the warp factor $h(\rho)$ cannot be analytically integrated in general, it can be if the integration constant $c_1$ is equal to $0$.
Indeed, introducing the \emph{incomplete gamma function}, defined as follows:
\begin{equation}
\Gamma[a,x] \equiv \int_x^\infty t^{a-1} e^{-t} dt \xrightarrow[x\to-\infty]{} e^{i2\pi a} e^{-x} \left(\frac{1}{x}\right)^{1-a} \Big\{ 1 + \mathcal{O}\Big(\frac{1}{x}\Big) \Big\}\;,
\end{equation} 
we can integrate 
\begin{equation} \begin{split}
h(\rho) &= - 27\pi N_c \int d\rho \frac{e^{-4\rho}}{(1-6\rho)^{2/3}} + c_2 =\\
&= \frac{9}{2}\pi N_c (\frac{3}{2e^2})^{1/3} \Gamma[\frac{1}{3},-\frac{2}{3}+4\rho] + c_2 \simeq \\
&\simeq \frac{27}{4} \pi N_c (-6\rho)^{-2/3} e^{-4\rho}  \:\;\text{for} \:\;\rho \to -\infty\;.
\end{split} \end{equation} 
The warp factor diverges for $\rho\to -\infty$, and the integration constant $c_2$ disappears in the IR. Moreover, if we integrate the proper line element $ds$ from a finite point to $\rho=-\infty$, we see that the throat has an \emph{infinite invariant length}.

The function $r(\rho)$ cannot be given as an analytic integral, but using the asymptotic behavior of $e^f$ for $\rho\to -\infty$ we can approximately integrate it: 
\begin{equation}
r(\rho) \simeq  6^{1/6} \Big[ (-\rho)^{1/6} e^\rho + \frac{1}{6} \Gamma[\frac{1}{6},-\rho] \Big] + c_3
\end{equation} 
in the IR. Fixing $r\to 0$ when $\rho\to -\infty$ we set $c_3=0$. We approximate further on
\begin{equation}\label{r(rho)}
r(\rho) \simeq 
 (-6\rho)^{1/6} e^\rho\;. 
\end{equation}  
Substituting $r$ in the asymptotic behavior of the functions appearing in the metric, we find that up to logarithmic corrections of relative order $1/|\log(r)|$:
\begin{equation} 
\begin{split}
e^{g(r)} &\simeq e^{f(r)} \simeq r \\
h(r) &\simeq \frac{27\pi N_c}{4} \frac{1}{r^4}\;.
\end{split} 
\end{equation}
Therefore the geometry approaches $AdS_5 \times T^{1,1}$ with logarithmic corrections in the IR limit $\rho\to -\infty$.
 
\subsubsection{UV Limit}
The solutions with backreacting flavors have a Landau pole in the ultraviolet ($\rho\to 0^-$), since the dilaton diverges (see (\ref{dilatonKW})). The asymptotic behaviors of the functions appearing in the metric are: 
\begin{align}
e^{2g} &\simeq (1+c_1)^{1/3}  \Big[ 1 - \frac{6\rho^2}{1+c_1} + \mathcal{O}(\rho^3) \Big] \\
e^{2f} &\simeq - 6\rho \, (1+c_1)^{-2/3} \Big[ 1 + 6\rho + \mathcal{O}(\rho^2) \Big] \\
h &\simeq c_2 + 27\pi N_c (1+c_1)^{-2/3} \Big[ -\rho -\frac{4}{1+c_1} \rho^3 + \mathcal{O}(\rho^4) \Big]\,\,.
\end{align}
Note that we have used (\ref{warp}) for  the warp factor.
One concludes  that $h(\rho)$ is monotonically decreasing with $\rho$; if it is positive at some radius, then it is positive  down to the IR. If the integration constant $c_2$ is larger than zero, $h$ is always positive and approaches $c_2$ at the Landau pole (UV). If $c_2=0$, then $h$ goes to zero at the pole. If $c_2$ is negative, then the warp factor vanishes at $\sub{\rho}{MAX}<0$ before reaching the pole (and the curvature diverges there). The physically relevant solutions seem to have $c_2>0$.

The curvature invariants, evaluated in string frame, diverge when $\rho\to 0^-$, indicating that the supergravity description cannot be trusted in the UV. For instance the Ricci scalar $R\sim (-\rho)^{-5/2}$ if $c_2 \neq 0$, whereas $R\sim (-\rho)^{-3}$ if $c_2 = 0$.
If $c_2<0$, then the Ricci scalar $R\sim (\sub{\rho}{MAX}-\rho)^{-1/2}$ when $\rho\to \sub{\rho}{MAX}^-$.

\subsubsection{IR Limit}

The IR ($\rho\to -\infty$) limit of the geometry of the flavored solutions is independent of the number of flavors, if we neglect logarithmic corrections to the leading term. Indeed, at the leading order, flavors decouple from the theory in the IR (see the discussion below eq. \eqref{betas}). The counterpart in our supergravity plus branes solution is evident when we look at the BPS system \eqref{BPSsystem}: when $\rho\to -\infty$ the $e^{\phi}$ term disappears from the system, together with all the backreaction effects of the D7-branes (see Appendix \ref{appAlt} for a detailed analysis of this phenomena), therefore the system reduces to the unflavored one.
\begin{itemize}
\item $c_1=0$ \\
The asymptotics of the functions appearing in the metric in the IR limit $\rho\to -\infty$ are:
\begin{align}
e^{g} &\simeq e^{f}\simeq (-6\rho)^{1/6} e^\rho \label{IR asymp1}\\
h &\simeq  \frac{27}{4} \pi N_c (-6\rho)^{-2/3} e^{-4\rho}\;. \label{IR asymp2}
\end{align}
Formula \eqref{ricci scalar} implies that the scalar curvature in string frame vanishes in the IR limit: $R^{(S)} \sim (-\rho)^{-1/2}\to 0$. An analogous but lengthier formula for the square of the Ricci tensor gives
\be \label{singul}
R^{(S)}_{MN} R^{(S)\,MN} = \frac{160}{9\pi^2} \, \frac{N_f}{N_c} (-\rho) + \cO(1) \quad \to \quad \infty \;,
\ee 

thus the supergravity description presents a singularity and some care is needed when computing observables from it.  The same quantities in Einstein frame have limiting behavior $R^{(E)} \sim (-\rho)^{-1/2}\to 0$ and $R^{(E)}_{MN} R^{(E)\,MN} \to 640/(27\pi N_c)$.

\item $c_1>0$\\
The asymptotics in the limit $\rho\to -\infty$ are:
\begin{align}
e^{g} &\simeq c_1^{1/6}\\
e^{f} &\simeq c_1^{-1/3} (-6\rho)^{1/2} e^{3\rho} \\
h &\simeq  27\pi N_c c_1^{-2/3} (-\rho)\;.
\end{align}
Although the radial coordinate ranges down to $-\infty$, the throat has a \emph{finite invariant length}.
The Ricci scalar in string frame is $R\sim(-\rho)^{-3} e^{-6\rho}\to -\infty$.

\item $c_1<0$\\
In this case the IR limit is $\rho\to\sub{\rho}{MIN}$. The asymptotics in this limit are:
\begin{align}
e^{g} &\simeq \big(-6\sub{\rho}{MIN} e^{6\sub{\rho}{MIN}} \big)^{1/6} (6\rho-6\sub{\rho}{MIN})^{1/6}   \\
e^{f} &\simeq \big(-6\sub{\rho}{MIN} e^{6\sub{\rho}{MIN}} \big)^{1/6} (6\rho-6\sub{\rho}{MIN})^{-1/3}   \\
h & \simeq  const.>0\;.
\end{align}
The throat has a \emph{finite invariant length}.
The Ricci scalar in string frame is $R\sim  (\rho-\sub{\rho}{MIN})^{-1/3} \to \infty$.
\end{itemize}
Using the criterion in \cite{Maldacena:2000mw}, that proposes the IR singularity to be physically acceptable if $g_{tt}$ is bounded near the IR problematic point, we observe that these
singular geometries are all acceptable. Gauge theory physics can be read from these supergravity backgrounds. We call them ``good singularities''.

\subsection{Detailed Study of the Dual Field Theory} \label{field theory analysis}

In this section we are going to undertake a detailed analysis of the dual gauge theory features, reproduced by the supergravity solution. The first  issue we want to address is what is the effect of the smearing on the gauge theory dual.

As we wrote above, the addition to the supergravity solution of one stack of localized noncompact D7-branes at $z_1=0$ put in the field theory flavors coupled through a superpotential term
\begin{equation}
W =\lambda \, \Tr(A_i B_k A_j B_l) \, \epsilon^{ij} \epsilon^{kl} + h_1 \, \tilde{q}^a A_1 Q_a + h_2 \,\tilde{Q}^a B_1 q_a \;,
\end{equation} 
where we explicitly wrote the flavor indices $a$. For this particular embedding the two branches are localized, say, at $\theta_1=0$ and $\theta_2=0$ respectively on the two spheres. One can exhibit a lot of features in common with the supergravity plus D7-branes solution:
\begin{itemize}
\item the theory has $U(N_f)\times U(N_f)$ flavor symmetry (the diagonal axial $U(1)_A$ is anomalous), each group corresponding to one branch of D7's;
\item putting only one branch there are gauge anomalies in QFT and a tadpole in SUGRA, while for two branches they cancel;
\item adding a mass term for the fundamentals the flavor symmetry is broken to the diagonal $U(N_f)$, while in SUGRA there are embeddings moved away from the origin for which the two branches merge.
\end{itemize}

The $SU(2) \times SU(2)$ part of the isometry group of the background without
D7's is broken by the presence of localized branes. It amounts to separate rotations of the two $S^2$ in the geometry and shifts the
location of the branches. Its action is realized through the
superpotential, and exploiting its action we can obtain the
superpotential for D7-branes localized in other places.
The two bifundamental doublets $A_j$ and $B_j$ transform as spinors of the respective $SU(2)$. So the flavor superpotential term for a configuration in which the two branches are located at $x$ and $y$ on the two spheres can be obtained by identifying two rotations that bring the north pole to $x$ and $y$. There is of course a $U(1)\times U(1)$ ambiguity in this. Then we have to act with the corresponding $SU(2)$ matrices $U_x$ and $U_y$ on the vectors $(A_1,A_2)$ and $(B_1,B_2)$ (which transform in the $(\mathbf{2},1)$ and $(1,\mathbf{2})$  representations) respectively, and select the first vector component. In summary we can write
\footnote{In case the two gauge couplings and theta angles are equal, we could appeal to the $\mathbb{Z}_2$ symmetry that exchanges them to argue $|h_1|=|h_2|$, but no more because of the ambiguities.}
\begin{equation}
W_f = h_1 \: \tilde q^x \Bigl[ U_x \binom{A_1}{A_2} \Bigr]_1 Q_x + h_2 \: \tilde Q^y \Bigl[ U_y \binom{B_1}{B_2} \Bigr]_1 q_y \;,
\end{equation}
where the notation $\tilde q^x$, $Q_x$ stands for the flavors coming from a first D7 branch being at $x$, and the same for a second D7 branch at $y$.

To understand the fate of the two phase ambiguities in the couplings $h_{1}$ and
$h_{2}$, we appeal to symmetries. The $U(1)$ action which gives $(q,\tilde q,Q,\tilde Q)$ charges $(1,-1,-1,1)$ is a symmetry explicitly broken by the flavor superpotential. The freedom of redefining the flavor fields acting with this $U(1)$ can be exploited to reduce to the case in which the phase of the two holomorphic couplings is the same. The $U(1)$ action with charges $(1,1,1,1)$ is anomalous with equal anomalies for both the gauge groups, and it can be used to absorb the phase ambiguity into a shift of the sum of Yang-Mills theta angles $\theta^{YM}_1+\theta^{YM}_2$ (while the difference holds steady). 
This is what happens for D7-branes on flat spacetime. The ambiguity we mentioned amounts to rotations of the transverse $\mathbb{R}^2$ space, whose only effect is a shift of $C_0$. As we show in the next section, the value of $C_0$ is our way of measuring the sum of theta angles through probe D(-1)-branes. Notice that if we put in our setup many separate stacks of D7's, all their superpotential $U(1)$ ambiguities can be reabsorbed in a single shift of $C_0$.

From a physical point of view, the smearing corresponds to put the D7-branes at different points on the two spheres, distributing each branch on one of the 2-spheres.
This is done homogeneously so that there is one D7 at every point of $S^2$. The non-anomalous flavor symmetry is broken from $U(1)_B\times SU(N_f)_R\times SU(N_f)_L$ (localized configuration) to $U(1)_B\times U(1)^{N_f-1}_V\times U(1)^{N_f-1}_A$ (smeared configuration).%
\footnote{The axial $U(1)$ which gives charges $(1,1,-1,-1)$ to one set of fields
$(q_x,\tilde q^x,Q_x, \tilde Q^x)$ coming from a single D7, is an anomalous symmetry.
For every D7-brane we consider, the anomaly amounts to a shift of the
same two theta angles of the gauge theory. So we can combine this $U(1)$
with an axial rotation of all the flavor fields, and get an anomaly
free symmetry. In total, from $N_f$ D7's we can find $N_f-1$ such anomaly
free axial $U(1)$ symmetries.}

Let us introduce a pair of flavor indices $(x,y)$ that naturally 
live on $S^2\times S^2$ and specify the D7. 
The superpotential for the whole system of smeared D7-branes is just the sum (actually an integral) over the indices $(x,y)$ of the previous contributions:
\begin{equation} \label{wflavors}
W = \lambda \, \Tr(A_i B_k A_j B_l) \, \epsilon^{ij} \epsilon^{kl} + h_1 \: \int_{S^2} d^2x \, \tilde q^x \Big[ U_x \binom{A_1}{A_2} \Big]_1 Q_x + h_2 \int_{S^2} d^2y \, \tilde Q^y \Big[ U_y \binom{B_1}{B_2} \Big]_1 q_y \;.
\end{equation} 
Again, all the $U(1)$ ambiguities have been reabsorbed in field redefinitions and a global shift of $\theta^{YM}_1+\theta^{YM}_2$.

In this expression the $SU(2)_A\times SU(2)_B$ symmetry is manifest: rotations of the bulk fields $A_j$, $B_j$ leave the superpotential invariant because they can be reabsorbed in rotations of the dummy indices $(x,y)$.
In fact, the action of $SU(2)_A\times SU(2)_B$ on the flavors is a subgroup of the broken $U(N_f)\times U(N_f)$ flavor symmetry. In the smeared configuration, there is a D7-brane at each point of the spheres and the group $SU(2)^2$ rotates all the D7's in a rigid way, moving each D7 where another was. So it is a flavor transformation contained in $U(N_f)^2$. By combining this action with a rotation of $A_i$ and $B_i$, we get precisely the claimed symmetry.

Even if written in an involved fashion, the superpotential \eqref{wflavors} does not spoil the features of the gauge theory. In particular, the addition of a flavor mass term still would give rise to the symmetry breaking pattern
\begin{equation}
U(1)_B\times U(1)^{N_f-1}_V\times U(1)^{N_f-1}_A \quad \to \quad U(1)^{N_f}_V \;. \nn
\end{equation}

\subsubsection{Holomorphic Gauge Couplings and $\beta$-functions}

In order to extract information on the gauge theory from the supergravity
solution, we need to know the holographic relations between the gauge
couplings, the theta angles and the supergravity fields. These formulae can be
properly derived only in the
orbifold $\mathbb{R}^{1,3} \times \mathbb{C}\times
\mathbb{C}^2/\mathbb{Z}_2$, where string theory can be quantized, by
considering fractional branes placed at the singularity.
The near-horizon geometry describing the IR dynamics on a stack of $N$
regular branes at the singularity is $AdS_5\times S^5/\mathbb{Z}_2$. The
dual gauge theory is an \Nugual{2} $SU(N)\times SU(N)$ SCFT with
bifundamental hypermultiplets.
In $\mathcal{N}=1$ language, an $\mathcal{N}=2$ vector multiplet
decomposes into a vector multiplet and a chiral multiplet in the adjoint
of the gauge group, whereas a bifundamental hypermultiplet decomposes into
two bifundamental chiral multiplets. Klebanov and Witten
\cite{Klebanov:1998hh} recognized that giving equal (but opposite) complex
mass parameters to the adjoint chiral superfields of this $\mathcal{N}=2$
SCFT, an RG flow starts whose IR fixed point is described by the gauge
theory dual to the $AdS_5\times T^{1,1}$ geometry.

\begin{figure}
\centering
\includegraphics[width=0.75\textwidth]{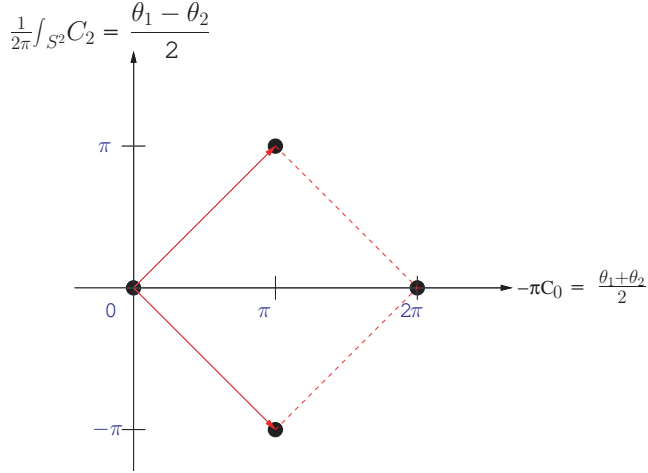}
\caption{Unit cell of the lattice of Yang-Mills $\theta$ angles and RR fields
integrals.}
\label{fig:lattice}
\end{figure}

In the \Nugual{2} orbifold theory, the holographic relations can be
derived exactly.
The result is the following:
\begin{align}
&\frac{4\pi^2}{g_1^2} + \frac{4\pi^2}{g_2^2}  = \frac{\pi e^{-\phi}}{g_s}
\label{g+}\\
&\frac{4\pi^2}{g_1^2} - \frac{4\pi^2}{g_2^2}  = \frac{e^{-\phi}}{g_s}
\bigg[\frac{1}{2\pi\alpha'}\int_{S_2} B_2 -\pi
\;\;\;(\mathrm{mod}\;2\pi)\bigg]  \label{g-} \\
&\theta_1^{YM} = -\pi C_0 +\frac{1}{2\pi} \int_{S_2} C_2
\;\;\;(\mathrm{mod}\;2\pi)\label{theta1}\\
&\theta_2^{YM} = -\pi C_0 -\frac{1}{2\pi} \int_{S_2} C_2
\;\;\;(\mathrm{mod}\;2\pi)\label{theta2}
\end{align}
where the integrals are performed over the 2-sphere that shrinks at the orbifold fixed point and could be blown-up.%
\footnote{ Actually, we would find the opposite sign in the $C_0$ term in the formulas
(\ref{theta1}) and (\ref{theta2}) for the $\theta$ angles. We are not sure about that sign. At any rate, with this minus sign the R-anomaly computation of the supergravity backgrounds of \cite{Ouyang:2003df} and of this paper match exactly the field theory computations.}
The ambiguity in \eqref{g-} is the $2\pi$ periodicity of
$\frac{1}{2\pi\alpha'}\int_{S_2} B_2$ which comes from the quantization
condition on $H_3$ (if fractional branes are absent). A shift of $2\pi$ amounts to move to a dual description of the gauge theory.%
\footnote{In the KW theory, this is Seiberg duality. Notice that the
periodicity must fail once flavor fields are added.}
The ambiguities of RR fields are more subtle: the periodicities in \eqref{theta1} and \eqref{theta2} correspond to the two kinds of fractional D(-1)-branes appearing in the theory.
The angles $\theta_1^{YM}$ and $\theta_2^{YM}$ come from the imaginary parts of the action of
the two kinds of fractional Euclidean D(-1) branes. Both of them are then
defined \emph{modulo} $2\pi$ in the quantum field theory:
\begin{equation}
(\theta_1^{YM} \,,\, \theta_2^{YM}) \equiv (\theta_1^{YM} + 2\pi \,,\, \theta_2^{YM}) \equiv (\theta_1^{YM} \,,\, \theta_2^{YM} + 2\pi)\;.
\label{theta ident}
\end{equation}
On the string theory side the periodicities exactly match: an
Euclidean fractional D(-1)-brane enters the functional
integral with a term $\exp\bigl\{-\frac{8\pi^2}{g_j^2}+i\theta_j^{YM}\bigr\}$.%
\footnote{We have written the complexified gauge coupling instead of the
supergravity fields for the sake of brevity: the use of the dictionary is
understood.}
Hence the imaginary part in the exponent is defined \emph{modulo} $2\pi$
in the quantum string theory.
The identification \eqref{theta ident} of the field theory translates on
the string side in:
\begin{equation}
(\pi C_0 \, , \, \frac{1}{2\pi}\int_{S^2} C_2)\equiv (\pi C_0+\pi \, , \,
\frac{1}{2\pi} \int_{S^2} C_2+\pi) \equiv (\pi C_0+\pi \, , \,
\frac{1}{2\pi}\int_{S^2} C_2-\pi) \;. \label{RR ident}
\end{equation}
The lattice is shown in figure \ref{fig:lattice}. The vectors of the unit
cell drawn in the figure are the ones defined by fractional branes.

From figure \ref{fig:lattice} and \eqref{RR ident} we can see that:
\begin{equation}
\pi C_0  \equiv \pi C_0 +2\pi\;.
\end{equation}
This is indeed the identification that arises from considering a regular
D(-1) brane, which can be seen as a linear superposition of the two kinds
of fractional D(-1)-branes. Notice that the closed string field $C_0$ in
this orbifold has periodicity $2$, differently from the periodicity $1$ in
flat space. This is due to the fact that in the orbifold the fundamental
physical objects are the fractional branes.

Usually in the literature the afore-mentioned holographic relations were
assumed to hold also in the conifold case.
Strassler remarked in \cite{Strassler:2005qs} that for the conifold
theory the formulae for the sum of the gauge couplings and the sum of
theta angles need to be corrected.
We expect that the formula for the sum of theta angles is
correct as far as anomalies are concerned, since anomalies
do not change in RG flows.
Instead the formula (\ref{g+}) may need to be corrected in the KW theory: in general
the dilaton could be identified with some combination of the gauge and
superpotential couplings.

Let us now make contact with our supergravity solution.
In the smeared solution, since $dF_1\neq0$ at every point, it is not
possible to define a scalar potential $C_0$ such that $F_1=dC_0$. We
by-pass this problem by restricting our attention to the non-compact
4-cycle defined by $\{\rho, \psi, \theta_1=\theta_2,\varphi_1=2\pi-\varphi_2\}$
\cite{Bertolini:2002yr}(note that it wraps the R-symmetry direction
$\psi$), so that we can pull-back on it and write
\begin{equation}
F_1^{eff}=\frac{N_f}{4\pi} \, d\psi
\end{equation}
and therefore
\begin{equation} \label{C0}
C_0^{eff}=\frac{N_f}{4\pi} (\psi-\psi_0)\;.
\end{equation}
Now we can identify:
\begin{align}
\frac{8\pi^2}{g^2} = \pi \, e^{-\phi} &= -\frac{3N_f}{4} \rho \label{g_YM 2} \\
\theta_1^{YM} + \theta_2^{YM} &= -\frac{N_f}{2} (\psi-\psi_0) \;, \label{theta_YM 2}
\end{align}
where we suppose for simplicity the two gauge couplings to be equal ($g_1=g_2\equiv g$). The
generalization to an arbitrary constant $B_2$ is straightforward since the
difference of the inverse squared gauge couplings does not run.
Although, as discussed above,  one cannot be sure of the validity of \eqref{g_YM 2}, we can try to extract some information.

Let us first compute the \hyph{\beta}function of the gauge couplings.
The identification \eqref{g+} allows us to define a ``radial''
\hyph{\beta}function that we can directly compute from supergravity
\cite{Olesen:2002nh}:
\begin{equation}\label{beta sugra}
\beta_{\frac{8\pi^2}{g^2}}^{(\rho)}\equiv \frac{\partial}{\partial\rho}
\frac{8\pi^2}{g^2}= \pi \frac{\partial e^{-\phi}}{\partial\rho} =
-\frac{3N_f}{4}\;.
\end{equation}
(Compare this result with eq. \eqref{betas}).
The physical \hyph{\beta}function defined in the field theory is of
course:
\begin{equation}\label{beta FT}
\beta_{\frac{8\pi^2}{g^2}}\equiv \frac{\partial}{\partial
\log\frac{\mu}{\Lambda}} \frac{8\pi^2}{g^2}\;,
\end{equation}
where $\mu$ is the subtraction scale and $\Lambda$ is a renormalization
group invariant scale.
In order to get the precise field theory \hyph{\beta}function from the
supergravity computation one needs the \emph{energy-radius} relation
$\rho=\rho\big( \frac{\mu}{\Lambda}\big)$, from which
$ \beta = \beta^{(\rho)} \: \partial \rho / \partial \log\frac{\mu}{\Lambda}$.
In general, for non-conformal duals, the radius-energy relation depends on
the phenomenon one is interested in and accounts for the scheme-dependence
in the field theory.

Even without knowing the radius-energy relation, there is some physical
information that we can extract from the radial \hyph{\beta}function
\eqref{beta sugra}. In particular, being the energy-radius relation
$\rho=\rho\big( \frac{\mu}{\Lambda}\big)$ monotonically increasing, the
signs of the two beta functions coincide.

In our case, using $r=\frac{\mu}{\Lambda}$ and eq. \eqref{r(rho)}, one gets matching between \eqref{betas} and \eqref{beta sugra}.

\subsubsection{R-symmetry Anomaly and Vacua}

Now we move to the computation of  the $U(1)_R$ anomaly. 
On the field theory side we follow the convention that the R-charge of the superspace Grassmann coordinates is $R[\vartheta]=1$. This fixes the R-charge of the gauginos $R[\lambda]=1$. Let us consider an infinitesimal R-symmetry transformation and calculate the $U(1)_R - SU(N_c) - SU(N_c) $ triangle anomaly. 
The anomaly coefficient in front of the instanton density of a gauge group is $\sum_f R_f T[\mathcal{R}^{(f)}]$, where the sum runs over the fermions $f$, $R_f$ is the R-charge of the fermion and $T[\mathcal{R}^{(f)}]$ is the Dynkin index of the gauge group representation $ \mathcal{R}^{(f)}$ the fermion belongs to, normalized as $ T[\mathcal{R}^{(fund.)}]=1$ and $ T[\mathcal{R}^{(adj.)}]=2 N_c$.
Consequently the anomaly relation in our theory is the following:
\begin{equation}\label{anomaly}
\partial_\mu J^\mu_R  = -\frac{N_f}{2} \, \frac{1}{32\pi^2} \big(F_{\mu\nu}^a \tilde{F}_a^{\mu\nu} + G_{\mu\nu}^a \tilde{G}_a^{\mu\nu}\big) \;,
\end{equation}
or in other words, under a $U(1)_R$ transformation of parameter $\varepsilon$, for both gauge groups the theta angles transform as
\begin{equation}\label{theta shift}
\begin{split}
\theta_i^{YM} \rightarrow \theta_i^{YM} - \frac{N_f}{2} \varepsilon \;.
\end{split}
\end{equation}

On the string/gravity side a $U(1)_R$ transformation of parameter $\varepsilon$ is realized (in our conventions) by the shift $\psi \rightarrow \psi +2 \varepsilon$.
This can be derived from the transformation of the complex variables \eqref{conifold}, which under a $U(1)_R$ rotation get $z_i\to e^{i\varepsilon}z_i$, or directly by the decomposition of the 10d spinor $\epsilon$ into 4d and 6d factors and the identification of the 4d supercharge with the 4d spinor.
By means of the dictionary \eqref{theta_YM 2} we obtain:
\begin{equation}\label{theta shift: gravity}
\theta_1^{YM} + \theta_2^{YM} \rightarrow \theta_1^{YM} + \theta_2^{YM} - 2 \,  \frac{N_f}{2} \varepsilon\;,
\end{equation}
in perfect agreement with \eqref{theta shift}.

The $U(1)_R$ anomaly is responsible for the breaking of the symmetry group, but usually a discrete subgroup survives. Disjoint physically equivalent vacua, not connected by other continuous symmetries, can be distinguished thanks to the formation of domain walls among them, whose tension could also be measured. We want to read the discrete symmetry subgroup of $U(1)_R$ and the number of vacua both from field theory and supergravity.
In field theory the $U(1)_R $ action has an extended periodicity (range of inequivalent parameters) $\varepsilon\in[0,8\pi)$ instead of the usual $2\pi$ periodicity, because the minimal charge is $1/4$. Let us remark however that when $\varepsilon$ is a multiple of $2\pi$ the transformation is not an R-symmetry, since it commutes with supersymmetry.
The global symmetry group contains the baryonic symmetry $U(1)_B$ as well, whose parameter we call $\alpha\in[0,2\pi)$, and the two actions $U(1)_R$ and $U(1)_B$ satisfy the following relation: $\mathcal{U}_R (4\pi)=\mathcal{U}_B (\pi)$.
Therefore the group manifold $U(1)_R \times U(1)_B $ is parameterized by 
$\varepsilon\in[0,4\pi)$, $\alpha\in[0,2\pi)$ (this parameterization realizes a nontrivial torus) and $U(1)_B$ is a true symmetry of the theory.
The theta angle shift \eqref{theta shift} allows us to conclude that the $U(1)_R$ anomaly breaks the symmetry according to $U(1)_R\times U(1)_B \to \mathbb{Z}_{N_f}\times U(1)_B $, where the latter is given by $\varepsilon = 4n\pi/N_f \; (n=0,1,\dots,N_f-1)$, $\alpha\in[0,2\pi)$.

Coming to the string side, the solution for the metric, the dilaton and the field strengths is invariant under arbitrary shifts of $\psi$. But the nontrivial profile of $C_0$, which can be probed by D(-1)-branes for instance, breaks this symmetry.
The presence of DBI actions in the functional integral tells us that the RR potentials are quantized, in particular $C_0$ is defined modulo integers. Taking the formula \eqref{C0} and using the periodicity $4\pi$ of $\psi$, we conclude that the true invariance of the solution is indeed $\mathbb{Z}_{N_f}$.

One can be interested in computing the domain wall tension in the field theory by means of its dual description in terms of a D5-brane with 3 directions wrapped on a 3-sphere (see \cite{Herzog:2002ih} for a review in the conifold geometry). It is easy to see that, as in Klebanov-Witten theory, this object is stable only at $r=0$ ($\rho\to -\infty$), where the domain wall is tensionless. 
\subsubsection{The UV and IR Behaviors} \label{irtheory} \label{uvtheory}

The supergravity solution allows us to extract the IR dynamics of the KW field theory with massless flavors. Really what we obtained is a class of solutions, parameterized by two integration constants $c_1$ and $c_2$.  Momentarily, we will say something about their meaning but  anyway some properties are independent of them.

The fact that the \hyph{\beta}function is always positive, with the only critical point at vanishing gauge coupling, tell us that the theory is irreparably driven to that point, unless the supergravity approximation breaks down before ($c_1<0$), for instance because of the presence of curvature singularities. Using the $\rho$ coordinate this is clear-cut. In cases where the string coupling falls to zero in the IR, the gravitational coupling of the D7 to the bulk fields also goes to zero and the branes tend to decouple. The signature of this is in equation \eqref{eqnf} of the BPS system: the quantity $e^\phi N_f$ can be thought of as the effective size of the flavor backreaction which indeed vanishes in the far IR. The upshot is that flavors can be considered as an ``irrelevant deformation'' of the $AdS_5\times T^{1,1}$ geometry.

The usual technique for studying deformations of an $AdS_5$ geometry is through the GKPW \cite{Gubser:1998bc,Witten:1998qj} formula in AdS/CFT. Looking at the asymptotic behavior of fields in the $AdS_5$ effective theory:%
\footnote{Notice that usually the GKPW prescription or the holographic renormalization methods are used when 
we may have flows starting from a conformal point in the UV. In this case, our conformal point is in the IR 
and one may doubt about the validity in this unconventional case. See Section 6 in the paper \cite{Skenderis:2006di}
for an indication that applying the prescription in an IR point makes sense, even when the UV 
geometry is very far away from $AdS_5 \times M_5$. We thank Kostas Skenderis for correspondence on this  issue.}
\begin{equation}
\delta \Phi = a\, r^{\Delta-4} + c\, r^{-\Delta} \;,
\label{asymptotic}
\end{equation}
we read, on the CFT side, that the deformation is $H=H_{CFT} + a\,\mathcal{O}$ with $c=\langle \mathcal{O} \rangle$ the VEV of the operator corresponding to the field $\Phi$, and $\Delta$ the quantum dimension of the operator $\cO$. Alternatively, one can compute the effective 5d action and look for the masses of the fields, from which the dimension is extracted with the formula:
\begin{equation}
\Delta = 2 + \sqrt{4 + m^2} \:,
\label{mass-dimension}
\end{equation}
with the mass expressed in units of inverse $AdS$ radius.
We computed the 5d effective action for the particular deformations $e^{f(r)}$, $e^{g(r)}$ and $\phi(r)$ and including the D7-brane action terms (the details are in Section 
\ref{generalizations}). After diagonalization of the effective K\"ahler potential, we got a scalar potential $V$ containing a lot of information. First of all, minima of $V$ correspond to the $AdS_5$ geometries, that is conformal points in field theory. The only minimum is formally at $e^\phi=0$, and has the $AdS_5\times T^{1,1}$ geometry. Then, expanding the potential at quadratic order the masses of the fields can be read; 
from here we deduce that 
we have operators of dimension 6 and 8 taking VEV, and a marginally irrelevant operator inserted.%
\footnote{To distinguish between a VEV and an insertion we have to appeal to the first criterium described in eq. (\ref{asymptotic}) and below. }

The operators taking VEV where already identified in \cite{Klebanov:2000nc,Benvenuti:2005qb}. The dimension 8 operator is $\Tr F^4$ and represents the deformation from the conformal KW solution to the non-conformal 3-brane solution. The dimension 6 operator is a combination of the operators $\Tr (\cW_\alpha \bar \cW^\alpha)^2$ and represents a relative metric deformation between the $S^2\times S^2$ base and the $U(1)$ fiber of $T^{1,1}$.
The marginally irrelevant insertion is the flavor superpotential, which would be marginal at the hypothetic $AdS_5$ (conformal) point with $e^\phi=0$, but is in fact irrelevant driving the gauge coupling to zero in the IR and to very large values in the UV. 
Let us add that the scalar potential $V$ can be derived from a superpotential $W$, from which in turn the BPS system (\ref{BPSsystem}) can be obtained.

Since in the IR the flavor branes undergo a sort of decoupling, the relevant deformations dominate and their treatment is much the same as for the unflavored Klebanov-Witten solution \cite{Klebanov:2000nc,Benvenuti:2005qb,Strassler:2005qs}. We are not going to repeat it here, and we will concentrate on the case $c_1=c_2=0$. The supergravity solution flows in the IR to the $AdS_5\times T^{1,1}$ solution (with corrections of relative order $1/|\log(r)|$). On one hand the R-charges and the anomalous dimensions tend to the almost conformal values:
\begin{equation} \begin{split}
R_{A,B} &= \frac{1}{2} \\
R_{q,Q} &= \frac{3}{4}
\end{split} \qquad \qquad \begin{split}
\gamma_{A,B} &= -\frac{1}{2} \\
\gamma_{q,Q} &= \frac{1}{4} \;.
\end{split} \end{equation}
Using the formula for the \hyph{\beta}function of a superpotential dimensionless coupling:
\begin{equation}
\beta_{\tilde h} = \tilde h \Big[ -3 + \sum\nolimits_\Phi \big( 1+ \frac{\gamma_\Phi}{2} \big) \Big] \;,
\end{equation} 
where $\Phi$ are the fields appearing in the superpotential term, we obtain that the total superpotential \eqref{wflavors} is indeed marginal. On the other hand the gauge coupling flows to zero. Being at an almost conformal point, we can derive the radius-energy relation through rescalings of the radial and Minkowski direction, getting $r=\mu/\Lambda$. Then the supergravity beta function coincides with the exact (perturbative) holomorphic \hyph{\beta}function (in the Wilsonian scheme):%
\footnote{Here it is manifest why the SUGRA \hyph{\beta}function computed in this context with probe branes matches the field theory one, even if this requires the absence of order $N_f/N_c$ corrections to the anomalous dimensions $\gamma_{A,B}$, which one does not know how to derive (the stress-energy tensor is linear in $N_f/N_c$). It is because those corrections are really of order $e^\phi N_f/N_c$, and in the IR $e^\phi\to 0$.}
\begin{equation}
\beta_g = - \frac{g^3}{16\pi^2} \Big[ 3N_c - 2N_c (1-\gamma_A) - N_f(1-\gamma_f) \Big] \;.
\end{equation} 

If we are allowed to trust the orbifold relation \eqref{g+} relating gauge coupling constants and dilaton, we conclude that the gauge coupling flows to zero in the IR. This fact could perhaps explain the divergence of the curvature invariants in string frame \cite{Itzhaki:1998dd}, as revealed by \eqref{singul}. The field theory would enter the perturbative regime at this point. However, it is hard to understand why the anomalous dimensions of the fields are large while the theory seems to become perturbative.
For this reason, we question the validity in the conifold case of the holographic relation \eqref{g+}, that can be derived only for the orbifold. In Appendix \ref{appAlt} we propose an alternative interpretation of the IR regime of our field theory, based on some nice observations made in \cite{Strassler:2005qs} about the KW field theory. We argue that the theory may flow to a strongly coupled fixed point, although the string frame curvature invariant is large, as in the Klebanov-Witten solution for small values of $g_s N_c$.

Contrary to the IR limit, the UV regime of the theory is dominated by flavors and we find the same kind of behavior for all values of the relevant deformations $c_1$ and $c_2$. The gauge couplings increase with the energy, irrespective of the number of flavors. At a finite energy scale that we conventionally fixed to $\rho=0$, the gauge theory develops a Landau pole, as told by the string coupling that diverges at that particular radius. This energy scale is finite, because $\rho=0$ is at finite proper distance from the bulk points $\rho<0$.

At the Landau pole radius the supergravity description breaks down for many reasons: the string coupling diverges as well as the curvature invariants (both in Einstein and string frame), and the $\psi$ circle shrinks. An UV completion must exist, and finding it is an interesting problem. One could think about obtaining a new description in terms of supergravity plus branes through various dualities. In particular T-duality will map our solution to a system of NS5, D4 and D6-branes, which could then be uplifted to M-theory. Anyway, T-duality has to be applied with care because of the presence of D-branes on a non-trivial background, and we actually do not know how to T-dualize the Dirac-Born-Infeld action. We leave this interesting problem for the future.

\section{Part II: Generalizations}
\label{generalizations}
In this section we are going to extend the smearing procedure of the
D7-brane, which was formulated in section \ref{sect2} for the particular case of
the  $AdS_5\times T^{1,1}$ space, to  the more general case of a
geometry of the type $AdS_5\times M_5$, where $M_5$ is a five-dimensional
compact manifold. Of course, the requirement of supersymmetry restricts
greatly the form of $M_5$. Actually, we will verify that, when $M_5$ is 
Sasaki-Einstein, the formalism of section \ref{sect2} can be easily generalized.
As a result of this generalization we will get a more intrinsic formulation 
of the smearing, which eventually could be further generalized to other
types of flavor branes in different geometries. 

First of all we are going to generalize the effect of the smearing on the Wess-Zumino term of the D7-brane action for a general geometry. 
Following the line of thought that led to the action \eqref{action} and as is clear from the fact that it is linear in $C_8$, the smearing can be modelled by means of the substitution:
\begin{equation} \label{WZaction-general}
S_{WZ}\,=\,T_7\,\,\sum_{N_f}\,\,\int_{{\cal M}_8}\,\,\hat C_8\,\,
\rightarrow\,\,T_7\,\,\int_{{\cal M}_{10}}\,
\Omega\wedge C_8\,\,,
\end{equation}
where $\Omega$ is a two-form which determines the distribution of the RR
charge of the D7-brane in the smearing and ${\cal M}_{10}$ is the full
ten-dimensional manifold. For a supersymmetric brane one
expects the charge density  to be equal to  the mass density and, thus, the
smearing of the DBI part of the D7-brane action should be also determined by
the form $\Omega$. Let us explain  in detail how this can be done. First of
all, let us suppose that $\Omega$ is {\it decomposable}, \ie\ that it can be
written as the wedge product of two one-forms. In that case, at an arbitrary
point,  $\Omega$ would determine an eight-dimensional orthogonal hyperplane,
which we are going to identify with the tangent space of the D7-brane
worldvolume. A general two-form $\Omega$ will not be decomposable. However,
it can be written as a finite sum of the type:
\beq
\Omega\,=\,\sum\nolimits_i \Omega^{(i)} \;,
\eeq
where each $\Omega^{(i)}$ is decomposable. At an arbitrary point, each of
the  $\Omega^{(i)}$'s is dual to an eight-dimensional hyperplane. Thus,
$\Omega$ will determine locally a collection of eight-dimensional
hyperplanes. In the smearing procedure, to each decomposable component of
$\Omega$ we associate the volume form of its orthogonal complement in ${\cal
M}_{10}$. Thus, the contribution of every $\Omega^{(i)}$ to the DBI action
will be proportional to the ten-dimensional volume element. Accordingly, let
us perform the following substitution:
\beq
S_{DBI} = -T_7 \; \sum_{N_f} \int_{{\cal M}_8} d^8\xi \, \sqrt{-\hat G_8} \;\; e^{\phi}
\quad \rightarrow \quad
-T_7 \int_{{\cal M}_{10}} d^{10}x \, \sqrt{-G} \;\; e^{\phi} \; \sum\nolimits_i
\big|\,\Omega^{(i)}\,\big| \;,
\label{DBIaction-general}
\eeq
where $\big|\,\Omega^{(i)}\,\big|$ is the modulus of $\Omega^{(i)}$ and
represents the mass density of the $i^{th}$ piece of $\Omega$ in the
smearing. There is a natural definition of $\big|\,\Omega^{(i)}\,\big|$ 
which is invariant under coordinate transformations. Indeed, let us suppose
that  $\Omega^{(i)}$ is given by:
\beq
\Omega^{(i)}\,=\,{1\over 2!}\,\,\sum_{M,N}\,\,\Omega^{(i)}_{MN}\,
dx^M\wedge dx^N\,\,.
\eeq
Then, $\big|\,\Omega^{(i)}\,\big|$ is defined as follows:
\beq
\big|\,\Omega^{(i)}\,\big|\,\equiv\,
\sqrt{{1\over 2!}\,\Omega^{(i)}_{MN}\,\Omega^{(i)}_{PQ}\,
G^{MP}\,G^{NQ}}\,\,.
\label{modulus}
\eeq

Notice that  $\Omega$ acts as a magnetic source for the
field strength $F_1$. Actually, from the equation of motion of $C_8$ one gets that
$\Omega$ is just the violation of the Bianchi identity for $F_1$, namely:
\beq
dF_1\,=\,-\Omega\,\,.
\label{Bianchi-general}
\eeq
For a supersymmetric configuration the form $\Omega$ is not arbitrary.
Indeed, eq. (\ref{Bianchi-general}) determines $F_1$ which, in turn, enters
the equation that determines the Killing spinors of the background.
On the other hand, $\Omega$ must come from the superposition (smearing) of \hyph{\kappa}symmetric
branes.
When the manifold $M_5$ is Sasaki-Einstein, we will show in Section \ref{SEBPS} that $\Omega$ 
can be determined in terms of the K\"ahler form of the K\"ahler-Einstein base
of $M_5$ and that the resulting DBI+WZ action is a direct generalization of
the result written in (\ref{action}).  We will also show that the existence
of Killing spinors implies that the functions appearing in the ansatz
satisfy a system of first-order differential equations analogous to that
written in  (\ref{BPSsystem}).

\subsection{General Smearing and DBI Action}

Here we will elaborate on the previous construction: writing the DBI action for a general smearing of supersymmetric D7-branes. We mean that in general on an \Nugual{1}  background there is a
continuous family of supersymmetric 4-manifolds%
\footnote{Even if we try to be general, we still stick to the case with vanishing $B_2$ background and vanishing $F_{MN}$ on the brane world-volume.}
that the D7-branes can wrap corresponding to quarks with the same
mass and quantum numbers.
All these configurations preserve the same four supercharges, so we can think of putting D7's arbitrarily distributed (with arbitrary density functions) on these manifolds. We want to write the DBI plus WZ action for this system.

Supersymmetry plays a key role. The fact that we can put D7's and not anti-D7's implies that the charge distribution completely specifies the system. For D7-branes the charge distribution is a 2-form $\Omega$, which can be localized (a ``delta form'' or current) or smooth (for smeared systems). The Bianchi identity reads $dF_1=-\Omega$ and is easily implemented through the WZ action \eqref{WZaction-general}: $S_{WZ} = T_7 \int \Omega \wedge C_8$.
Notice that a well defined $\Omega$ not only must be closed (which is charge conservation) but also exact. Moreover the supersymmetry of this class of solutions forces $\Omega$ to be a real (1,1)-form (with respect to the complex structure). Supersymmetry also guides us in writing the DBI action, because the energy distribution must be equal to the charge distribution. But there is a subtlety here, because the energy distribution is not a 2-form, and some more careful analysis is needed.

Let us start considering the case of a single D7-brane localized on $\mathcal{M}_8$. We can write its DBI action as a bulk 10d integral by using a localized distribution 2-form $\Omega$ such that
\begin{equation} \label{parallel smearing}
\int_{\mathcal{M}_8} d^8\xi \,e^\phi \, \sqrt{-\hat G_8} = \int d^{10}x \,  \,e^\phi\,\sqrt{-G} \, |\Omega| \;.
\end{equation}
$\Omega$ is the Poincar\'e dual to $\mathcal{M}_8$. It can be (locally) written as $\Omega = \delta^{(2)}(\mathcal{M}_8) \sqrt{-\hat G_8}/\sqrt{-G} \: \alpha\wedge\beta$, through a properly normalized delta function and the product of two 1-forms (in general not separately globally defined) orthogonal to the 8-submanifold.%
\footnote{This orthogonality does not need a metric. A 1-form is a linear function from the tangent space to $\mathbb{R}$, and its kernel is a 9d hyperplane. The 8d hyperplane, tangent to the submanifold, orthogonal to the two 1-forms, is the intersection of the two kernels.}
In particular it is decomposable.

The decomposability of a 2-form can be established through Pl\"ucker's relations, and the minimum number of decomposable pieces needed to write a general 2-form is half of its rank as a matrix%
\footnote{The rank of an antisymmetric matrix is always even.}.
So the decomposability of a 2-form at a point means that it is dual to one 8d hyperplane at that point; in general a 2-form is dual to a collection of 8d hyperplanes.

If we do a parallel smearing of our D7-brane we get a smooth charge distribution 2-form, non-zero at every point. This corresponds to put a lot of parallel D7's and go to the continuum limit. Being the smearing parallel, we never have intersections of branes and the 2-form is still decomposable. As a result \eqref{parallel smearing} is still valid. If instead we construct a smeared system with intersection of branes, the charge distribution $\Omega$ is no longer decomposable. Every decomposable piece corresponds to one 8d hyperplane, tangent to one of the branes at the intersection. Since energy is additive, the DBI action is obtained by summing the moduli of the decomposable pieces (and not just taking the modulus of $\Omega$). Every brane at the intersection defines its 8d hyperplane and gives its separate contribution to the DBI action and to the stress-energy tensor. We simply sum the separate contributions because of supersymmetry: the D7's do not interact among themselves due to the cancellation of attractive/repulsive forces.
Notice that in doing the smearing of bent branes, one generically obtains unavoidable self-intersections.

Summarizing, given the splitting of the charge distribution 2-form into decomposable pieces $\Omega = \sum\nolimits_k \Omega^{(k)}$, the DBI action reads
\begin{equation}
S_{DBI} = -T_7 \int d^{10}x \, \sqrt{-G} \: e^\phi \, \sum\nolimits_k \bigl| \Omega^{(k)} \bigr| \;. 
\end{equation}
The last step is to provide a well defined and coordinate invariant way of splitting the charge distribution $\Omega$ in decomposable pieces.
It turns out that {\it the splitting in the minimal number of pieces compatible with supersymmetry is almost unique}.

In our setup, $\Omega$ lives on the internal 6d manifold, which is complex and $SU(3)$-structure. This means that the internal geometry has an integrable complex structure $\mathcal{I}$ and a non-closed K\"ahler form $\mathcal{J}$ compatible with the metric: $\mathcal{J}_{ab} = g_{ac} \mathcal{I}_b^{\phantom{b}c}$. We can always find a vielbein basis that diagonalizes the metric and block-diagonalizes the K\"ahler form:
\begin{equation} \begin{aligned}
\label{diagonal cplx}
g &= \sum\nolimits_a e^a \otimes e^a \\
\mathcal{J} &= e^1\wedge e^2 + e^3\wedge e^4 + e^5\wedge e^6 \;.
\end{aligned} \end{equation}
This pattern is invariant under the structure group $SU(3)$ (without specifying the holomorphic 3-form, it is invariant under $U(3)$), as is also clear by expressing them in local holomorphic basis: $e^{z_i} \equiv e^{2i-1}+i\,e^{2i}$, $\bar e^{\bar z_i} \equiv e^{2i-1}-i\,e^{2i}$, with $i=1,2,3$. One gets the canonical expressions: $g = \sum_i e^{z_i} \otimes_S \bar e^{\bar z_i}$ and $\mathcal{J} = \frac{i}{2} e^{z_i} \wedge \bar e^{\bar z_i}$.

In our class of solutions, the supersymmetry equations force the charge distribution to be a real $(1,1)$-form with respect to the complex structure (see \cite{Grana:2005jc}). Notice that such a property is shared with $\mathcal{J}$. The dilatino equation is $e^\phi \bar F_1 ^{(0,1)} = i \bar\partial \phi$ (which without sources amounts to the holomorphicity of the axio-dilation $\tau = C_0 + i\, e^{-\phi}$). From this one gets
\begin{equation}
\Omega = -dF_1 = 2i\, e^{-\phi} \big( \partial\phi \wedge \bar\partial\phi - \partial\bar\partial \phi \big) \;.
\end{equation} 
It's manifest that $\Omega$ is $(1,1)$ and $\Omega^* = \Omega$. Going to complex components $\Omega = \Omega_{l\bar k} e^{z_l} \wedge \bar e^{\bar z_k}$, the reality condition translates to the matrix $\Omega_{l\bar k}$ being anti-hermitian. Thus it can be diagonalized with an $SU(3)$ rotation of vielbein that leaves \eqref{diagonal cplx} untouched, and the eigenvalues are imaginary.

Going back to real vielbein and summarizing, there is always a choice of basis which satisfies the diagonalizing condition \eqref{diagonal cplx} and in which the charge distribution can be written as the sum of three real (1,1) decomposable pieces:
\begin{equation}
\Omega = -\lambda_1 \, e^1\wedge e^2 - \lambda_2 \, e^3\wedge e^4 - \lambda_3 \, e^5\wedge e^6 \;.
\end{equation}
Supersymmetry forces the eigenvalues $\lambda_a$ to be real and, as we will see, positive. Moreover, as inferred by the previous construction, the splitting is unique as long as the three eigenvalues $\lambda_a$ are different, while there are embiguities for degenerate values, but different choices give the same DBI action.

We conclude noticing that, in order to extract the eigenvalues $|\lambda_k|=|\Omega^{(k)}|$ it is not necessary to construct the complex basis: one can simply compute the eigenvalues of the matrix $(\Omega)_{MP}g^{PN}$ in any coordinate basis. But in order to compute the stress-energy tensor, the explicit splitting into real (1,1) decomposable pieces is in general required.

\subsection{The BPS Equations for Any Sasaki-Einstein Space}
\label{SEBPS}
Let us now explain in detail the origin of the system of first-order differential equations (\ref{BPSsystem}). As already explained in section \ref{sect2}, the system  (\ref{BPSsystem}) is a consequence of supersymmetry. Actually, it turns out that it can be derived in the more general situation that corresponds to having  smeared D7-branes in a space of the type 
$AdS_5\times M_5$,  where $M_5$ is a five-dimensional Sasaki-Einstein (SE) manifold. Notice that the
$T^{1,1}$ space considered up to now is a SE manifold. In general,  a SE manifold can be represented as a one-dimensional bundle over a four-dimensional K\"ahler-Einstein (KE) space. Accordingly, 
we will write the $M_5$ metric  as follows
\beq
ds^2_{SE}=ds^2_{KE}+(d\tau+A)^2\,\,,
\eeq
where $\partial / \partial \tau$ is a Killing vector and $ds^2_{KE}$ stands for  the metric of the  KE space with K\"ahler form $J=dA\,/\,2$. In the case of the $T^{1,1}$ manifold the KE base is just $S^2\times S^2$, where the $S^2$'s are  parametrized by the angles $(\theta_i,\varphi_i)$ and the fiber $\tau$ is parametrized by the angle $\psi$.

Our ansatz  for ten-dimensional metric in Einstein frame will correspond to a deformation of the standard $AdS_5\times M_5$. Apart from the ordinary warp factor $h(r)$,  we will introduce some squashing between the one form dual to the Killing vector and the KE base, namely:
\beq  \label{metric}
ds^2\,=\,\Big[\,h(r)\,\Big]^{-{1\over 2}}\,dx^2_{1,3}\,+\,
\Big[\,h(r)\,\Big]^{{1\over 2}}\,\Big[\,
dr^2\,+\,e^{2g(r)}\,ds^2_{KE}\,+\,e^{2f(r)}\,
\big(\,d\tau+A)^2\,\Big] .
\eeq
Notice that, indeed, the ansatz (\ref{metric}) is of the same type  as the one considered in eq. (\ref{configuration}) for the deformation of $AdS_5\times T^{1,1}$.  In addition our background  must have a Ramond-Ramond five form:
\beq  \label{F5}
F_5\,=\,K(r)\,dx^0\wedge\cdots dx^4\wedge dr\,+\,{\rm Hodge\,\,dual} ,
\eeq
and a  Ramond-Ramond one-form $F_1$ which violates Bianchi identity. Recall that this violation, which we want to be compatible with supersymmetry,  is a consequence of having a smeared D7-brane source in our system. Our proposal for $F_1$ is the following:
\beq  \label{F1}
F_1\,=\,C\,(d\tau+A)\,\, ,
\eeq
where $C$ is a constant which should be related to the number of 
flavors. Moreover, the violation of 
the Bianchi
identity is the following\footnote{We are considering that $J={1 \over 2}J_{ab}dx^a \wedge dx^b$ and that the Ricci tensor of the KE space satisfies $R_{ab}=\,6\,g_{ab}$.}:
\beq
dF_1\,=\,2 \, C\,\, J .
\label{dF_1=J}
\eeq
Notice that eq. (\ref{dF_1=J}) corresponds to taking $\Omega=-2CJ$ in our general expression 
(\ref{Bianchi-general}). 
To proceed with 
this proposal
we should try to solve the Killing spinor equations by imposing the appropriate
projections.  Notice that the
ansatz is compatible with the K\"ahler structure of the KE  base and
this is usually related to supersymmetry.

Before going ahead, it may be useful for the interested reader to make contact with the explicit case studied in the previous section, namely the Klebanov-Witten model. In that case the KE base is
\beq
ds^2_{KE}\,=\,{1 \over 6} \sum_{i=1,2} ( d\theta^2_{i}\,+\,\sin^2{\theta_{i}}\,d\varphi^2_{i} )
\eeq
whereas the one form dual to the Killing vector $\partial / \partial \tau$ is $d\tau=d\psi / 3$ and the form $A$ reads
\beq
A\,=\,{1 \over 3}\Big( \cos{\theta_{1}}\,d\varphi_{1} \,+\, \cos{\theta_{2}}\,d\varphi_{2}\Big)\,\,.
\eeq
Moreover, the constant $C$ was set to ${{3\,N_{f}} \over {4\pi}}$ in that case. 

Let us choose the following frame for the ten-dimensional metric:
\begin{equation} \begin{aligned}
\hat{e}^{x^{\mu}} &=\,\big[\,h(r)\,\big]^{-{1\over 4}}\,dx^{\mu} \,, &
\hat{e}^{r} &=\,\big[\,h(r)\,\big]^{{1\over 4}}\,dr\,\,, \\
\hat{e}^0 &=\,\big[\,h(r)\,\big]^{{1\over 4}}\,e^{f(r)}\,
(d\tau+A) \qquad\qquad &
\hat{e}^a &=\,\big[\,h(r)\,\big]^{{1\over 4}}\,e^{g(r)}\,e^a 
\,\,,
\label{10vielbein}
\end{aligned} \end{equation}
where $e^a  \quad a=1,\ldots,4$ is the one-form basis for the KE space such that 
$ds^2_{KE}\,=\,e^a\, e^a $. In the Klebanov-Witten model the basis taken in (\ref{T11frame}) corresponds
to:
\begin{equation} \begin{aligned}
e^{1} &=\sin{\theta_{1}}\,d\varphi_{1} \,\,, \qquad\qquad &
e^{2} &=\,\,d\theta_{1} \,\,, \\
e^{3} &=\sin{\theta_{2}}\,d\varphi_{2} \,\,, &
e^{4} &=\,\,d\theta_{2} \,\,.
\end{aligned} \end{equation}
Let us write the five-form $F_5={\cal F}_5+{}^*{\cal F}_5$ of eq. (\ref{F5})
in frame components:
\bear
&&{\cal F}_5\,=\,K(r)\,\big[\,h(r)\,\big]^{{3\over 4}}\,
\hat{e}^{x^{0}}\wedge\cdots \wedge \hat{e}^{x^{3}}\wedge \hat{e}^{r} \,\,,\rc\rc
&&{}^*{\cal F}_5\,=\,-K(r)\,\big[\,h(r)\,\big]^{{3\over 4}}\,
\hat{e}^{0}\wedge\cdots \wedge \hat{e}^{4}\,=\,-
Kh^2\,e^{4g+f}\,(d\tau+A)\wedge e^1 \wedge \dots \wedge e^4 .\rc
\eear
The equation $dF_5=0$ immediately implies:
\beq \label{Bian}
Kh^2e^{4g+f}\,=\,{\rm constant}\,=\,{(2\pi)^4 N_c\over {Vol(M_5)}}\,\,,
\eeq
where the constant has been obtained by imposing the quantization condition (\ref{F5quantization})
for a generic $M_5$.  It will also be useful in what follows to write the one-form $F_1$ in frame components:
\beq
F_1\,=\,C\,h^{-{1 \over 4}}\,e^{-f}\,\hat{e}^{0} .
\label{F1frame}
\eeq
Let us list the non-zero components of the spin connection:
\begin{equation} \begin{aligned}
\hat{\omega}^{x^{\mu}r} &=\,-{1\over 4}\,h'\,h^{-{5\over 4}}\,\,
\hat{e}^{x^{\mu}}\,\,,\qquad\qquad (\mu=0,\cdots, 3)\,\,, \\
\hat{\omega}^{ar} &=\,\Big[\,{1\over 4}\,{h'\over h}\,+\,g'\,\,\Big]\,
h^{-{1\over 4}}\,\,\hat{e}^{a}\,\,,\qquad\qquad (a=1,\cdots, 4)\,\,, \\
\hat{\omega}^{0r} &=\,\Big[\,{1\over 4}\,{h'\over h}\,+\,f'\,\,\Big]\,
h^{-{1\over 4}}\,\,\hat{e}^{0}\,\,\,\,, \\
\hat{\omega}^{0}_{\,\,\,\, a} &=\, e^{f-2g}  h^{-{1\over 4}}\, J_{ab}\,\hat{e}^b\,\,
\,\,, \\
\hat{\omega}^{ab} &=\,\omega^{ab}\,-\,
e^{f-2g}  h^{-{1\over 4}}\, J^{ab}\hat{e}^0 \,\, ,
\label{spinconection}
\end{aligned} \end{equation}
where $\omega^{ab}$ are components of the spin connection of the KE base.

Let us now study under which conditions our ansatz preserves some amount of supersymmetry. To address this point we must look at the supersymmetric variations of the dilatino ($\lambda$) and gravitino ($\psi_M$).  These variations have been collected in appendix \ref{SUSYapp}, for both the Einstein and string frame. We have written them in eq. (\ref{complexvariations}) for the particular case in which the three-forms of supergravity are zero. Recall that the variations written in (\ref{complexvariations}) correspond to the Einstein frame and we have used a complex spinor notation. 

It is quite obvious from the form of our ansatz for $F_1$ in (\ref{F1frame})  that the equation resulting from the dilatino variation is:
\beq
\big ( \phi'\, - \,i\,e^{\phi}\,C\,e^{-f}\,\Gamma_{r0} \big )\, \epsilon\,=\,0\,\,.
\label{BPSdilatino}
\eeq
In eq. (\ref{BPSdilatino}), and in what follows, the indices of the $\Gamma$-matrices refer to the vielbein components (\ref{10vielbein}). 

Let us move on to the more interesting case of the gravitino transformation. The space-time and the radial components of the equation do not depend on the structure of the internal space and always yield the following two equations:
\bear
&&h'\,+\,K\,h^2\,=\,0 \,\, , \rc\rc
&&\partial_r \epsilon \,-\, {1 \over 8}\,K\,h\, \epsilon\,=\,0\,\, .
\label{hBPS}
\eear
To get eq. (\ref{hBPS}) we have imposed  the D3-brane projection
\beq \label{proj3}
\Gamma_{x^0x^1x^2x^3}\, \epsilon\,=\, -i \,\epsilon\,,
\eeq
and we have used the fact that the ten-dimensional spinor is chiral with chirality
\beq \label{chiral}
\Gamma_{x^0\ldots x^3r01234}\,\epsilon\,=\,\,\epsilon \,\, .
\eeq
It is a simple task to integrate the second  differential equation  in (\ref{hBPS}):
\beq \label{spinor}
\epsilon\,=\,h^{-{1 \over 8}} \hat{\epsilon}\,\, ,
\eeq
where $\hat{\epsilon}$ is a spinor which can only 
depend on the coordinates of the Sasaki-Einstein space.

In order to study the variation of the SE components of the gravitino it is useful to write the covariant derivative along the SE directions
in terms of the covariant derivative in the KE space. The covariant derivative, written as a one-form for those components, $\hat{D}\equiv d \, + \, {1 \over 4} \, \hat{\omega}_{IJ}\, \Gamma^{IJ}$, is given by
\bear
&&\hat{D}\,=\, D\, -\,{1 \over 4}\,J_{ab}\,h^{-{1 \over 4}}\,e^{f-2g}\,\Gamma^{ab}\,\hat{e}^0\,-\,{1 \over 2}\,J_{ab}\,h^{-{1 \over 4}}\,e^{f-2g}\,\Gamma^{0b}\,\hat{e}^a\,+\,  \rc\rc
&&+\, {1 \over 2}\,h^{-{1 \over 4}}\,\big ( {1 \over 4}\,{h' \over h}\,+\,g' \big )\,\Gamma^{ar}\,\hat{e}^a\,+\,{1 \over 2}\,h^{-{1 \over 4}}\,\big ( {1 \over 4}\,{h' \over h}\,+\,f' \big )\,\Gamma^{0r}\,\hat{e}^0 \,\, ,
\eear
where $D$ is the covariant derivative in the internal KE space.

The equation for the SE components of the gravitino transformation is
\beq
\hat{D}_I \, \epsilon\, - \, {1 \over 8}\,K\,h^{{3 \over 4}}\, \Gamma_{rI}\,\epsilon\,+\,{i \over 4}\,e^{\phi}\,F^{(1)}_I\,\epsilon\,=\,0.
\eeq
This equation can be split  into a part coming from the coordinates in the KE space and a part coming from the coordinate which parameterizes the Killing vector. For this purpose, it is convenient to represent the frame one-forms $e^a$ and the fiber one-form $A$ in a coordinate basis of the KE space
\begin{equation} \begin{aligned}
e^a &=\,E^a_m \, dy^m\, , \\
A &=\,A_m\,dy^m\, ,
\end{aligned} \end{equation}
with $y^m \quad m=1,\ldots,4$ a set of space coordinates in the KE space.

After a bit of algebra one can see that the equation obtained for the space coordinates $y^m$ is simply
\begin{equation} \begin{aligned}  \label{eq1}
& D_m \, \epsilon \, -\,{1 \over 4}\,J_{ab}\,e^{2(f-g)}\,A_m \, \Gamma^{ab}\,\epsilon \,-\,{1 \over 2}\,J_{ab}\,h^{-{1 \over 4}}\,e^{f-2g}\,E^a_m\, \Gamma^{0b}\,\epsilon \,+\,  \\
& +\, {1 \over 2}\, h^{-{1 \over 4}}\,\big ( {1 \over 4}\,{h' \over h}\,+\,g' \big )\,E^a_m\,\Gamma^{ar}\,\epsilon\,+\,{1 \over 2}\,\big ( {1 \over 4}\,{h' \over h}\,+\,f' \big )\, e^f \, A_m \, \Gamma^{0r}\,\epsilon \,-\, \\
& -\, {1 \over 8}\, K \, h^{{3 \over 4}}\, \big ( E^a_m\, \Gamma^{ra}\,+\,h^{{1 \over 4}}\,e^f\,A_m\,\Gamma^{r0} \big )\,\epsilon \,+\, {i \over 4}\,e^{\phi}\,C\,A_m\,\epsilon\,=\,0\,\, ,
\end{aligned} \end{equation}
whereas the equation obtained for the fiber coordinate $\tau$ is given by
\begin{equation} \begin{aligned}  \label{eq2}
& {\partial \epsilon \over \partial \tau} \, -\,{1 \over 4}\,J_{ab}\,e^{2(f-g)} \, \Gamma^{ab}\,\epsilon \,+\,{1 \over 2}\,\big ( {1 \over 4}\,{h' \over h}\,+\,f' \big )\, e^f  \, \Gamma^{0r}\,\epsilon \, -\,\\
& -\, {1 \over 8}\, K \, h \, e^f \,\Gamma^{r0} \,\epsilon \,+\, {i \over 4}\,e^{\phi}\,C\,\epsilon\,=\,0\,\, .
\end{aligned} \end{equation}

Let us now solve these equations for the spinor $\epsilon$. First of all, let us consider the dilatino equation (\ref{BPSdilatino}).  Clearly, this equation implies that the spinor must be an eigenvector of the matrix $\Gamma_{r0}$. Accordingly, let us require that $\epsilon$ satisfies 
\beq
\Gamma_{r0}\,\epsilon=\,-\,i \,\epsilon \,\, .
\label{rprojection}
\eeq
Moreover, a glance at eqs. (\ref{eq1}) and (\ref{eq2}) reveals that $\epsilon$ must also be an eigenvector of the matrix $J_{ab}\Gamma^{ab}$. Actually, by combining eqs. (\ref{proj3}) , (\ref{chiral})  and (\ref{rprojection}) one easily obtains  that
\beq
\Gamma_{12}\epsilon\,=\,\Gamma_{34}\epsilon\,\,.
\label{12=34}
\eeq
To simplify matters, let us assume that we have chosen the one-form basis $e^a$ of the KE in such a way that the K\"ahler two-form $J$ takes the canonical form:
\beq
J\,=\, e^1\wedge e^2\,+\,e^3\wedge e^4\,\,.
\label{canonicalJ}
\eeq
In this basis, after using the condition (\ref{12=34}),  one trivially gets:
\beq
J_{ab}\Gamma^{ab}\,\epsilon\,=\,4\Gamma_{12}\,\epsilon\,\,.
\eeq
Thus, in order to diagonalize $J_{ab}\Gamma^{ab}$, let us impose the projection
\beq
\Gamma_{12}\,\epsilon\,=\, -i\epsilon\,\,,
\label{Gamma12}
\eeq
which implies 
\beq
\Gamma_{34}\,\epsilon\,=\,- i\epsilon\,\,,
\qquad\qquad
J_{ab}\Gamma^{ab}\,\epsilon\,=\,-4i\epsilon\,\,.
\eeq

Let us now use the well-known fact that any KE space admits a covariantly constant spinor $\eta$ satisfying:
\beq
D_m\,\eta\,=\,-{3 \over 2}\,i\,A_m\,\eta\,\, ,
\label{cov-eta}
\eeq
from which one can get a Killing spinor of the five-dimensional SE space as:
\beq
\hat{\epsilon}\,=\,e^{-i\,\,{3 \over 2}\tau}\,\eta\,\, .
\label{hatepsilon}
\eeq
Actually, in the KE frame basis we are using, $\eta$ turns out to be a constant spinor which satisfies the conditions $\Gamma_{12}\,\eta\,=\, \Gamma_{34}\,\eta\,=\, -i\eta$. Let us now insert the SE Killing spinor $\hat{\epsilon}$ of eq.  (\ref{hatepsilon}) in our ansatz (\ref{spinor}), \ie\ we take the solution of  our SUSY equations to be:
\beq
\epsilon\,=\,h^{-{1 \over 8}}\,e^{-{3 \over 2}\,i\tau}\,\eta\,\, .
\label{spinorsol}
\eeq
By plugging (\ref{spinorsol}) into eqs. (\ref{eq1}) and (\ref{eq2}), and using the projections imposed to $\epsilon$ and (\ref{cov-eta}), one can easily see that eqs. (\ref{eq1}) and (\ref{eq2}) reduce to the following  two differential equations:
\bear
&&{1 \over 4} {h' \over h}\,+\,g'\,+\,{1 \over 4}\,K\,h\,-\,e^{f-2g}\,=\,0\,\, , \rc\rc
&&{1 \over 4} {h' \over h}\,+\,f'\,+\,{1 \over 4}\,K\,h\,+2\,e^{f-2g}\,-\,3\,e^{-f}\,+ \,{C \over 2}e^{\phi-f}\,=\,0\,\, .
\eear
By combining all equations obtained so far in this section we arrive at a system of first-order BPS equations for the deformation of any space of the form $AdS_5\times M_5$:
\bear  \label{BPS}
&& \phi'\, -  \, C \,e^{\phi-f}\,=\,0\,\, , \rc\rc
&&h'\,+\,{(2\pi)^4 N_c\over {Vol(M_5)}}\,\, e^{-f-4g}\,=\,0 \,\, , \rc\rc
&&\,g'\,-\,e^{f-2g}\,=\,0\,\, , \rc\rc
&&\,f'\,+2\,e^{f-2g}\,-\,3\,e^{-f}\,+\,{C \over 2}e^{\phi-f}\,=\,0\,\, .
\eear
Notice that, indeed, this system reduces to the one written in eq. (\ref{BPSsystem}) for the conifold, if we take into account that for this later case the constant C is $3N_f/(4\pi)$ and $Vol(T^{1,1})=16 \pi^3 /27$. 

It is now a simple task to count the supersymmetries of the type (\ref{spinorsol})
preserved by our background: it is just thirty-two divided by the number of
independent algebraic projection imposed to the constant spinor $\eta$. As a
set of independent projections one can take the ones written in eqs.
(\ref{proj3}), (\ref{rprojection}) and (\ref{Gamma12}). It follows that our deformed background
preserves four supersymmetries generated by  Killing spinors of the type
displayed in eq. (\ref{spinorsol}).

\subsection{The BPS and Einstein Equations} \label{BPS-Einstein}

In this section we will prove that the BPS system implies the fulfilment of
the second-order Euler-Lagrange equations of motion for the combined
gravity plus brane system. To begin with, let us consider the equation of
motion of the dilaton, which can be written as:
\beq
{1 \over {\sqrt{-G}}}\partial_M \Big ( G^{MN}\,
\sqrt{-G}\,\,\partial_N\phi \Big
)\,=\,e^{2\phi}\,F^2_{1}\,-\,
{2\kappa^2_{10}\over \sqrt{-G}}\,\,
{\delta\over \delta \phi}\,\,S_{DBI}\,\,,
\label{dilatoneq-general}
\eeq
where $G_{MN}$ is the ten-dimensional metric. Using the DBI action (\ref{DBIaction-general}) for the smeared D7-branes configuration, we find:
\begin{equation}
-\frac{2\kappa_{10}^2}{\sqrt{-G}} \frac{\delta}{\delta \phi} S_{DBI} = e^\phi \, \sum\nolimits_i
\big|\Omega^{(i)}\big|\;.
\end{equation}
The charge density distribution is 
$\Omega=-2CJ$ (see eq. (\ref{dF_1=J})).  
Recall that the K\"ahler form $J$ of the KE base manifold has the canonical
expression (\ref{canonicalJ}). It follows that $\Omega$ has two decomposable components
given by:
\begin{equation} \begin{aligned}
\Omega^{(1)} &=\,-2C\,e^1\wedge e^2\,=\,-2C\,h^{-{1\over 2}}\,e^{-2g}\,
\hat e^1\wedge \hat e^2\,\,, \\
\Omega^{(2)} &=\,-2C\,e^3\wedge e^4\,=\,-2C\,h^{-{1\over 2}}\,e^{-2g}\,
\hat e^3\wedge \hat e^4\,\,,
\label{dFi}
\end{aligned} \end{equation}
where the $\hat e^a$ one-forms have been defined in (\ref{10vielbein}).
Therefore, the moduli of the $\Omega^{(i)}$'s can be straightforwardly
computed:
\beq
\big|\Omega^{(1)}\big|\,=\,\big|\Omega^{(2)}\big|\,=\,
2|\,C\,|\,h^{-{1\over 2}}\,e^{-2g}\,\,.
\label{|dFi|}
\eeq
By using the explicit form of the metric, our ansatz for $F_1$ and the previous formulae (\ref{|dFi|}) one can
convert eq. (\ref{dilatoneq-general}) into the following:
\beq
\phi''\,+\,(4g'+f')\,\phi'\,=\,C^2\,e^{2\phi-2f}\,+\,4\,|C|\,\,e^{\phi-2g}\,\,.
\label{dil-eq2}
\eeq
It is now a simple exercise to verify that eq. (\ref{dil-eq2}) holds if the
functions $\phi$, $g$ and $f$ solve the first-order BPS system (\ref{BPS})
and the constant $C$ is non-negative. In what follows we shall assume that
$C\ge 0$.

To check the Einstein equation we need to calculate the Ricci tensor. 
In flat coordinates the components of the Ricci tensor can be computed by
using the spin connection. The expression of the curvature two-form in terms
of the spin connection is
\beq
 R_{\hat M\hat N}\,=\,d\hat\omega_{\hat M\hat N}\,+\,
\hat\omega_{\hat M\hat P}\wedge
\hat\omega^{\hat P}_{\,\, \hat N}\,\,  ,
\eeq
with the curvature two-form defined as follows:
\beq
R^{\hat M}_{\,\, \hat N}\,=\,{1 \over 2}\,
R^{\hat M}_{\,\, \hat N \hat P\hat Q}\,e^{\hat P}
\wedge e^{\hat Q}\,\,  .
\eeq 
By using the values of the different components of the ten-dimensional spin
connection written in (\ref{spinconection}) we can easily obtain  
the Riemann tensor and, by  simple contraction of indices, we arrive at the
following flat components of the Ricci tensor:
\begin{equation} \begin{aligned}
R_{x^ix^j} &=\,h^{-{1 \over 2}}\,\eta_{x^ix^j}
\Bigg(\,\, {1 \over 4}{h'' \over h}\,-\,{1 \over 4}\Bigg({h' \over
h}\Bigg)^2\,+\,{1 \over 4}{h' \over h}f'\,+\,{h' \over h}g' \Bigg)\,\,  ,  \\
R_{rr} &=\,h^{-{1 \over 2}}\,\Bigg( -{1 \over 4}{h'' \over h}\,-\,{1
\over 4}\Bigg({h' \over h}\Bigg)^2\,-\,{1 \over 4}{h' \over h}f'
\,-\,{h' \over h}g'
\,-\,f''\,-\,(f')^2\,-\,4\,g'' \,-\,4(g')^2\,
\Bigg)\,\,  ,  \\
R_{00} &=\,h^{-{1
\over 2}}\,\Bigg( -{1 \over 4}{h'' \over h}\,+\,{1 \over 4}\Bigg({h' \over
h}\Bigg)^2\,-\,{1 \over 4}{h' \over h}f'\,-\,{h' \over h}g'
\,-\,f''\,-\,(f')^2\,-\,4\,g'\,f'\,+\,4\,e^{2f-4g} \,\Bigg)\,\,  ,  \\
R_{aa} &=\,h^{-{1 \over 2}}\,\Bigg( -{1 \over 4}{h'' \over h}\,+\,{1 \over
4}\Bigg({h' \over h}\Bigg)^2\,-\,{1 \over 4}{h' \over h}f'\,-\,{h' \over h}g'
\,-\,g''\,-\,\\
&\qquad\qquad\qquad\qquad\qquad\qquad\qquad\qquad
-\,4\,(g')^2\,-\, g'\,f'\,-\,2\,e^{2f-4g}\,+\,6\,e^{-2g}\, \Bigg)\,\,  , \\
R_{\hat M \hat N} &=\,0\,\,,
\qquad \qquad M \neq N \,\, .
\label{Ricci-components}
\end{aligned} \end{equation}
From these values it is straightforward to find 
the expression of the scalar curvature, which  is simply
\begin{multline} \label{scalar}
R\,=\,-h^{-{1 \over 2}}\,\Bigg(\, {1 \over 2}{h'' \over h}+
{1 \over 2}{h' \over h}f'+2\,{h' \over h}g'\,
+\,8\,g''+20\,(g')^2\,+\, \\
+8\,g'\,f'+2\,f''+2\,(f')^2+4\,e^{2f-4g}-24\,e^{-2g} \Bigg)\,\,  .
\end{multline}
Let us  evaluate  the different
contributions to the right-hand side of Einstein's equations.  The
contributions from the five- and one-forms have been written in the first
equation in (\ref{eqs2nd}) and is immediately computable from our ansatz 
of eqs. (\ref{F5}) and (\ref{F1}). On the other hand, the contribution of
the DBI part of the action is just
\beq
T_{MN}\,=\,{2\kappa_{10}^2 \over \sqrt{-G}}\,\,
{\delta S_{DBI} \over \delta G^{MN}}\,\,.
\eeq
By using our expression (\ref{DBIaction-general}) of $S_{DBI}$, with
$\Omega=-dF_1$, together with the definition (\ref{modulus}), one easily
arrives at the following expression of the stress-energy tensor of the
D7-brane:
\beq
T_{\hat M\hat N} = -\frac{e^\phi}{2} \,
\Big[ \eta_{\hat M\hat N} \, \sum_i \,
\big|\,\Omega^{(i)}\,\big| - 
\sum_i \, {1\over \big|\,\Omega^{(i)}\,\big|}\,\,
(\Omega^{(i)})_{\hat M\hat P}\,\,
(\Omega^{(i)})_{\hat N\hat Q} \, \eta^{\hat P\hat Q} \, \Big] \;,
\label{TMN}
\eeq
where we have used that $2\kappa_{10}^2 T_7\,=\,1$ and we have written the
result in flat components. By using in (\ref{TMN}) the values given in eqs. 
(\ref{dFi}) and (\ref{|dFi|}) of $dF_1^{(i)}$ and its modulus, we arrive
at the simple result:
\bear
&&T_{x^ix^j}\,=\,-2C\,h^{-{1\over 2}}\,e^{\phi-2g}\,\,\eta_{x^ix^j}
\,\,,\rc\rc
&&T_{rr}\,=\,T_{00}\,=\,-2C\,h^{-{1\over 2}}\,e^{\phi-2g}\,\,,\rc\rc
&&T_{ab}\,=\,-C\,h^{-{1\over 2}}\,e^{\phi-2g}\,\delta_{ab}\,\,,
\qquad\qquad (a,b=1,\cdots, 4)\,\,,
\label{TMNcomponents}
\eear
where the indices refer to our vielbein basis (\ref{10vielbein}). As a check
of this result one can explicitly verify that the result of eq. (\ref{Tmunu}) for
the conifold, when written in flat indices, reduces to the simple expressions
written in (\ref{TMNcomponents}).

With all this information we can write, component by component, the set of
second order differential equations for $h$, $g$, $f$ and $\phi$ that are
equivalent to the Einstein equations. One can then verify, after some
calculation, that these equations are satisfied if $\phi$ and the functions
of our ansatz solve the first-order system (\ref{BPS}). Therefore, we have
succeeded in proving that the background obtained from the supersymmetry
analysis is a solution of the equations of motion of the supergravity plus
Born-Infeld system. Notice that the SUSY analysis determines $F_1$, \ie\ the
RR charge distribution of the smeared D7-branes. What we have just proved is
that eq. (\ref{TMN}) gives the correct energy-momentum distribution
associated to the charge distribution $\Omega=-dF_1$ of smeared
flavor branes.

To finish this section let us write the  DBI action in a different, and very
suggestive, fashion. It turns out that, for our ansatz, the on-shell  DBI action 
 can be written as the
integral of a ten-form and the corresponding expression is very similar to
the one for the WZ term (eq. (\ref{WZaction-general})). Actually, we show
below that
\beq
S_{DBI} = - T_7 \, \int_{{\cal M}_{10}} \, e^{\phi} \,
\Omega\wedge \Omega_8\,\,,
\label{SDBI-forms}
\eeq
where $\Omega_8$ is an eight-form which, after performing the wedge product
with the smearing two-form $\Omega$, gives rise to a volume form of the
ten-dimensional space. Let us factorize in $\Omega_8$ the factors coming
from the Minkowski directions:
\beq
\Omega_8\,=\,h^{-1}\,d^{4}x \wedge \Omega_4\, ,
\label{Omega_8}
\eeq 
where $\Omega_4$ is a four-form in the internal space. Actually, one can
check that $\Omega_4$ can be written as:
\beq \label{cal}
\Omega_4\,=\,{1 \over 2}\,\,{\cal J} \wedge {\cal J},
\eeq
where ${\cal J}$ is the following two-form:
\beq
{\cal J}\,=\,h^{{1 \over 2}}\,e^{2g}\,J\,+\,h^{{1 \over 2}} 
e^{f}\,dr\wedge(d\tau\,+\,A)\, .
\label{calJ}
\eeq
To verify this fact, let us recall that $\Omega = 2CJ$ and thus
\beq
\Omega \wedge \Omega_8 = C\,h^{-1}\,
d^{4}x\,\wedge J\wedge {\cal J} \wedge {\cal J} \;.
\eeq
Taking into account that ${1\over 2}\,\,J\wedge
J$ is the volume form of the KE base of $M_5$, we readily get:
\beq
d^{4}x\,\wedge\,J\wedge {\cal J} \wedge {\cal J}\,=\,4e^{-2g}\,
h^{{1\over 2}}\,\,\sqrt{-G}\,\,d^{10}x\,\,,
\eeq
from where one can easily prove that eq. (\ref{SDBI-forms})  gives
the same result as in equation (\ref{DBIaction-general}) with
$\Omega=-dF_1$.

\subsection{A Superpotential and the BPS Equations }

It is interesting to obtain the  system of first-order BPS 
equations (\ref{BPS}) by  using
an alternative approach, namely by deriving them from a superpotential. 
Generically, let us consider a one-dimensional classical mechanics system in
which  $\eta$ is the ``time" variable and 
${\cal A}(\eta)$, $\Phi^m(\eta)$ ($ m=1,2\ldots)$ are the generalized
coordinates.  Let us assume that the Lagrangian of this system takes the
form:
\beq \label{lagran}
 L\,=\,e^{\cal A} \Big [ \kappa \,\,(\partial_{\eta}{\cal A})^2\,-\,{1 \over
2}G_{mn}(\Phi)\,\partial_{\eta}\Phi^m
\,\partial_{\eta}\Phi^n\,-\,V(\Phi) \Big ]\,\, ,
\eeq
where $\kappa$ is a constant and $V(\Phi)$ is some potential, which we 
assume that is independent of the coordinate ${\cal A}$. 
If one can find a superpotential $W$ such that:
\beq \label{potential}
V(\Phi)\,=\,{1 \over 2}G^{mn}\,{{\partial W} \over 
{\partial\Phi^m}}\,{{\partial W} \over {\partial\Phi^n}}\,-\,{1
\over {4\kappa}}\,\,W^2\,\,  ,
\eeq
then the equations of motion are automatically satisfied 
by the solutions of the first order system:
\beq \label{eqs}
{{d\,{\cal A}} \over {d\eta}}\,=\,-{1 \over {2\kappa}}W \,\,  ,  \qquad
{{d\,\Phi^m} \over {d\eta}}\,=\, G^{mn}\,{{\partial W} \over {\partial
\Phi^n}} \,\,  .
\eeq
 Let us now show how we can recover our system (\ref{BPS})
from this formalism.
 The first step is to look for an effective Lagrangian for the dilaton and
the functions of our ansatz
whose equations of  motion are the same as those obtained from the Einstein
and dilaton equations of Type IIB supergravity. One can see that this
lagrangian is:
 \beq
 L_{eff}\,=\,h^{{1\over 2}}\,e^{4g+f}\, \Big 
[ R\,-\,{1\, \over 2}\,h^{-{1\over 2}}\,
(\phi')^2\,-\,{Q^2\over 2}\,\,h^{-{5\over 2}}\,e^{-8g-2f}\,-\,
{C^2\over 2}\,\,h^{-{1\over 2}}\,e^{2\phi-2f}
\,-\,4\,C\,h^{-{1\over 2}}\,e^{\phi-2g}\, \Big ]\,\, ,
 \eeq
where $R$ is the scalar curvature as written in (\ref{scalar}) and
$Q$ is the constant
\beq
Q\,\equiv\,{(2\pi)^4 N_c\over {Vol(M_5)}}\,\,.
\label{Q}
\eeq
The Ricci scalar (\ref{scalar})  contains second derivatives. Up to total derivatives $L_{eff}$ takes the form:
 \bear
 L_{eff}&=&e^{4g+f} \Bigg[  -{1 \over 2}\Bigg({h'
\over h}\Bigg)^2\,+\,12\,(g')^2\,+\,8\,
g'\,f'\,-\,4\,e^{2f-4g}\,+\,24\,e^{-2g}\,-\,{1 \over 2}(\phi')^2\,\,- \rc\rc
&&\qquad\qquad
-{Q^2\over 2}\,\,h^{-2}\,e^{-8g-2f}\,
-\,{C^2 \over 2}\,e^{2(\phi-f)}\,-\,4\,C\,e^{\phi-2g}
\Bigg]\,\, .
\label{Leff-in-r}
 \eear
We want to pass from  the lagrangian (\ref{Leff-in-r}) to that
in eq. (\ref{lagran}). With that purpose in mind let us perform 
the following redefinition of fields:
\beq \label{fields}
e^{{3\over 4}\,{\cal A}}\,=\,
h^{{1 \over 2}}e^{4g+f}\,\,,
\qquad  \qquad 
e^{2\tilde{g}}\,=\,h^{{1 \over 2}}e^{2g} , \qquad  \qquad  
e^{2\tilde{f}}\,=\,h^{{1 \over 2}}e^{2f} .
 \eeq
In addition,  we need to do the following  change of the radial variable
\footnote{The change of the Lagrangian under that change of radial
variable is $\hat{ L}_{eff}={{dr} \over {d\eta}}\,\,
 L_{eff}$.} 
\beq \label{radial}
{{dr} \over {d\eta}}\,=\,
e^{{{\cal A}\over 4}-{8\over 3}\tilde g\,
-\,{2\over 3}\tilde f}\,\,.
\eeq
Once we have done the previous redefinitions, the Lagrangian we obtain
is:
\beq
 \hat{  L}_{eff}\,=\,e^{{\cal A}} 
\Bigg[\,\,{3\over 4}\,\, (\dot{{\cal A}})^2\,-\,{28 \over
3}\,(\dot{\tilde{g}})^2\,-\,{4
\over 3}(\dot{\tilde{f}})^2\,-\,{8 \over 3}\,
{\dot{\tilde{g}}}\,{\dot{\tilde{f}}}\,-\,{1 \over
2}(\dot{\phi})^2\,-\,V(\tilde{g},\tilde{f},\phi) \Bigg]\,\,,
\label{Leff-in-eta}
\eeq
where the dot means derivative with respect to $\eta$ and 
$V(\tilde{g},\tilde{f},\phi)$ is the following potential:
\beq
V(\tilde{g},\tilde{f},\phi)\,=\,e^{-{2 \over
3}(4\tilde{g}+\tilde{f})} \Bigg(
\,4\,e^{2\tilde{f}-4\tilde{g}}-24\,e^{-2\tilde{g}}
\,+\,{Q^2 \over 2}\,e^{-2(4\tilde{g}+\tilde{f})}
\,+\,{C^2 \over 2}\, e^{2(\phi-\tilde{f})}
\,+\,4\,C\,e^{\phi-2\tilde{g}}\,\,
\Bigg)\,\,.
\label{tilde-pot}
\eeq
The above lagrangian has the desired form (see eq. (\ref{lagran}))
 and we can identify the constant $\kappa$ and the elements of the
kinetic matrix $G_{mn}$ as:
\beq
 \kappa\,=\,{3 \over 4}\,\, , 
\quad G_{\tilde{g}\tilde{g}}\,=\,{56 \over 3}\,\, , \quad
G_{\tilde{f}\tilde{f}}\,=\,{8 \over 3}\,\, , \quad
G_{\tilde{g}\tilde{f}}\,=\,{8 \over 3}\,\, , 
\quad G_{\phi \phi}\,=\,1\,\, .
\label{constants}
\eeq
One can now check that, given the above expression of  the
potential,  the following superpotential
\beq
W\,=\,e^{-{1 \over 3}(4\tilde g+\tilde f)} \Big [ Q\,
e^{-4\tilde g-\tilde f}\,-\,4\,e^{\tilde f-2\tilde g}\,-\,6e^{-\tilde f}\,+\,Ce^{\phi-\tilde f}  \Big ]\,\, 
\eeq
satisfies  eq. (\ref{potential}) for the values of $\kappa$ and 
$G_{mn}$ written in eq. (\ref{constants}). It is now immediate to write
the first-order differential equations that stem from this superpotential. 
Explicitly we obtain:
\bear
&&\dot{{\cal A}}\,=\,-\,{2 \over 3}\,\, W \,\, , \rc\rc
&&\dot{\tilde{g}}\,=\,{1 \over 4}e^{-{1 \over 3}(4\tilde{g}+\tilde{f})}\, 
\Big [ - Q e^{-4\tilde{g}-\tilde{f}}\,+\,4\,e^{\tilde{f}-2\tilde{g}}\,
\Big ]\,\, , \rc\rc &&\dot{\tilde{f}}\,=\,{1 \over 4}\,e^{-{1 \over
3}(4\tilde{g}+\tilde{f})}\, \Big [ - Q
e^{-4\tilde{g}-\tilde{f}}\,-\,8\,e^{\tilde{f}-2\tilde{g}}\,+\,12\,e^{-\tilde{f}}\,-\,2\,C\,e^{\phi-\tilde{f}}
\Big ] \,\, , \rc\rc 
&&\dot{{\phi}}\,=\,C\,e^{\phi-{4 \over
3}(\tilde{g}+\tilde{f})}\,\, .
\eear
In order to verify that this system is equivalent to the one obtained from
supersymmetry, let us write down explicitly these equations in terms
of  the old radial variable (see eq. (\ref{radial})) and fields (see eqs.
(\ref{fields})). One gets:
\bear
&&{h' \over h}\,+\,8\,g'\,+\,2\,f'\,=\,-  Q\,h^{-1}\,
e^{-4g-f}\,+\,4\,e^{f-2g}\,+\,6e^{-f}\,-\,Ce^{\phi-f}  \,\, , \rc\rc
&&{1 \over 4}{h' \over h}\,+\,g'\,=\, e^{f-2g}\,-\,
{1 \over 4}Q\,h^{-1}\,e^{-4g-f} \,\, , \rc\rc
&&{1 \over 4}{h' \over h}\,+\,f'\,=\, 3\,e^{-f}\,-\,2\,e^{f-2g}
\,-\,{1 \over 4}Q\,h^{-1}\,e^{-4g-f}\,-\,{1 \over 2}C\,e^{\phi-f} 
\,\, , \rc\rc &&\phi'\,=\,C\,e^{\phi-f}\,\, ,
\eear
which  are nothing else than a combination of the system of BPS
equations written in (\ref{BPS}).

Let us now use the previous results to study the 5d effective action
resulting from the compactification along $M_5$ of our solution. The fields
in this effective action are the functions $\tilde f$ and $\tilde g$, which
parameterize the deformations along the fiber and the KE base of $M_5$ 
respectively, and the dilaton. Actually, in terms of the new radial variable
$\eta$ introduced in (\ref{radial}), the ten-dimensional metric can be written as:
\beq
ds^2\,=\,e^{-{2\over 3}\,\,(\,\tilde{f}\,+\,4\,\tilde g\,)}\,\,
\Big[\,e^{{{\cal A}\over 2}}\,dx^{\mu}dx_{\mu}\,+\,d\eta^2\,\Big]\,+\,
e^{2\tilde g}\,ds^2_{KE}\,+\,e^{2\tilde f}\,(d\tau+A)^2\;.
\eeq
The corresponding analysis for the unflavored theory was
performed in \cite{Klebanov:2000nc,Benvenuti:2005qb}. 
For simplicity, let us work in units in which the $AdS_5$ radius
$L$ is one. Notice that the quantity $Q$ defined in (\ref{Q}) is just
$Q=4L^4$. Thus, in these units $Q=4$. To make contact with the analysis of
refs. \cite{Klebanov:2000nc,Benvenuti:2005qb}, let us introduce new fields
$q$ and $p$ which, in terms of $\tilde{f}$ and $\tilde{g}$ are defined as
follows%
\footnote{The function $p$ is called $f$ in refs.
\cite{Klebanov:2000nc,Benvenuti:2005qb}.}:
\beq
q\,=\,{2 \over {15}}\,(\,\tilde{f}\,+\,4\,\tilde g\,)\,\, , 
\qquad\qquad
p\,=\,-\,{1 \over {5}}\,(\,\tilde{f}\,-\,\tilde g\,)\,\, .
\label{pq-scalars}
\eeq
In terms of these new fields,  the potential (\ref{tilde-pot}) turns out to
be
\beq
V(p,q,\phi) = 4\,e^{-8q-12p}\,-\,24\,
e^{-8q-2p}\,+\,{C^2 \over
2}\,e^{2\phi-8q+8p}\,+\,8
\,e^{-20q}\,+\,4\,C\,e^{\phi-8q-2p} \;,
\label{pq-pot}
\eeq
and the effective lagrangian (\ref{Leff-in-eta}) can be written as:
\beq
L_{eff}\,=\,\sqrt{-g_5}\,\Big[\,R_5\,-\,{1\over 2}\,\dot \phi^2\,-\,20\,\dot
p^2\,-\,30\,\dot q^2\,-V\,\Big]\,\,,
\eeq
where $g_5\,=\,-e^{2{\cal A}}$ is the
determinant of the five-dimensional metric 
$ds^2_5\,=\,e^{{\cal A}\over 2}\, \,dx^{\mu}dx_{\mu}\,+\,d\eta^2$ 
and 
$R_5\,=\,-[2\,\ddot {\cal A}\,+{5\over 4}\,\dot{\cal A}^2\,\Big]$ is its
Ricci scalar. One can check that the minimum of the potential (\ref{pq-pot})
occurs only at $p=q=e^{\phi}=0$, which corresponds to the
conformal $AdS_5\times M_5$ geometry.  Moreover, by expanding $V$ around
this minimum at second order we find out that the fields $p$ and $q$ defined
in (\ref{pq-scalars}) diagonalize the quadratic potential. The corresponding
masses are $m_p^2=12$ and $m_q^2=32$. By using these values in the
mass-dimension relation (\ref{mass-dimension}), we get:
\begin{equation} \begin{aligned}
m_p^2 &= 12 & \qquad\Longrightarrow\qquad \Delta_p &= 6 \\
m_q^2 &= 32 & \qquad\Longrightarrow\qquad \Delta_q &= 8 \;.
\end{aligned} \end{equation}
These scalar modes $p$ and $q$ are dual to the dimension 6 and 8 operators
discussed in section \ref{sect2}.

\subsection{General Deformation of the Klebanov-Witten Background}
\label{generalKWdef}

In this section we will explore the possibility of having a more general
flavor deformation of the $AdS_5\times T^{1,1}$ background. Notice that, as 
$T^{1,1}$ is a $U(1)$ bundle over $S^2\times S^2$, there exists the
possibility of squashing with different functions each of the two $S^2$'s of
the KE base. In the unflavored case this is precisely the type of
deformation that occurs when the singular conifold is substituted by its
small resolution. For this reason, it is worth considering this type of
metric also in our flavored background. To be precise, 
let us adopt the following ansatz for the metric, five-form and one-form:
\begin{equation} \begin{aligned}
ds^2 &= h^{-1/2} dx_{1,3}^2 + 
h^{1/2} \bigg\{ dr^2 +
\frac{1}{6} \sum_{i=1,2} e^{2g_i}( d\theta_i^2 + \sin^2 \theta_i \,
 d\varphi_i^2)  + \frac{e^{2f}}{9} (d\psi + \sum_{i=1,2} \cos\theta_i \,
d\varphi_i)^2 \bigg\} \\
F_5 &= (1+\ast) \, d^4x \wedge K\, dr \\
F_1 &= \frac{C}{3} ( d\psi + \cos\theta_2\, d\varphi_2 + 
\cos\theta_1\, d\varphi_1 ) \;,
\end{aligned} \end{equation}
where $C=3N_f/4\pi$, all functions depend on $r$ and $g_1(r)$ and $g_2(r)$
are, in general,  different (if $g_1=g_2=g$ we recover our ansatz
(\ref{configuration})).  The equation $dF_5=0$ immediately implies:
\beq \label{Bian1}
Kh^2e^{2g_1+2g_2+f}\,=\,27\pi N_c\,\equiv\,Q\,\,,
\eeq
which allows to eliminate the function $K$ in favor of the other functions
of the ansatz. By following the same steps as in the $g_1=g_2$ case and requiring that the background preserve four supersymmetries, 
we  get a system of first-order BPS equations  for this
kind of deformation, namely:
\begin{equation} \begin{aligned}
\phi' &=  \, C \,e^{\phi-f}\,\, , \\
h' &=\,-Q\,e^{-f-2g_1-2g_2} \,\, , \\
g_i' &=\,e^{f-2g_i}\,\,,\qquad\qquad (i=1,2)\,\, ,\\
f' &=\,3\,e^{-f}\,-\,e^{f-2g_1}\,-\,e^{f-2g_2}\,-\,
{C \over 2}\,\,e^{\phi-f}\,\,.
\label{BPSg1g2}
\end{aligned} \end{equation}
Notice that, as it should, the system (\ref{BPSg1g2})
reduces to eq. (\ref{BPS}) when $g_1=g_2$.

It is not difficult to integrate this system of differential equations
by following the same method that was employed for the $g_1=g_2$ case. First
of all, we change the radial coordinate:
\beq
dr\,=\, e^f \,d\rho\,\,,
\label{rho-generalKW}
\eeq
what allows us to get a new system:
\begin{equation} \begin{aligned} \label{sol}
\dot{\phi} &=  \, C \,e^{\phi}\, \,\, , \\
\dot{h} &=\,-Q\,e^{-2g_1-2g_2}\,\, , \\
\dot{g}_i &=\,e^{2f-2g_i}\,\, , 
\qquad\qquad (i=1,2)\,\,, \\
\dot f &=\,3\,-\,e^{2f-2g_1}\,-\,e^{2f-2g_2}\,-\,
{C \over 2}\,\,e^{\phi}\,\,,
\end{aligned} \end{equation}
where now the derivatives are taken with respect to the new variable $\rho$.

The equation for the dilaton in (\ref{sol}) can be integrated immediately,
with the result:
\beq
e^{\phi}\,=\,-{1\over C}\,\,{1\over \rho}\,\,,
\qquad\qquad (\rho<0)\,\,,
\label{dil-g1g2}
\eeq
where we have absorbed an integration constant in a shift of the radial
coordinate.  Moreover, by combining the equations for $g_1$ and $g_2$ one
easily realizes that the combination $e^{2g_1}\,-\,e^{2g_2}$ is constant.
Let us write:
\beq
e^{2g_1}\,=\,e^{2g_2}\,+\,a^2\,\,.
\label{g1g2}
\eeq
On the other hand, by using the solution for $\phi(r)$ just found and the
equations for the $g_i$'s in (\ref{sol}),  the first-order equation for $f$
can be rewritten as:
\beq
\dot{f}\,=\, 3\,-\, \dot{g}_1 \,-\, \dot{g}_2\, +\,
 \frac{1}{2\rho}\,\,,
\eeq
which can be integrated immediately, to give:
\beq
e^{2f+2g_1+2g_2}\,=\,-c\rho e^{6\rho}\,\,,
\label{f-and-gs}
\eeq
with $c$ being an integration constant. This constant can be absorbed by
performing a suitable redefinition. In order to make contact with the case
in which $g_1=g_2$ let us take $c=6$. Then, by combining  (\ref{f-and-gs})
with  the equation of $g_2$, we get
\beq
e^{4g_2+2g_1}\,\dot g_2\,=\,e^{2g_1+2g_2+2f}\,=\,-6\rho e^{6\rho}\,\,,
\eeq
which, after using the relation (\ref{g1g2}), can be integrated with the
result
\beq
e^{6g_2}\,+\,{3\over 2}\,a^2\,e^{4g_2}\,=\,(1-6\rho)\,e^{6\rho}\,+\,c_1\,\,.
\label{cubic}
\eeq
Notice that, indeed, for $a=0$ this equation reduces to the $g_1=g_2$
solution (see eq. (\ref{egKW})). Moreover, by combining eqs. (\ref{g1g2}) and 
(\ref{f-and-gs}) the expression of $f$ can be
straightforwardly written in terms of $g_2$, as follows:
\beq
e^{2f}\,=\,-{6\rho e^{6\rho}\over e^{4g_2}+a^2\,e^{2g_2}}\,\,.
\eeq
It is also easy to get the expression of the warp factor $h$:
\beq
h(\rho)\,=\, - Q \,\int\,
\frac{d\rho}{e^{4g_2} \,+\, a^2\, e^{2g_2}}
\,+\,c_2\,\,.
\eeq
Thus, the full solution is determined in terms of $e^{2g_2}$ which, in turn,
can be obtained from (\ref{cubic}) by solving a cubic algebraic equation. In
order to write the  explicit value of $e^{2g_2}$, let us define the function:
\beq
\xi(\rho)\,\equiv\,(1-6\rho)\,e^{6\rho}\,+\,c_1\,\,.
\eeq
Then, one has:
\beq
 e^{2g_2}\,=\, {1\over 2}\,\,\Bigg[\,-a^2\,+\,
{a^4\over \big[\,\zeta(\rho)\,\big]^{{1\over 3}}}\,+\,
\big[\,\zeta(\rho)\,\big]^{{1\over 3}}\,\,\Bigg]\,\,,
\eeq
where the function $\zeta(\rho)$ is defined in terms of $\xi(\rho)$ as:
\beq
\zeta(\rho)\,\equiv\,4\,\xi(\rho)\,-\,a^6\,+\,
4\,\sqrt{\xi(\rho)^2\,\,-\,{a^6\over 2}\,\xi(\rho)}\,\,.
\eeq
In expanding these functions in series near the UV ($\rho \to 0$) one gets a similar behavior to the one discussed in section \ref{solKW}. Very interestingly, in the IR of the field theory, that is when $\rho \to - \infty$, we get a behavior that is ``softened'' respect to what we found in section \ref{solKW}. This is not unexpected, given the deformation parameter $a$. Nevertheless, the solutions are still singular. Indeed, the dilaton was not affected by the deformation $a$.

\subsection{Massive Flavors}
In the ansatz we have been using up to now we have assumed that the density
of RR charge of the D7-branes is independent of the holographic coordinate.
This is, of course, what is expected for a flavor brane configuration which
corresponds to massless quarks. On the contrary, in the massive quark case,
a supersymmetric D7-brane has a non-trivial profile in the radial direction
\cite{Arean:2004mm} and, in particular ends at some non-zero value of the
radial coordinate. These massive embeddings have free parameters which could be
used to smear the D7-branes. It is natural to think that the corresponding
charge and mass distribution of the smeared flavor branes will depend on the
radial coordinate in a non-trivial way.

It turns out that there is a simple modification of our ansatz for $F_1$
which gives rise to a charge and mass distribution with the characteristics
required to represent smeared flavor branes with massive quarks. Indeed, let
us simply substitute in (\ref{BPS}) the constant $C$ by a function $C(r)$. In
this case:
\begin{equation} \begin{aligned}
F_1 &=\,C(r)\,(d\tau\,+\,A)\,\,, \\
dF_1 &=\,2\,C(r)\,J\,+\,C'(r) dr \wedge (d\tau\,+\,A)\, .
\end{aligned} \end{equation}
Notice that the SUSY analysis of sect. \ref{SEBPS} remains unchanged since
only $F_1$, and not its derivative, appears in the supersymmetric
variations of the dilatino and gravitino. The final result is just the same
system (\ref{BPS}) of first-order BPS equations, where now one has to
understand that 
$C$ is a prescribed function of $r$, which encodes the non-trivial profile
of the D7-brane. Notice that $C(r)$ determines the running of the dilaton
which, in turn, affects the other functions of the ansatz.

A natural question to address here is whether or not the solutions of the
modified BPS system solve the equations of motion of the supergravity plus
branes system. In order to check this fact, let us write the DBI term of
the action, following our prescription (\ref{DBIaction-general}). Notice
that, in the present case, $\Omega = -dF_1$ is the sum of three decomposable
pieces:
\beq
\Omega\,=\,\Omega^{(1)}\,+\,\Omega^{(2)}\,+\,\Omega^{(3)}\,\,,
\eeq
where $\Omega^{(1)}$ and $\Omega^{(2)}$ are just the same as in eq. \eqref{dFi},
while $\Omega^{(3)}$ is given by:
\beq
\Omega^{(3)}\,=\, -C'(r)\,dr\wedge (d\tau+A)\,=\,
-h^{-{{1 \over 2}}}\,e^{-f}\,C'(r)\,\,\, \hat{e}^{r}\wedge \hat{e}^{0}\,\,.
\eeq
The modulus of this new piece of $\Omega$ can be straightforwardly computed,
namely:
\beq
|\,\Omega^{(3)}\,|\,=\,h^{-{{1 \over 2}}}\,e^{-f}\,
|\,C'(r)\,|\,\,.
\eeq
By using this result, together with the one in (\ref{|dFi|}), one readily
gets the expression of the DBI terms of the action of the smeared D7-branes:
\beq \label{DBIM}
S_{DBI}\,=\,-\,T_7 \int_{{\cal M}_{10}}\,h^{-{{1 \over 2}}}\,
e^{\phi}\,
\Big(4\,|\,C(r)\,|\,e^{-2g}\,+\,|\,C'(r)\,|\,e^{-f}\,\Big)
\,\sqrt{-G}\,\,d^{10}x\,.
\eeq
From this action it is immediate to find the equation of motion of the
dilaton, \ie:
\beq
\phi''\,+\,(4g'+f')\,\phi'\,=\,
C^2\,e^{2\phi-2f}\,+\,4\,|C|\,\,e^{\phi-2g}\,+\,
e^{\phi-f}\,|\,C'\,|\,\,.
\label{dil-eq-massive}
\eeq
It can be verified that the first-oder BPS equations (\ref{BPS}) imply the
fulfilment of eq. (\ref{dil-eq-massive}), provided the functions
$C(r)$ and $C'(r)$ are non-negative. Notice that now, when computing
the second derivative of $\phi$ from the BPS system (\ref{BPS}) with 
$C=C(r)$, a new term containing $C'(r)$ is generated. It is easy to verify
that this new term matches precisely the last term on the right-hand side of
(\ref{dil-eq-massive}). 

It remains to verify the fulfilment of Einstein's equation. The
stress-energy tensor of the brane can be computed from eq. (\ref{TMN}), where
now the extra decomposable piece of $dF_1$  must be taken into account. The
result one arrives at, in the  vielbein basis (\ref{10vielbein}), is a direct
generalization of (\ref{TMNcomponents}):
\bear
&&T_{x^ix^j}\,=\,-\,e^{\phi}\,\,h^{-{1 \over 2}}\,
\Big [ 2\,|\,C(r)\,|\,e^{-2g}  \,+\,{1 \over 2}\, |\,C'(r)\,|\,e^{-f}\,\Big ]\,
\,\eta_{x^ix^j} \,\,,
\qquad (i,j=0,\ldots,3)\,\, , \rc\rc 
&&T_{ab}\,=\,-\,e^{\phi}\,
h^{-{1 \over 2}} \Big [ |\,C(r)\,|\,e^{-2g}  \,+\,{1 \over 2}\,
|\,C'(r)\,|\,e^{-f}\,\Big ]\,\,\delta_{ab}\,\,,
\, \qquad\qquad (a,b=1,\ldots,4)\,\, , \rc\rc 
&&T_{rr}\,=\,T_{00}\,=\,-\,2\,|\,C(r)\,|\,\,h^{-{1 \over 2}} \,
e^{\phi-2g}  \,\, .
\label{TMNcomponents-massive}
\eear
As happened for the equation of motion of the dilaton, one can verify that
the extra pieces on the right-hand side of (\ref{TMNcomponents-massive})
match precisely those generated by the second derivatives appearing in the
expression (\ref{Ricci-components}) of the Ricci tensor if 
$C(r)$ and $C'(r)$ are non-negative. As a consequence,
the first-order equations (\ref{BPS}) with a function $C(r)$ also imply  the
equations of motion for the ten-dimensional metric $g_{MN}$. 
It is also interesting to point out that, if $C(r)$ and $C'(r)$ are non-negative, 
$S_{DBI}$ can also be written in the form (\ref{SDBI-forms}), where
$\Omega_8$ is exactly the same eight-form as in eqs. (\ref{Omega_8}) and
(\ref{cal}). 

Notice that, if the function $C(r)= 3N_f(r)/4\pi$ has a Heaviside-like shape  ``starting'' at some finite value of the radial coordinate, then our BPS equations and solutions will be the ones given in section \ref{solKW}  for values of the radial coordinate bigger than  the ``mass of the flavor''. However,  below that radial value the solution will be the one of Klebanov-Witten (or deformations of it, see appendix B), with a non-running dilaton. Aside from decoupling in the field theory, this is clearly indicating
that the addition of massive flavors ``resolves'' the singularity.
Physically this behavior is expected and makes these massive flavor more interesting.

\section{Summary, Future Prospects and Further Discussion }

In this paper we followed the method of \cite{Casero:2006pt} 
to construct a dual to the field theory defined by the Klebanov and Witten 
after  $N_f$ flavors of quarks and antiquarks have been added to both gauge groups. In 
Section \ref{sect2} of this work, we 
wrote BPS equations describing the dynamics of this system and found 
solutions to this first order system, that of course solves also all the 
second order equations of motion. We analyzed the solutions to the BPS system and learnt 
that, even when singular, the character of the singularity permits to get 
field theory conclusions from the supergravity perspective.

We proposed a formulation for the dual field theory to these 
solutions, constructing a precise 4-dimensional superpotential.
We studied these solutions making many matchings with  field theory 
expectations that included the R-symmetry breaking and Wilsonian beta 
function. Also, using the well known (supergravity) 
superpotential approach, 
we learnt that our field theory, aside from being deformed by a marginal 
(then turned irrelevant) operator, 
modifies its dynamics by giving VEV to operators of dimension six and 
eight.
We explained how to change relations between couplings and $\theta$-angles 
in the theory, from the perspective of our solutions. We believe that 
these many checks should encourage other physicists to study this 
background more closely.

In Section \ref{generalizations} of this paper, we presented a careful account of the many 
technical details 
regarding the derivation of the results in section \ref{sect2} summarized above. But 
most interestingly, Section \ref{generalizations}  is not only about technical details. Indeed, 
using the logic and intuitions developed in Section \ref{sect2}, we generalized the 
approach described there for {\it any} five dimensional manifold that can 
be written as a Sasaki-Einstein space (a one-dimensional fibration over a  
K\"ahler-Einstein space). It is surprising that the same structure of BPS 
eqs and  ten-dimensional superpotential repeats for all the manifolds 
described above. This clearly points to some ``universality'' of the 
behavior of 4-dimensional \Nugual{1} SCFT's with flavors.

We have added some brief comments about what happens when we take the 
number of flavors $N_f=0$ in 
our BPS eqs (see Appendix \ref{appendixb}). It is interesting to recover some solutions studied in the 
past from this perspective since it puts into context
previous analysis. Again, the careful study of this ``unflavored'' 
solutions might be of interest to many physicists.
We shortly commented on the possibility of adding to the dynamics of the 
4-d field theory fundamentals with mass, presenting a 
general context to do this. We 
will exploit this procedure in the future to get a better understanding of 
our singular backgrounds, make contact with field theory results and study 
many other interesting problems. 

All the results described above not only 
motivate a more detailed analysis of this approach from a field 
theoretical 
viewpoint, 
but also emphasize the need for a deeper geometrical study, that 
clearly 
will reveal interesting underlying structure.

\subsection{Future Directions}
Many things can be done following the results of this paper.
It is natural to extend the treatment to the case of the 
Klebanov-Tseytlin and Klebanov-Strassler solutions. The result is likely to be interesting, since the 
fundamentals and the KT cascade ``push in different directions'' in the RG 
flow. One might find a fine-tuned situation in which the IR dynamics is 
different from the one in the Klebanov-Strassler model.

Other things that immediately come to mind are to study the dynamics of 
moving strings in this backgrounds, details related to dibaryons, flavor 
symmetry breaking, etc. Even when technically involved, it should 
be nice to understand the backreaction of probes where the worldvolume 
fields have been
turned on, since some interesting problems may be addressed.

The formalism developed to deal with configurations of IIB dual to 
massive 
fundamentals seems useful in different contexts.
Needless to say, the approach adopted here is immediately generalizable to 
the case of type IIA backgrounds. Duals to \Nugual{1} field theories have been 
constructed and it seems natural to apply our methods in those cases.

Finding black hole solutions in our geometries is not an elementary task; 
but it should not be very difficult. The interest of this problem resides in 
the fact that this will produce a ``well-defined'' black hole background 
where to study, among other things, plasmas that include the dynamics of 
color and flavor at strong coupling. This is a very well defined problem 
that we believe of much interest.

On the field theory side, it should be interesting to understand in more 
detail how the smearing procedure affects the superpotential. We gave 
a possible answer and detailed study can uncover interesting subtleties. 
Here again, similar ideas can be extended to other situations in type IIA 
and type IIB. Getting a better handle on the field theory interpretation 
of our ``generalized'' approach of Part II seems also interesting. 
Indeed, understanding in detail what is the ``universality'' that produces 
the same dynamics for a large class of \Nugual{1} SCFT's with flavor would be nice.
\subsection{Further Discussion}
Let us finish this paper with some discussions that might be of interest 
for the reader. The first point we want to address is what could be the 
application of these results to Physics. Indeed, it is not easy to find an 
interesting physical system displaying a Landau pole (without a UV 
completion, like QED has, for example). Of course, as explained 
above, this paper is a first step in a  more detailed study of a 
cascading field theory with flavors, that with no doubt has applications 
in Physics. Nevertheless, one can find some interesting problems already 
at this stage.

As described above, finding a black hole in our geometry, might be a good 
simulation of the Physics of a strongly coupled quark-gluon plasma. Even more, 
since we would be only 
interested in effects in the hydrodynamics regime, using the IR of this 
black hole solutions should be enough to learn about Physics at RHIC, for 
example.
One can also think that our paper starts the study of the different phases 
of this generalized \Nugual{1} SQCD obtained from Klebanov-Witten-like models.

Let us change the subject of the discussion and go back to
our procedure, that was well explained in the introduction of this paper.
The reader may remember the difference between a weakly gauged 
symmetry and a global symmetry, let us now connect this to 
supergravity. One important distinction between the 
approaches for finding string duals to field theories with 
fundamentals is that in the approach where the solution consists only of 
supergravity fields, the field theory will have this global symmetry 
weakly 
gauged. On the contrary, in our case backreacting with the Born-Infeld 
action, the  symmetry will be global. We can see this clearly in the fact 
that the BI action has the freedom to add worldvolume gauge fields (and 
scalars), hence introducing a gauged symmetry in the bulk, dual to a 
global 
symmetry in the boundary. In a (complicated) reduction of our Type IIB 
plus Born-Infeld action to five dimensions, we would see some $SU(N_f)$ 
gauge fields (as many as branches of flavor branes we added) that would 
enter in the holographic formulas to compute field theory correlators.

It is interesting to notice that depending on the physical situation we
want to work with, we should choose the approach used here or the 
complementary one of finding a solution purely in supergravity.
Indeed, for situations where we do not want to take into account the 
``flavor degrees of freedom'' of the extra branes, but what we want is to 
introduce some operator in the dual field theory (like a giant graviton, 
a domain wall or a Wilson line) we should work within the purely 
supergravity approach \cite{Gomis:2006sb}. Indeed, if we are thinking about the presence of an 
operator (say in $N=4$ SYM), there should be no ``flavor degrees of 
freedom'' in the solutions.

Finally, we would like to comment on the smearing procedure. One way in 
which we can think about it is to realize that usually (unless they are 
D9 branes) the 
``localized'' flavor branes will break part of the isometries of the 
original background dual to the unflavored field theory.
The ``smeared'' flavor branes on the other hand reinstate these isometries 
(global symmetries of the field theory dual). In some sense the flavor 
branes are `deconstructing' these dimensions (or these global groups) for 
the field theory of interest. In the case in which we have a finite number 
of flavors, these manifolds become fuzzy, while for $N_f\to \infty$, we recover 
the full invariance.

\section*{Acknowledgments}
Some physicists helped us with discussions, conversations and correspondence, 
to understand and present the results of this paper in a  better way.
We would like to thank Daniel Are\'an, Adi Armoni, Sergio Benvenuti, Gaetano Bertoldi, Matteo Bertolini, Nadav Drukker, Jose D. Edelstein, Davide Forcella, Tim Hollowood, Carlos Hoyos, Emiliano Imeroni, Harald Ita, Prem Kumar, Javier Mas,  Paolo Merlatti, Asad Naqvi, Angel Paredes, Rodolfo Russo, Kostas Skenderis and Alberto Zaffaroni for their very valuable input. 
The  work of FC and AVR was
supported in part by MCyT and  FEDER  under grant
FPA2005-00188,  by Xunta de Galicia (Conselleria de Educacion and grant PGIDIT06PXIB206185PR)
and by  the EC Commission under  grant MRTN-CT-2004-005104.

\vskip 1cm
\renewcommand{\theequation}{\rm{A}.\arabic{equation}}
\setcounter{equation}{0}
\appendix
\section{Appendix: SUSY Transformations  in String and Einstein frame}
\label{SUSYapp}
\medskip
The supersymmetry transformations of Type IIB supergravity were found long ago
in ref. \cite{SUSYIIB}. Here we will follow the conventions of the appendix A of 
\cite{Martucci}, where they are written in string frame. Let us recall them:
\begin{equation} \begin{aligned}
\delta_{\epsilon}  \lambda^{(s)} &= {1 \over 2} \Big ( \Gamma^{(s)\,M} \partial_M\phi\,+\,{1 \over 2}{1 \over 3!}H_{MNP}\,\Gamma^{(s)\,MNP}\,\sigma_3 \Big )\epsilon^{(s)}\,-\,{1 \over 2}e^{\phi} \Big ( F_M^{(1)}\,\Gamma^{(s)\,M}\,(i\sigma_2)\,+ \\
&\qquad +\,{1 \over 2}{1 \over 3!}F_{MNP}^{(3)}\,\Gamma^{(s)\,MNP}\,\sigma_1\Big )\epsilon^{(s)} , \\
\delta_{\epsilon}  \psi^{(s)}_M &= \nabla^{(s)}_M\epsilon^{(s)}\,+\,{1 \over 4} {1 \over 2!}H_{MNP}\,\Gamma^{(s)\,NP}\,\sigma_3 \epsilon^{(s)}\,+\,{1 \over 8}e^{\phi} \Big ( F_N^{(1)}\,\Gamma^{(s)\,N}\,(i\sigma_2)\,+ \\
&\qquad +\,{1 \over 3!}F_{NPQ}^{(3)}\,\Gamma^{(s)\,NPQ}\,\sigma_1\,+\,{1 \over 2}{1 \over 5!}F_{NPQRT}^{(5)}\,\Gamma^{(s)\,NPQRT}\,(i\sigma_2)\Big )\Gamma^{(s)}_M\epsilon^{(s)} \,\,  , \label{susyIIB}
\end{aligned} \end{equation}
where the superscript $s$ refers to the string frame, 
$\sigma_i \quad i=1,2,3$ are the Pauli matrices, $H$ is the NSNS three-form and
$F^{(1)}$, $F^{(3)}$ and $F^{(5)}$ are the RR field strengths.  In (\ref{susyIIB}) $\epsilon$ is 
a doublet of Majorana-Weyl spinors of positive chirality.

We can study how these equations change under a rescaling of the metric like
\beq
g^{(s)}_{MN}\,=\,e^{{\phi \over 2}}\,g_{MN} .
\eeq
In doing that it is useful to follow Section 2 of \cite{Strominger}. Under the above change for the metric, there are some quantities which also change:
\bear
&&\Gamma^{(s)}_{M}\,=\,e^{{\phi \over 4}}\,\Gamma_{M}\,\, , \rc
&&\epsilon^{(s)}\,=\,\,e^{{\phi \over 8}}\,\epsilon\,\, , \rc
&&\lambda^{(s)}\,=\,\,e^{-{\phi \over 8}}\,\lambda\,\, , \rc
&&\psi_M\,=\,e^{-{\phi \over 8}} \Big ( \psi^{(s)}_{M} \,-\, {1 \over 4} \Gamma^{(s)}_M \, \lambda^{(s)} \Big ) \,\, .
\eear
The equation for the dilatino in the new frame can be easily obtained whereas in doing the same for the gravitino equation we will use that
\beq
\nabla^{(s)}_M\epsilon^{(s)}\,=\,e^{{\phi \over 8}} \Big [ \nabla_M\epsilon \,+\, {1 \over 8}\Gamma_M^{\,\,N}(\nabla_N \phi)\,+\,{1 \over 8}(\nabla_M\phi)\Big ]\,\, .
\eeq
After some algebra with gamma-matrices, the SUSY transformations in Einstein frame we obtain are the following ones:
\bear \label{Eframe}
\delta_{\epsilon}  \lambda&=&{1 \over 2}\Gamma^M \big (  \partial_M\phi-e^{\phi}F_M^{(1)}(i\sigma_2) \big )\epsilon\,+{1 \over 4}{1 \over 3!} \Gamma^{MNP}\big (e^{-{\phi \over 2}} H_{MNP}\sigma_3 -e^{{\phi \over 2}}F_{MNP}^{(3)} \sigma_1 \big )\epsilon   ,  \rc\rc
\delta_{\epsilon}  \psi_M&=&\nabla_M\epsilon+{1 \over 4}e^{\phi}F_M ^{(1)}(i\sigma_2)\epsilon-{1 \over 96}\big (e^{-{\phi \over 2}} H_{NPQ}\sigma_3 +e^{{\phi \over 2}}F_{NPQ}^{(3)} \sigma_1 \big )\big ( \Gamma_M^{\,\, NPQ}-9\delta^N_M\Gamma^{PQ} \big )\epsilon+\,\, \rc\rc
&&+{1 \over 16}{1 \over 5!}F_{NPQRT}^{(5)}\Gamma^{NPQRT}(i\sigma_2)\Gamma_M\epsilon   .  
\eear
In order to write the expression of the SUSY transformations, it is convenient to change the notation used for the spinor. Up to now we have considered the double spinor notation, namely the two
Majorana-Weyl spinors $\epsilon_1$ and $\epsilon_2$ form a two-dimensional vector 
$\begin{pmatrix}\epsilon_1\cr\epsilon_2\end{pmatrix}\,\,$.  We can rewrite the double spinor in complex notation as\footnote{Notice that there is an ambiguity in the choice of the relation between complex and real spinors.} $\epsilon\,=\,\epsilon_1\,-\,i\epsilon_2$. It is then
straightforward to find the following rules to pass from complex to real spinors:
\beq
\epsilon^*\,\leftrightarrow\,\sigma_3\,\epsilon\,\,,
\,\,\,\,\,\,\,\,\,\,\,\,\,\,\,\,\,\,\,
-i\epsilon^*\,\leftrightarrow\,\sigma_1\,\epsilon\,\,,
\,\,\,\,\,\,\,\,\,\,\,\,\,\,\,\,\,\,\,
i\epsilon\,\leftrightarrow\, i\sigma_2\,\epsilon\,\,,
\label{rule}
\eeq
where the Pauli matrices act on the two-dimensional vector 
$\begin{pmatrix}\epsilon_1\cr\epsilon_2\end{pmatrix}$.

\section{Appendix: The Unflavored Solutions}
\label{appendixb}
\renewcommand{\theequation}{\thesection.\arabic{equation}}
\setcounter{equation}{0}
In this appendix we will study our BPS system of linear ordinary differential equations \eqref{eqng}-\eqref{eqnh} and we will find its general solution in the absence of D7-branes.
Not only we will recover the solution describing a stack of D3-branes placed at the apex of the real Calabi-Yau cone over a generic Sasaki-Einstein 5-manifold $M_5$, preserving (at least) four supercharges, and its near-horizon limit $AdS_5\times M_5$ dual to the (at least) \Nugual{1} superconformal gauge theory describing the IR dynamics on the stack of D3-branes, but also the solution describing D3-branes smeared homogeneously on a blown-up 4-cycle inside the Calabi-Yau, discussed in the paper \cite{PandoZayas:2001iw} for the case of the conifold (more precisely a $\mathbb{Z}_2$ orbifold of it) and then in full generality in \cite{Benvenuti:2005qb} for all Calabi-Yau cones. We will also study the unflavored  limit of  the general deformation of the KW model analyzed  in section \ref{generalKWdef} and we will show that it gives rise to the two-parameter metrics found
in ref. \cite{PandoZayas:2001iw}.

Let us look at our BPS system of linear ordinary differential equations (\ref{eqng}-\ref{eqnh}). We will sometimes refer to the case of the conifold for the sake of simplicity. The generalization to any Sasaki-Einstein is straightforward, the only difference being the normalization in (\ref{eqnh}) and in the RR 5-form field strength, related to the volume of the Sasaki-Einstein base.

First of all notice that $N_f$ must be set to zero in the system of equations and not in our solution, since when we solved the equation for the dilaton (\ref{eqnphi}) we supposed that $N_f\neq 0$. This allowed us to get \eqref{dilatonKW} after shifting the radial variable.

It is easy to show that the most general solution to the BPS system when $N_f=0$, up to redefinition of the coordinates, is the following:
\begin{align}
\phi(\rho) &= \phi_0 \\
e^{g(\rho)} &= \big[ e^{6\rho} + c_1 \big]^{1/6} \\
e^{f(\rho)} &=  e^{3\rho} \big[ e^{6\rho} + c_1 \big]^{-1/3} \\
h(\rho) &= c_2 - 4 L^4 \int d\rho \, e^{-4g(\rho)}  \\
r(\rho) &= \int d\rho \, e^{f(\rho)} \;.
\label{unflav1}
\end{align}
$L$ is the common radius of $AdS_5$ and the Sasaki-Einstein $M_5$ in the solution dual to the superconformal theory, and is fixed by the number of D3 branes and the volume of the Sasaki-Einstein manifold. For $T^{1,1}$: $L^4=\frac{27}{4}\pi N_c $.

The real integration constant $c_1$ discriminates different classes of solutions.

If $c_1=0$ then we recover the D3-branes solution (with nonzero $c_2$, that can be fixed to 1) or its near-horizon $AdS$ solution (with $c_2=0$):
\begin{align}
e^f &= e^g = e^\rho =r\\
h &= c_2 + \frac{L^4}{r^4}\;.
\end{align}

If $c_1>0$ the solution describes $N_c$ smeared D3-branes on the blown-up 4-cycle of the Calabi-Yau \cite{PandoZayas:2001iw, Benvenuti:2005qb}.
Indeed, let us consider the change of radial coordinate:
\begin{equation}
\label{change>}
\big[1+c_1 e^{-6\rho}\big]^{-1}= 1- \frac{b^6}{r^6} \equiv k(r)\;,
\end{equation}
with $r>b$. If we further identify
\begin{equation}
b^2= c_1^{1/3} \;,
\end{equation}
it follows that
\begin{align}
e^{2g} &=r^2\\
e^{2f} &=r^2 k(r)\\
e^{2f} d\rho^2 &= \frac{dr^2}{k(r)}\;,
\end{align}
so that the 6-dimensional metric, which is Calabi-Yau, is
\begin{equation}
\label{6metric>}
ds_6^2 = [k(r)]^{-1} dr^2 + \frac{k(r) r^2}{9} (d\psi + \sum_{i=1,2} \cos \theta_i d\varphi_i)^2 +
\frac{r^2 }{6}\sum_{i=1,2}(d\theta_i^2+\sin^2\theta_i d\varphi_i^2)\;,
\end{equation}
that describes a deformation of the Calabi-Yau where a K\"ahler-Einstein 4-cycle is blown up at $r=b$. In order for the resolved Calabi-Yau to be smooth, an orbifolding along the $U(1)$ fiber parameterized by $\psi$ is usually needed. For the case of the deformation of the conifold the orbifold action is $\mathbb{Z}_2$, so that $\psi$ ranges from $0$ to $2\pi$.
 The 10-dimensional metric of the solution is then:
\begin{equation}
\label{10metric>}
ds_{10}^2 = [h(r)]^{-1/2} dx_{1,3}^2 + [h(r)]^{1/2} ds_6^2 \;,
\end{equation}
with the warp factor\footnote{The additive integration constant in $h$ is omitted in order to asymptote to $AdS_5\times X_5$ for large values of $r$.}
\begin{equation}
\label{warp>}
h(r)=-2\frac{L^4}{b^4}\bigg[ \frac{1}{6} \log \frac{(\tilde{r}^2-1)^3}{\tilde{r}^6-1}+\frac{1}{\sqrt{3}}\Big(\frac{\pi}{2}-\arctan \frac{2\tilde{r}^2+1}{\sqrt{3}} \Big) \bigg]\;,\quad\tilde{r}=\frac{r}{b}\;.
\end{equation}
The gauge theory dual to this local K\"ahler deformation of the Calabi-Yau cone is a deformation of the superconformal theory due to the insertion of a VEV of a dimension 6 operator, which
is a combination of the operators $\Tr (\cW_\alpha \bar \cW^\alpha)^2$, $\cW_\alpha$ being the gluino superfield \cite{Benvenuti:2005qb}.
The orbifold action is needed to have a dual field theory whose mesonic branch of the moduli space is (the symmetric product of $N_c$ copies of) the resolved Calabi-Yau.

A similar analysis can be done for the solutions with $c_1<0$, but in that case the 6-dimensional transverse space happens to have a curvature singularity and cannot be described as an algebraic variety. Therefore the supergravity solution is not expected to describe a dual supersymmetric gauge theory.

Let us now study the unflavored limit of the general deformation of the KW solution
of section \ref{generalKWdef}. Recall that, in this case, the metric depends on three
functions ($f$, $g_1$  and $g_2$) and the warp factor $h$. In terms of the variable
$\rho$ introduced in (\ref{rho-generalKW}) the metric can be written as:
\beq
ds^2 = h^{-1/2} dx_{1,3}^2 + 
h^{1/2}\,e^{2f} \bigg\{ d\rho^2 +
\frac{1}{6} \sum_{i=1,2} e^{2g_i-2f}( d\theta_i^2 + \sin^2 \theta_i \,
 d\varphi_i^2)  + \frac{1}{9} (d\psi + \sum_{i=1,2} \cos\theta_i \,
d\varphi_i)^2 \bigg\}\,\,.
\eeq
The unflavored limit of the BPS system (\ref{BPSg1g2}) amounts to taking $C=0$. As in the previous case, the solution of section \ref{generalKWdef} is not valid in this limit (see
eq. (\ref{dil-g1g2})) and one has to take $C=0$ in the system (\ref{BPSg1g2}) and integrate it
again following the same steps as in section \ref{generalKWdef}. The result can be written in terms of the function $e^{2g_2(\rho)}$, which is the solution of the following
cubic equation:
\beq
e^{6g_2}\,+\,{3\over 2}\,a^2\,e^{4g_2}\,=\,e^{6\rho}\,+\,c_1\,\,,
\label{eg2-unflavor}
\eeq
where $c_1$ is an integration constant. In terms of $e^{2g_2(\rho)}$ the other
functions of the ansatz can be written as:
\begin{equation} \begin{aligned}
e^{2g_1} &= e^{2g_2}\,+\,a^2\,\,, \\
e^{2f} &= {e^{6\rho}\over e^{4g_2}+a^2\,e^{2g_2}}\,\,, \\
h(\rho) &= -Q\,\int{d\rho\over e^{4g_2}+a^2\,e^{2g_2}}\,+\,c_2\,\,,
\end{aligned} \end{equation}
where $Q$ is given by (\ref{Bian1}) and $c_2$ is a new integration constant. 
Let us now perform the following change of radial variable
\beq
e^{2g_2(\rho)}\,=\,{r^2\over 6}\,\,.
\eeq
Taking into account (\ref{eg2-unflavor}), the inverse relation between these two 
radial variables is:
\beq
e^{6\rho}\,=\,{1\over 216}\,\,\big[\,r^6\,+\,9a^2r^4\,-\,b^6\,\big]\,\,,
\eeq
where we have redefined the constant $c_1$ as:
\beq
b^2\,\equiv\,6\,(c_1)^{{1\over 3}}\,\,.
\eeq
By using these relations, one can readily prove that:
\bear
e^{2g_1}&=&{1\over 6}\,\,(r^2\,+\,6a^2)\,\,,\rc\rc
e^{2f}&=&{r^2\over 6}\,\kappa(r)\,\,,
\eear
where the function $\kappa(r)$ is defined as follows:
\beq
\kappa(r)\,\equiv\,{r^6\,+\,9a^2\,r^4\,-\,b^6\over
r^6\,+\,6a^2r^4}\,\,.
\label{kappa(r)}
\eeq
It is also easy to verify that:
\beq
d\rho\,=\,{dr\over r \kappa(r)}\,\,.
\eeq
Using these results and redefining the warp factor as 
$h(r)^{-{1\over 2}}\to h(r)^{-{1\over 2}}/6$, we get a metric of the form:
\beq
ds^2\,=\,\big[\,h(r)\,\big]^{-{1\over 2}}\,
\Big[\,dx^2_{1,3}\,\Big]\,+\,
\big[\,h(r)\,\big]^{{1\over 2}}\,ds^2_6\,\,,
\eeq
with $ds^2_6$ given by:
\bear
ds^2_6\,&=&\,\big[\,\kappa(r)\,\big]^{-1}\,dr^2\,+\,
{r^2\over 9}\,\,\kappa(r)\,
(\,d\psi\,+\sum_{a=1,2}\,\cos\theta_i\,d\varphi_i\,)^2\,+\,\rc\rc
&&+\,{1\over 6}\,(\,r^2\,+\,6a^2\,)\,
(\,d\theta_1^2\,+\,\sin^2\theta_1\,d\varphi_1^2\,)\,+\,
{1\over 6}\,r^2\, (\,d\theta_2^2\,+\,\sin^2\theta_2\,d\varphi_2^2\,)\,\,,
\eear
while  the warp factor $h$  can be represented as:
\beq
h(r)\,=\,-Q\,\int{rdr\over r^6\,+\,9a^2\,r^4\,-\,b^6}\,+\,c_2\,\,.
\eeq
This is the solution with two K\"ahler deformations found in ref. \cite{PandoZayas:2001iw}: the
$a$ constant parameterizes global deformations, while the $b$ parameter corresponds to 
local deformations. 

\section{Appendix: Alternative Interpretation of the IR \\ Regime} \label{appAlt}

Here we put an alternative description of the IR theory as deduced from supergravity, which honestly we could not discard. It mainly arises from the analysis of the Klebanov-Witten model at small values of the string coupling, and it is based on the non-validity of the orbifold relations \eqref{g+}-\eqref{theta2} for all values of the parameters in the KW model, that was extensively pointed out in \cite{Strassler:2005qs}. In the whole analysis that will follow, we will consider for clarity only the case of equal gauge couplings $g_1=g_2\equiv g$.

\begin{figure}[ht]
\begin{center}
\includegraphics[height=0.24\textheight]{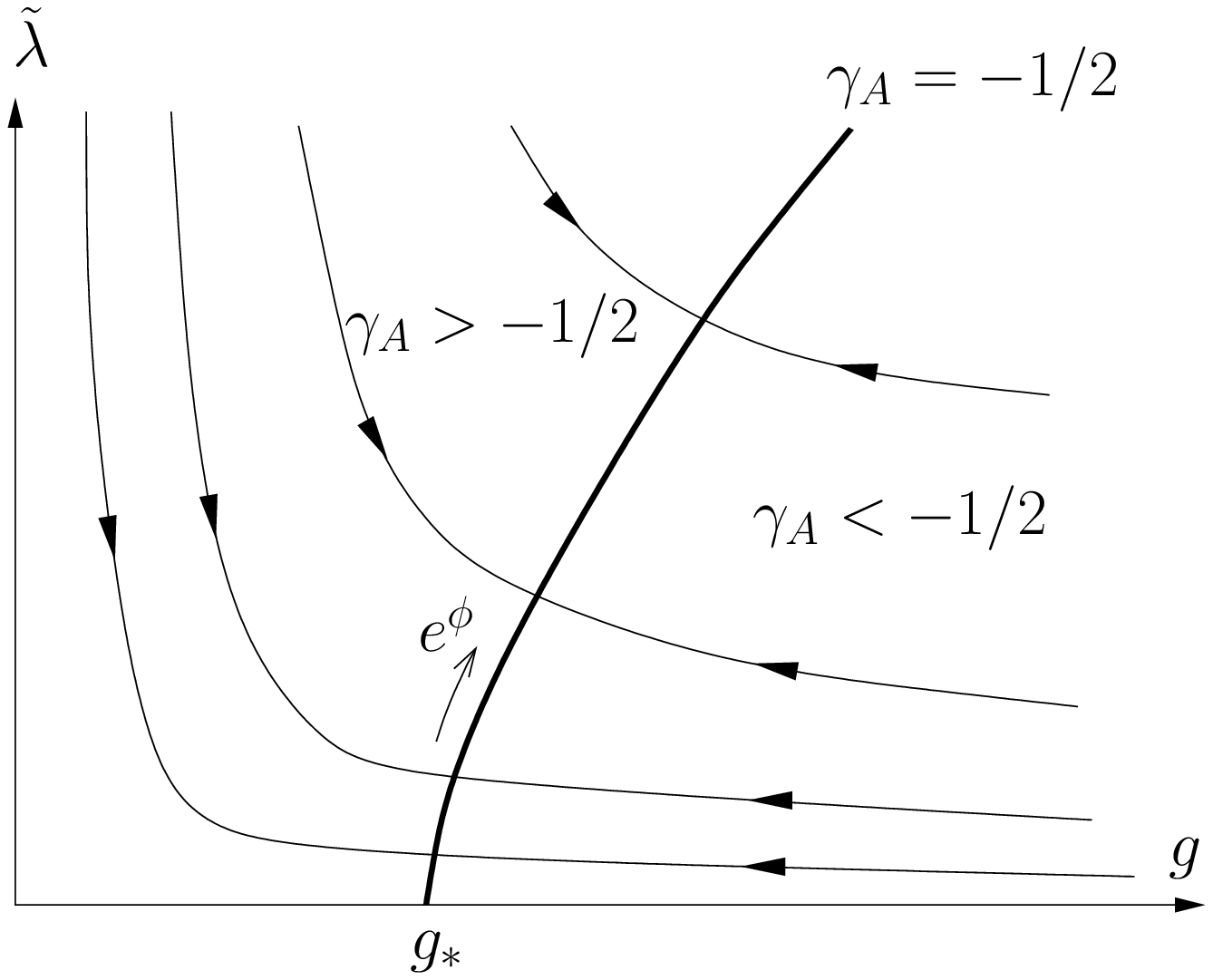}
\hspace{1cm}
\includegraphics[height=0.24\textheight]{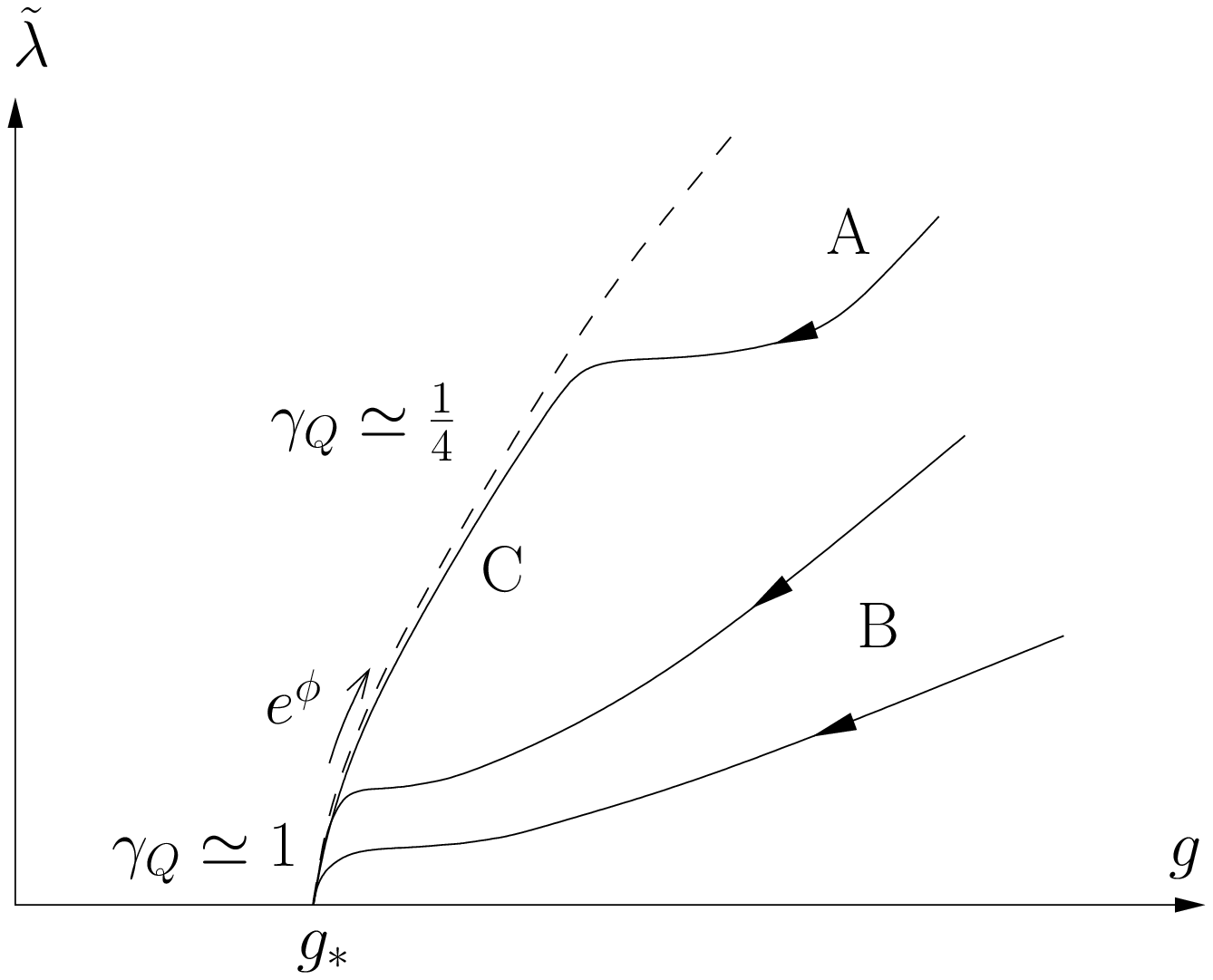}  \
\parbox[t]{0.42\textwidth}{\caption{RG flow phase space for the Klebanov-Witten
model. \label{fixed line KW}}}
\hspace{0.02\textwidth}
\parbox[t]{0.54\textwidth}{\caption{Klebanov-Witten model with flavors. The A-C
flow has backreacting D7's in the A piece and then follows the KW line in the C
piece; it corresponds to $N_f \ll N_c$. The B flow is always far from the KW
line, and corresponds to $N_f \gtrsim N_c$. \label{flow KW flavors}}}
\end{center}
\end{figure}

The curve of conformal points in the Klebanov-Witten model is obtained by requiring the anomalous dimension of the fields $A,B$ to be $\gamma_A(g,\tilde \lambda)=-1/2$, which assures $\beta_g=\beta_{\tilde\lambda}=0$ ($\tilde\lambda$ is the dimensionless coupling from the quartic superpotential). The qualitative shape of the curve is depicted in Figure \ref{fixed line KW}, as well as some possible RG flows. The important feature is that there is a minimum value $g_*>0$ that fixed points can have (due to the perturbative $\beta_g$ being negative, so that $g=0$ is an unstable IR point).
One way to determine this curve of fixed points is to apply the $a$-maximization procedure originally spelled in \cite{Intriligator:2003jj} by using Lagrange multipliers enforcing the marginality constraints \cite{Kutasov:2003ux}, and then express the Lagrange multipliers in terms of the gauge and superpotential couplings. 
This computation for the Klebanov-Witten model was done in \cite{Benvenuti:2005wi}.%
\footnote{We thank Sergio Benvenuti for making us aware of this method and of the literature on the subject.}
One can show that the curve of fixed points does not pass through the origin of the space of Lagrange multipliers, which is mapped into the origin of the space of couplings (free theory). 
In a particular scheme the curve of fixed points is an arc of hyperbola with the major axis along $\tilde\lambda=0$. The exact shape of the curve is scheme-dependent, due to scheme-dependence of the relation between Lagrange multipliers and couplings: we choose a scheme in which the Lagrange multipliers are quadratic in the couplings. This choice fixes a conic section, and it is such a hyperbola because the one-loop anomalous dimensions of the chiral superfields get a negative contribution from gauge interactions and a positive contribution from superpotential interactions.
The conclusion that the curve of conformal points does not pass through the origin of the space of coupling constants is physical.

The family of KW SUGRA solutions describes the fixed curve. It is parameterized
by $e^\phi$ that can take arbitrary values. For sufficiently large values of
it, we can trust the orbifold formula:
\begin{equation} \label{orbifold form}
\frac{g^2}{8\pi} = e^\phi \qquad \qquad \text{for} \qquad e^\phi \, N_c \gtrsim
1 \;.
\end{equation}
The 't Hooft coupling $g^2 N_c$ is large (at least of order 1, so the theory is
strongly coupled and the anomalous dimensions are of order 1) and the string
frame curvature $R_S\sim 1/(e^\phi N_c)$ is small.
For smaller values $e^\phi N_c \lesssim 1$, \eqref{orbifold form} cannot be
correct: it would give small 't Hooft coupling while the gauge theory is always
strongly coupled. The bottom end of the line corresponds to:
\begin{equation}
\{e^\phi \to 0\} \qquad \leftrightarrow \qquad \{g=g_*,\tilde\lambda=0\} \;,
\end{equation}
and the SUGRA curvature is large even if the field theory is still strongly
coupled. 
Anyway some quantities, for instance the quantum dimension of $A,B$, are
protected and do not depend on the coupling, so they can be computed in SUGRA
even for small values of $e^\phi N_c$.

\

The supergravity solution of our system with D7-branes is in the IR quite
similar to the KW geometry: the IR asymptotic background is $AdS_5\times
T^{1,1}$ (with corrections), but with running dilaton. The field theory is thus
deduced to be close to KW fixed line, but running along it as $e^\phi\to 0$ in
the IR. Moreover, $e^\phi$ controls the gravitational backreaction of the
D7-branes (as well as the gauge coupling), and as soon as $e^\phi N_f \lesssim
1$ the branes behave as probes. In this regime, we expect the quantities
computable from the background to be equal to the KW model ones: in particular
$\gamma_A=-1/2$.

We can distinguish different regimes, starting from the UV to the IR. Depending
on the values of $N_c$ and $N_f$ they can be either well separated or not
present at all. A section of the space of couplings and some RG flows are drawn
in Figure \ref{flow KW flavors}, but one should include the third orthogonal
direction $h$ which is not plotted.

\begin{itemize}
\item For $1<e^\phi$ we are in the Landau pole regime, and the dilaton (string coupling $e^\phi$) is large.

\item For $\frac{1}{N_f}<e^\phi<1$ we are in a complicated piece of the flow, quite far from the KW fixed line, as in the type A-B flows of Figure \ref{flow KW flavors}. In particular the D7-branes are backreacting. In this regime our SUGRA solution is perfectly behaved (as long as $\frac{1}{N_c}<e^\phi$).

\item For $\frac{1}{N_c}<e^\phi<\frac{1}{N_f}$ (this regime exists for $N_f<N_c$) we are in a region with almost probe D7-branes%
\footnote{The dual in field theory of the D7's being probes is that graphs with flavors in the loops are suppressed with respect to gauge fields in the loops, since $N_f < N_c$.%
}, so we are close to the KW line, but with large 't Hooft coupling, so we can trust \eqref{orbifold form}. We can expect the energy/radius relation to be quite similar to the conformal one, thus we can compute the gauge \hyph{\beta}function and deduce the flavor anomalous dimensions $\gamma_Q$. Apart from corrections, we get:
\begin{equation}
\gamma_A \simeq -\frac{1}{2} \qquad R_A \simeq \frac{1}{2} \qquad\qquad\qquad
\gamma_Q \simeq \frac{1}{4} \qquad R_Q \simeq \frac{3}{4} \;.
\end{equation}
The R-symmetry is classically preserved but anomalous as in supergravity. The various \hyph{\beta}functions are computed to be
\begin{equation}
\beta_g = \frac{3}{4} N_f \frac{g^3}{16\pi^2} \qquad\qquad \beta_{\tilde\lambda} \simeq 0 \qquad\qquad \beta_h \simeq 0 \;.
\end{equation} 
We want to stress that this regime in {\it not} conformal, and in fact the theory flows along the KW fixed line, as in the type C flow of Figure \ref{flow KW flavors}. The smaller is $N_f/N_c$, the longer is this piece of the flow.
For $N_f\gtrsim N_c$ this regime does not exist, and the theory follows the type B flows of Figure \ref{flow KW flavors}.

\item For $e^\phi < \text{Min}(\frac{1}{N_c} , \frac{1}{N_f})$ we are close to the end of the KW fixed line, and the gauge coupling is close to $g_*$. Again the D7's are almost probes. The string frame curvature is large, as in the KW model at small $g_s N_c$. Since the gauge coupling cannot go below $g_*$, its \hyph{\beta}function vanishes even if the string coupling continues flowing to zero. We get in field theory:
\begin{gather}
\gamma_A \simeq -\frac{1}{2} \qquad R_A \simeq \frac{1}{2} \qquad\qquad\qquad
\gamma_Q \simeq 1 \qquad R_Q \simeq \frac{3}{4} \\
\beta_g \simeq 0 \qquad\qquad \beta_{\tilde\lambda} \simeq 0 \qquad\qquad \beta_h = \frac{3}{4} h \;.
\end{gather}
All the flows accumulate at the point $\{g=g_*,\tilde\lambda=0\}$ of Figure \ref{flow KW flavors}, but the theory is {\it not} conformal. In fact the coupling $h$ always flows to smaller values, and the theory moves ``orthogonal'' to the figure. For this reason $\gamma_Q$ and $R_Q$ do not satisfy the relation of superconformal theories.

\item The end of the flow is the superconformal point with $h=0$ (and $g=g_*$), which should correspond to $e^\phi=0$ and cannot be described by supergravity. Without the cubic superpotential one can construct a new anomaly free R-symmetry with $R_Q=1$, by combining the previous one ($R_Q=3/4$) with the anomalous axial symmetry which gives charge 1/4 to every flavor. This satisfies known theorems on superconformal theories. Moreover, the fact that $h\to 0$ in the far infrared realizes in field theory the incapability of resolving the D7 separation at small energies, and the flavor symmetry $S(U(N_f) \times U(N_f))$ is restored.
\end{itemize}

\begin{figure}[ht]
\begin{center}
\includegraphics[height=0.10\textheight]{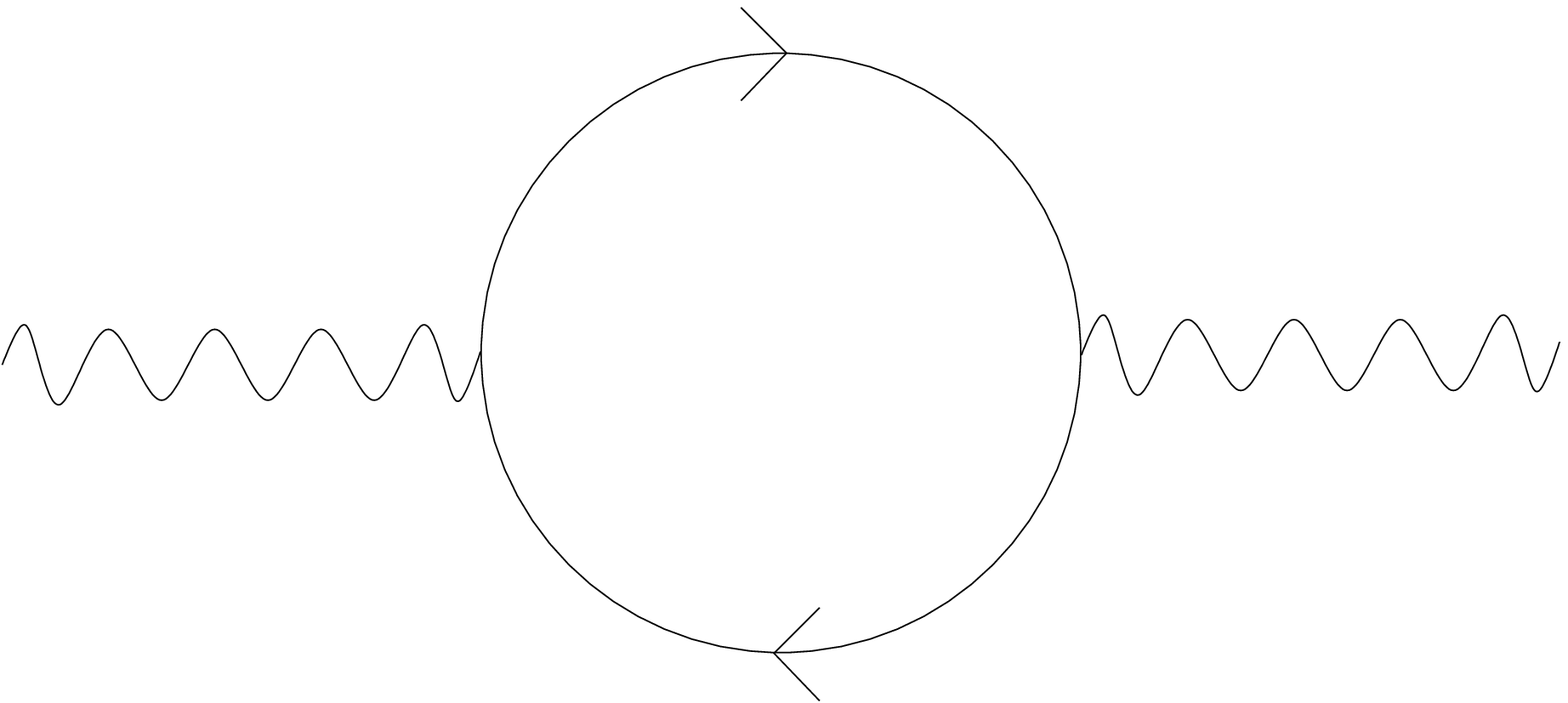}
\hspace{2cm}
\includegraphics[height=0.08\textheight]{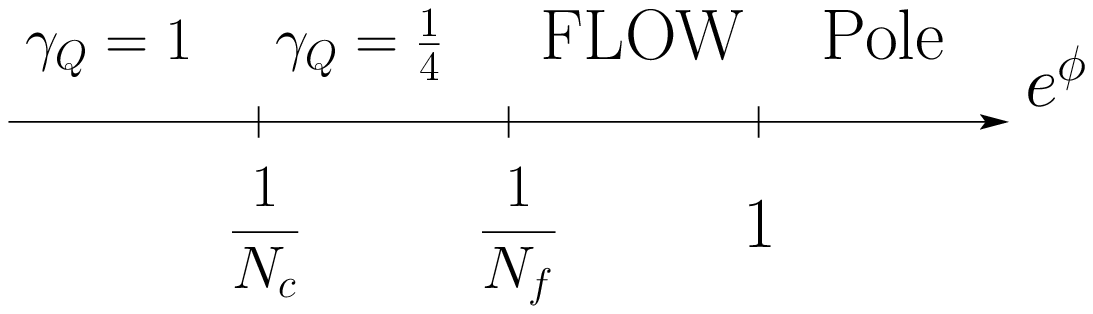} \
\parbox[t]{0.48\textwidth}{\caption{Flavor 1-loop correction to the gauge
propagator. \label{loop}}}
\hspace{0.02\textwidth}
\parbox[t]{0.48\textwidth}{\caption{Regimes of KW with flavors for $N_f<N_c$.
\label{phases}}}
\end{center}
\end{figure}

Note that when $N_f \gtrsim N_c$ and the D7-branes are probes (this is the
regime $e^\phi < \frac{1}{N_f} < \frac{1}{N_c}$ and $g=g_*$)
one could think hard to see in field theory a suppression of graphs with flavors
in the loops, with respect to gauge fields in the loops. Consider the gauge
propagator at 1-loop with flavors (Figure \ref{loop}). It is of order $g_*^2
N_f$, not suppressed with respect to the graph with gauge fields in the loop of
order $g_*^2 N_c$. But if we sum all the loops with flavors, we must obtain the
flavor contribution to the \hyph{\beta}function, which for $g \simeq g_*$ and
so $\gamma_Q \simeq 1$ is indeed very small.

A summary of the phase space for $N_f<N_c$ is in Figure \ref{phases}. The
computation in \cite{Ouyang:2003df} is valid in the region $\frac{1}{N_c}
<e^\phi< \frac{1}{N_f}$ of the phase space.


\newpage
\begin{thebibliography}{99}
\bibitem{Maldacena:1997re}
  J.~M.~Maldacena,
  ``{\it The large N limit of superconformal field theories and 
supergravity},''
  Adv.\ Theor.\ Math.\ Phys.\  {\bf 2}, 231 (1998)
  [Int.\ J.\ Theor.\ Phys.\  {\bf 38}, 1113 (1999)];
  hep-th/9711200.

\bibitem{Gubser:1998bc}
  S.~S.~Gubser, I.~R.~Klebanov and A.~M.~Polyakov,
  ``{\it Gauge theory correlators from non-critical string theory},''
  Phys.\ Lett.\ B {\bf 428}, 105 (1998);
  hep-th/9802109.

\bibitem{Witten:1998qj}
  E.~Witten,
  ``{\it Anti-de Sitter space and holography},''
  Adv.\ Theor.\ Math.\ Phys.\  {\bf 2}, 253 (1998);
  hep-th/9802150.

\bibitem{Itzhaki:1998dd}
  N.~Itzhaki, J.~M.~Maldacena, J.~Sonnenschein and S.~Yankielowicz,
  ``{\it Supergravity and the large N limit of theories with sixteen
  supercharges},''
  Phys.\ Rev.\ D {\bf 58}, 046004 (1998);
  hep-th/9802042.

\bibitem{Girardello:1999bd}
  L.~Girardello, M.~Petrini, M.~Porrati and A.~Zaffaroni,
  Nucl.\ Phys.\ B {\bf 569}, 451 (2000);
  hep-th/9909047.
  L.~Girardello, M.~Petrini, M.~Porrati and A.~Zaffaroni,
  JHEP {\bf 9905}, 026 (1999);
  hep-th/9903026.
  L.~Girardello, M.~Petrini, M.~Porrati and A.~Zaffaroni,
  JHEP {\bf 9812}, 022 (1998);
  hep-th/9810126.
  D.~Z.~Freedman, S.~S.~Gubser, K.~Pilch and N.~P.~Warner,
  Adv.\ Theor.\ Math.\ Phys.\  {\bf 3}, 363 (1999);
  hep-th/9904017.
  D.~Z.~Freedman, S.~S.~Gubser, K.~Pilch and N.~P.~Warner,
  JHEP {\bf 0007}, 038 (2000);
  hep-th/9906194.
  J.~Polchinski and M.~J.~Strassler,
  hep-th/0003136.
J.~Babington, D.~E.~Crooks and N.~J.~Evans,
  Phys.\ Rev.\ D {\bf 67}, 066007 (2003);
  hep-th/0210068.
J.~Babington, D.~E.~Crooks and N.~J.~Evans,
  JHEP {\bf 0302}, 024 (2003);
  hep-th/0207076.

\bibitem{Klebanov:1998hh}
  I.~R.~Klebanov and E.~Witten,
  ``{\it Superconformal field theory on threebranes at a Calabi-Yau  
singularity},
  Nucl.\ Phys.\ B {\bf 536}, 199 (1998)
  [arXiv:hep-th/9807080].


\bibitem{Klebanov:1999rd}
  I.~R.~Klebanov and N.~A.~Nekrasov,
  ``{\it Gravity duals of fractional branes and logarithmic RG flow},''
  Nucl.\ Phys.\ B {\bf 574}, 263 (2000)
  [arXiv:hep-th/9911096].

\bibitem{Klebanov:2000nc}
  I.~R.~Klebanov and A.~A.~Tseytlin,
  ``{\it Gravity duals of supersymmetric SU(N) $\times$ SU(N+M) gauge theories},''
  Nucl.\ Phys.\ B {\bf 578}, 123 (2000)
  [arXiv:hep-th/0002159].

\bibitem{Klebanov:2000hb}
  I.~R.~Klebanov and M.~J.~Strassler,
  ``{\it Supergravity and a confining gauge theory: Duality cascades and
  $\chi$SB-resolution of naked singularities},''
  JHEP {\bf 0008}, 052 (2000)
  [arXiv:hep-th/0007191].

\bibitem{Gubser:2004qj}
  S.~S.~Gubser, C.~P.~Herzog and I.~R.~Klebanov,
  {\it``Symmetry breaking and axionic strings in the warped deformed 
conifold},''
  JHEP {\bf 0409}, 036 (2004)
  [arXiv:hep-th/0405282].

\bibitem{Dymarsky:2005xt}
  A.~Dymarsky, I.~R.~Klebanov and N.~Seiberg,
  ``On the moduli space of the cascading SU(M+p) x SU(p) gauge theory,''
  JHEP {\bf 0601}, 155 (2006)
  [arXiv:hep-th/0511254].



\bibitem{Butti:2004pk}
  A.~Butti, M.~Grana, R.~Minasian, M.~Petrini and A.~Zaffaroni,
  ``{\it The baryonic branch of Klebanov-Strassler solution: A supersymmetric
  family of SU(3) structure backgrounds},''
  JHEP {\bf 0503}, 069 (2005)
  [arXiv:hep-th/0412187].

\bibitem{Strassler:2005qs}
  M.~J.~Strassler,
  ``{\it The duality cascade},''
  arXiv:hep-th/0505153.

\bibitem{Karch:2002sh}
  A.~Karch and E.~Katz,
  ``{\it Adding flavor to AdS/CFT},''
  JHEP {\bf 0206}, 043 (2002)
  [arXiv:hep-th/0205236].
  


\bibitem{Klebanov:2004ya}
  I.~R.~Klebanov and J.~M.~Maldacena,
  ``Superconformal gauge theories and non-critical superstrings,''
  Int.\ J.\ Mod.\ Phys.\ A {\bf 19}, 5003 (2004)
  [arXiv:hep-th/0409133].


                                                                              
\bibitem{Bigazzi:2005md}
  F.~Bigazzi, R.~Casero, A.~L.~Cotrone, E.~Kiritsis and A.~Paredes,
  ``{\it Non-critical holography and four-dimensional CFT's with 
fundamentals},''
  JHEP {\bf 0510}, 012 (2005)
  [arXiv:hep-th/0505140].

\bibitem{Casero:2006pt}
  R.~Casero, C.~Nunez and A.~Paredes,
  ``{\it Towards the string dual of N = 1 SQCD-like theories},''
  Phys.\ Rev.\ D {\bf 73}, 086005 (2006)
  [arXiv:hep-th/0602027].

\bibitem{Paredes:2006wb}
  A.~Paredes,
  ``{\it On unquenched N = 2 holographic flavor},''
  arXiv:hep-th/0610270.

\bibitem{Murthy:2006xt}
  S.~Murthy and J.~Troost,
  ``{\it D-branes in non-critical superstrings and duality in N = 1 gauge theories
  with flavor},''
  JHEP {\bf 0610}, 019 (2006)
  [arXiv:hep-th/0606203].



\bibitem{Arean:2004mm}
  D.~Arean, D.~E.~Crooks and A.~V.~Ramallo,
  ``{\it Supersymmetric probes on the conifold},''
  JHEP {\bf 0411}, 035 (2004)
  [arXiv:hep-th/0408210].

\bibitem{Ouyang:2003df}
  P.~Ouyang,
  ``{\it Holomorphic D7-branes and flavored N = 1 gauge theories},''
  Nucl.\ Phys.\ B {\bf 699}, 207 (2004)
  [arXiv:hep-th/0311084].
T.~S.~Levi and P.~Ouyang,
  ``{\it Mesons and flavor on the conifold},''
  arXiv:hep-th/0506021.

\bibitem{Kuperstein:2004hy}
  T.~Sakai and J.~Sonnenschein,
 ``Probing flavored mesons of confining gauge theories by supergravity,''
  JHEP {\bf 0309}, 047 (2003)
  [arXiv:hep-th/0305049].
S.~Kuperstein,
  ``{\it Meson spectroscopy from holomorphic probes on the warped deformed
  conifold},''
  JHEP {\bf 0503}, 014 (2005)
  [arXiv:hep-th/0411097].

\bibitem{Gursoy:2005cn}
  U.~Gursoy and C.~Nunez,
  ``{\it Dipole deformations of N = 1 SYM and supergravity backgrounds with 
  U(1) $\times$ U(1) global symmetry},''
  Nucl.\ Phys.\ B {\bf 725}, 45 (2005)
  [arXiv:hep-th/0505100].
R.~P.~Andrews and N.~Dorey,
  ``{\it Deconstruction of the Maldacena-Nunez compactification},''
  Nucl.\ Phys.\ B {\bf 751}, 304 (2006)
  [arXiv:hep-th/0601098].




\bibitem{'tHooft:1973jz}
  G.~'t Hooft,
  ``{\it A Planar Diagram Theory for Strong Interactions},''
  Nucl.\ Phys.\ B {\bf 72}, 461 (1974).

\bibitem{Veneziano:1976wm}
  G.~Veneziano,
  ``{\it Some Aspects Of A Unified Approach To Gauge, Dual And Gribov Theories},''
  Nucl.\ Phys.\ B {\bf 117}, 519 (1976).


\bibitem{Greene:1989ya}
  B.~R.~Greene, A.~D.~Shapere, C.~Vafa and S.~T.~Yau,
  ``{\it Stringy Cosmic Strings And Noncompact Calabi-Yau Manifolds},''
  Nucl.\ Phys.\ B {\bf 337}, 1 (1990).

\bibitem{Aharony:1998xz}
  O.~Aharony, A.~Fayyazuddin and J.~M.~Maldacena,
  ``{\it The large N limit of N = 2,1 field theories from three-branes in
  F-theory},''
  JHEP {\bf 9807}, 013 (1998)
  [arXiv:hep-th/9806159].


\bibitem{Grana:2001xn}
  M.~Grana and J.~Polchinski,
  ``{\it Gauge / gravity duals with holomorphic dilaton},''
  Phys.\ Rev.\ D {\bf 65}, 126005 (2002)
  [arXiv:hep-th/0106014].



\bibitem{Bertolini:2001qa}
  M.~Bertolini, P.~Di Vecchia, M.~Frau, A.~Lerda and R.~Marotta,
  ``{\it N = 2 gauge theories on systems of fractional D3/D7 branes},''
  Nucl.\ Phys.\ B {\bf 621}, 157 (2002)
  [arXiv:hep-th/0107057].



\bibitem{Burrington:2004id}
  B.~A.~Burrington, J.~T.~Liu, L.~A.~Pando Zayas and D.~Vaman,
  ``{\it Holographic duals of flavored N = 1 super Yang-Mills: Beyond the 
  probe approximation},''
  JHEP {\bf 0502}, 022 (2005)
  [arXiv:hep-th/0406207].


\bibitem{Kirsch:2005uy}
  I.~Kirsch and D.~Vaman,
  ``{\it The D3/D7 background and flavor dependence of Regge trajectories},''
  Phys.\ Rev.\ D {\bf 72}, 026007 (2005)
  [arXiv:hep-th/0505164].

\bibitem{Cherkis:2002ir}
  S.~A.~Cherkis and A.~Hashimoto,
  JHEP {\bf 0211}, 036 (2002)
  [arXiv:hep-th/0210105].
  B.~S.~Acharya, F.~Denef, C.~Hofman and N.~Lambert,
  [arXiv:hep-th/0308046].
M.~Gomez-Reino, S.~Naculich and H.~Schnitzer,
  Nucl.\ Phys.\ B {\bf 713}, 263 (2005)
  [arXiv:hep-th/0412015].
R.~Clarkson, A.~M.~Ghezelbash and R.~B.~Mann,
  JHEP {\bf 0404}, 063 (2004)
  [arXiv:hep-th/0404071].
R.~Clarkson, A.~M.~Ghezelbash and R.~B.~Mann,
  JHEP {\bf 0408}, 025 (2004)
  [arXiv:hep-th/0405148].
A.~M.~Ghezelbash and R.~B.~Mann,
  JHEP {\bf 0410}, 012 (2004)
  [arXiv:hep-th/0408189].
  J.~Erdmenger and I.~Kirsch,
  JHEP {\bf 0412}, 025 (2004)
  [arXiv:hep-th/0408113].
H.~J.~Schnitzer,
 [ arXiv:hep-th/0612099].
  
    
  
\bibitem{Candelas:1989js}
  P.~Candelas and X.~C.~de la Ossa,
  ``{\it Comments on conifolds},''
  Nucl.\ Phys.\ B {\bf 342}, 246 (1990).
  
  
  


\bibitem{PandoZayas:2001iw}
  L.~A.~Pando Zayas and A.~A.~Tseytlin,
  ``{\it 3-branes on spaces with R $\times$ S(2) $\times$ S(3) topology},''
  Phys.\ Rev.\ D {\bf 63}, 086006 (2001)
  [arXiv:hep-th/0101043].
 
\bibitem{Benvenuti:2005qb}
  S.~Benvenuti, M.~Mahato, L.~A.~Pando Zayas and Y.~Tachikawa,
  ``{\it The gauge / gravity theory of blown up four cycles},''
  arXiv:hep-th/0512061.

\bibitem{Morrison:1998cs}
  D.~R.~Morrison and M.~R.~Plesser,
  ``{\it Non-spherical horizons. I},''
  Adv.\ Theor.\ Math.\ Phys.\  {\bf 3}, 1 (1999)
  [arXiv:hep-th/9810201].

\bibitem{Maldacena:2000mw}
  J.~M.~Maldacena and C.~Nunez,
  ``{\it Supergravity description of field theories on curved manifolds and a no  go
  theorem},''
  Int.\ J.\ Mod.\ Phys.\ A {\bf 16}, 822 (2001)
  [arXiv:hep-th/0007018].

\bibitem{Olesen:2002nh}
  P.~Olesen and F.~Sannino,
  ``{\it N = 1 super Yang-Mills from supergravity: The UV-IR connection},''
  arXiv:hep-th/0207039.
  

\bibitem{Herzog:2002ih}
  C.~P.~Herzog, I.~R.~Klebanov and P.~Ouyang,
  ``{\it D-branes on the conifold and N = 1 gauge / gravity dualities},''
  arXiv:hep-th/0205100.
  
  
\bibitem{Skenderis:2006di}
  K.~Skenderis and M.~Taylor,
  ``{\it Holographic Coulomb branch vevs},''
  JHEP {\bf 0608}, 001 (2006)
  [arXiv:hep-th/0604169].
  
  
\bibitem{Bertolini:2002yr}
  M.~Bertolini and P.~Merlatti,
  ``{\it A note on the dual of N = 1 super Yang-Mills theory},''
  Phys.\ Lett.\ B {\bf 556}, 80 (2003)
  [arXiv:hep-th/0211142].

  
  

\bibitem{Grana:2005jc}
  M.~Grana,
  ``{\it Flux compactifications in string theory: A comprehensive review},''
  Phys.\ Rept.\  {\bf 423}, 91 (2006)
  [arXiv:hep-th/0509003].
  
  
  
\bibitem{Gomis:2006sb}
  J.~Gomis and F.~Passerini,
  JHEP {\bf 0608}, 074 (2006)
  [arXiv:hep-th/0604007].
  J.~Gomis and C.~Romelsberger,
  JHEP {\bf 0608}, 050 (2006)
  [arXiv:hep-th/0604155].
  O.~Lunin,
  JHEP {\bf 0606}, 026 (2006)
  [arXiv:hep-th/0604133].
  S.~Yamaguchi,
  JHEP {\bf 0605}, 037 (2006)
  [arXiv:hep-th/0603208].
  O.~Lunin and J.~M.~Maldacena,
  JHEP {\bf 0505}, 033 (2005)
  [arXiv:hep-th/0502086].
  
  

\bibitem{SUSYIIB}
  J.~H.~Schwarz,
  ``{\it Covariant Field Equations Of Chiral N=2 D = 10 Supergravity},''
  Nucl.\ Phys.\ B {\bf 226}, 269 (1983).

\bibitem{Martucci}
  L.~Martucci, J.~Rosseel, D.~Van den Bleeken and A.~Van Proeyen,
  ``{\it Dirac actions for D-branes on backgrounds with fluxes},''
  Class.\ Quant.\ Grav.\  {\bf 22}, 2745 (2005)
  [arXiv:hep-th/0504041].

  
\bibitem{Strominger}
  A.~Strominger,
  ``{\it Superstrings with Torsion},''
  Nucl.\ Phys.\ B {\bf 274}, 253 (1986).


\bibitem{Intriligator:2003jj}
  K.~Intriligator and B.~Wecht,
  ``{\it The exact superconformal R-symmetry maximizes a},''
  Nucl.\ Phys.\ B {\bf 667}, 183 (2003)
  [arXiv:hep-th/0304128].
  
  
\bibitem{Kutasov:2003ux}
  D.~Kutasov,
  ``{\it `New results on the 'a-theorem' in four dimensional supersymmetric field
  theory},''
  arXiv:hep-th/0312098.

\bibitem{Benvenuti:2005wi}
  S.~Benvenuti and A.~Hanany,
  JHEP {\bf 0508}, 024 (2005)
  [arXiv:hep-th/0502043].
  S.~Benvenuti and M.~Kruczenski,
  JHEP {\bf 0610}, 051 (2006)
  [arXiv:hep-th/0505046].
  

  
\end{thebibliography}
\end{document}